\documentclass[11pt]{article}
\pdfoutput=1

\usepackage{jheppub}

\usepackage{amsmath, amssymb} 
\usepackage{mathtools}
\usepackage{esint}
\usepackage{bm}
\usepackage{dsfont}
\usepackage{braket}
\usepackage{eqparbox}
\usepackage{enumitem}
\hypersetup{
	colorlinks,
	urlcolor=Maroon,
	linkcolor=Maroon,
	citecolor=Maroon
	}

\numberwithin{equation}{section}

\usepackage[dvipsnames,table]{xcolor}
\usepackage{soul} 
\usepackage{calc}
\usepackage{subcaption}
\usepackage{tikz}
\usepackage{tkz-euclide} 
\usetikzlibrary{arrows.meta}
\tikzset{hidden/.style = {thick, dashed}}
\usepackage{scalefnt}
\usepackage[font=small,labelfont=bf]{caption}
\usepackage{graphicx}

\usepackage[numbers, sort&compress]{natbib}

\usepackage{comment} 
\usepackage[colorinlistoftodos]{todonotes}

\usepackage{hyperref}
\hypersetup{colorlinks=true, linktoc=page, linkcolor=Maroon, citecolor=Maroon}

\newcommand{\<}{\langle}
\renewcommand{\>}{\rangle}
\newcommand{\Tr}{\text{Tr}}

\newcommand{\be}{\begin{equation}}
\newcommand{\ee}{\end{equation}}
\newcommand{\bea}{\begin{eqnarray}}
\newcommand{\eea}{\end{eqnarray}}
\newcommand{\reef}[1]{(\ref{#1})}

\newcommand{\ca}{\mathcal{A}}
\newcommand{\pa}{\partial}
\newcommand{\eps}{\epsilon}
\newcommand{\lra}{\leftrightarrow}
\usetikzlibrary{decorations}

\usepackage{multirow}

\title{String Theory from Maximal Supersymmetry}

\date{\today}

\author[a]{Henriette Elvang,}
\author[b]{Aidan Herderschee,}
\author[a]{Roger Morales}

\affiliation[a]{
    Leinweber Institute for Theoretical Physics, Randall Laboratory of Physics\\
    University of Michigan, Ann Arbor\\
    450 Church St, Ann Arbor, MI 48109-1040, USA}
\affiliation[b]{Institute for Advanced Study\\
Einstein Drive,
Princeton, NJ 08540, USA}

\emailAdd{elvang@umich.edu}
\emailAdd{aidanh@ias.edu}
\emailAdd{rmespasa@umich.edu}

\preprint{LITP-25-18}

\abstract{
We explore the space of non-gravitational, maximally supersymmetric, planar 4d effective field theories (EFTs) that have $\mathcal{N}=4$ super Yang–Mills (SYM) at leading order. In the weakly-coupled regime, we examine the combined constraints of $\mathcal{N}=4$ supersymmetry and $SU(4)$ R-symmetry together with the requirement of standard tree-level factorization on the massless poles of the  4-, 5-, and 6-point EFT scattering amplitudes.
When imposing a parity condition on the 6-scalar amplitudes, we find highly nontrivial nonlinear constraints on the 4-point Wilson coefficients. When these novel constraints are combined with positivity, the resulting bounds on the 4-point Wilson coefficients converge to the values of the open string Veneziano amplitude. Our results strongly suggest that supersymmetry, R-symmetry,  
the parity requirement on the 6-scalar amplitudes, and positivity are sufficient to single out this unique UV completion at tree level. Our findings, moreover, highlight the power of higher-point amplitudes in constraining EFT data and imply that the space of consistent quantum field theories is even more restricted than previously suggested by causality or swampland-based approaches.
}

\begin{document}

\maketitle

\addtocontents{toc}{\protect\setcounter{tocdepth}{2}}

\newpage 
\section{Introduction}
\label{sec:intro}

The computational tractability and phenomenological relevance make weakly coupled effective field theories (EFTs) a fruitful starting point for studying the rich landscape of quantum field theories. 
The standard lore of EFTs instructs us to include all local higher-derivative operators compatible with the symmetries, up to a given order in the derivative expansion, and to equip the operators with couplings (the Wilson coefficients) that, a priori, can take any values. The Wilson coefficients encode the unknown ultraviolet (UV) physics --- and, conversely, any known UV physics has a particular imprint on the Wilson coefficients.  
From the low-energy perspective alone, the space of EFTs may seem vast. However, it has become increasingly clear over the past two decades that the space of EFTs is far more constrained than once believed. 
From a top-down perspective, insights from the Swampland program indicate that the subset of gravitational EFTs that admit a consistent UV completion is dramatically smaller than naively expected \cite{Vafa:2005ui,Ooguri:2006in,Arkani-Hamed:2006emk}. 
From a bottom-up perspective, basic principles like unitarity and analyticity imply powerful two-sided bootstrap bounds on Wilson coefficients of both gravitational and non-gravitational EFTs \cite{Adams:2006sv,Arkani-Hamed:2020blm,Caron-Huot:2020cmc,Tolley:2020gtv}. 
Despite these constraints, the remaining parameter space is still very large and an improved understanding of additional fundamental restrictions is necessary to isolate EFTs with consistent physical UV completions.

In this paper, we restrict our attention to non-gravitational, weakly-coupled maximally supersymmetric 4d planar EFTs that reduce to $\mathcal{N}=4$ super Yang–Mills (SYM) in the leading-order low-energy limit. 
Our analysis has two parts: 
\begin{itemize}
\item In part 1, we work purely within the low-energy EFT without making assumptions about the properties, or even the existence, of a UV completion. Working order-by-order in the low-energy expansion, we solve the combined constraints of  $\mathcal{N}=4$ supersymmetry (SUSY) and $SU(4)$ R-symmetry, assuming that the single-trace $4$-, $5$- and $6$-point EFT amplitudes have only standard tree-level factorization on massless poles. When combined with the  condition that 6-scalar amplitudes are parity-even (see more details in Section \ref{intro:New_constraints}), we find novel and highly restrictive {\em nonlinear relations among the Wilson coefficients} associated with the EFT operators of the form $\Tr D^{2k}F^4$. 
\item In part 2, we combine these nonlinear constraints with standard S-matrix bootstrap 
assumptions on the 4-point amplitude: unitarity, analyticity, and that $A_4/s^2$ goes to zero at large $|s|$ for fixed $u<0$ or $t<0$. The numerical bootstrap bounds restrict the Wilson coefficients to a non-convex region that converges towards a single curve in the space of Wilson coefficients.  The numerics strongly indicate that the bounds will converge towards a unique solution, namely the open superstring 4-point tree amplitude, i.e.~the Veneziano amplitude~\cite{Veneziano:1968yb} with kinematic prefactors appropriate for four-gluon scattering. The curve is parameterized by the string-tension  $\alpha'$ relative to the assumed (arbitrary, but non-zero) gap in the bootstrap spectrum.
\end{itemize}
In part 1 we carry out a bottom-up construction of a particular 6-scalar NMHV amplitude in the general EFT context and subject it to a SUSY Ward identity that relates it to its five cyclic permutations. Every pole term in the 6-scalar amplitude is a function of Mandelstam variables only, so the additional assumption that the scalar amplitude is parity-even comes down to the absence of local contact terms involving Levi–Civita contractions, e.g.~$\eps_{\mu\nu\rho\sigma} p_i^\mu p_j^\nu p_k^\rho p_l^\sigma$. When such local terms are present, our SUSY Ward identity does not give rise to nonlinear constraints, however, excluding the parity-odd terms imposes nonlinear constraints on the 4-point Wilson coefficients.

The combined results of part 1 and 2 is numerical evidence that the open superstring tree-level amplitude is the only unitary UV completion of a {\em tree-level} $\,\mathcal{N}=4$ SYM EFT with the parity-assumption. We compare our analysis with other recent bootstraps of the Veneziano amplitude in Section \ref{sec:compare}. Another possible UV completion of the $\mathcal{N}=4$ EFT is the Coulomb branch of $\mathcal{N}=4$ SYM. The leading correction to its 4-point amplitude comes from a 1-loop box of massive $W$-multiplets, and the resulting Wilson coefficients are incompatible with our nonlinear constraints; consistency therefore requires the 6-scalar amplitudes of the Coulomb branch to contain Levi–Civita-bearing local terms, so the parity-even assumption is violated and our nonlinear constraints do not apply.

We now provide an overview of the analysis and results in the two parts of the paper.

\subsection{Part 1: New Constraints on Wilson Coefficients from Maximal SUSY}
\label{intro:New_constraints}

Consider a 4d gluon EFT  with a standard kinetic term $\Tr F^2$. The single-trace higher-derivative terms in the Lagrangian can be written schematically as\footnote{We are not including operators of the form
$\Tr D^{2k}F^3$ with $k\ge 1$   because they can be converted to operators with four or more fields via field redefinitions. }
\bea
 \nonumber
 \mathcal{L}&=&  
 -\frac{1}{4} \Tr F^2 + \Tr F^3 + \Tr F^4 + \Tr D^2 F^4 + \Tr D^4 F^4 + \Tr D^6 F^4 + \dots
 \\
 \nonumber
 \text{number of operators} 
 &=&
 \hspace{2cm}~1 \hspace{1.2cm}3
 \hspace{1.4cm}5 \hspace{1.6cm}8 \hspace{1.6cm}10
 \\
 \label{gluonEFT}
 \text{with lin. } \mathcal{N}=4 \text{ SUSY} 
 &=& 
 ~ \hspace{2cm} 0 \hspace{1.2cm}1
 \hspace{1.4cm}1 \hspace{1.6cm}2 \hspace{1.6cm}~2
 \\[-1mm]
 \nonumber
  \text{effective couplings}
  &=&  ~ 
  \hspace{3.2cm} a_{0,0} \hspace{1cm}a_{1,0}
 \hspace{1.2cm}a_{2,0} \hspace{1.3cm}a_{3,0}
 \hspace{1.6cm}
 \\[-2mm]
   \nonumber
\phantom{\text{eff.~couplings}}
&\phantom{=}& \!\!\!   
\hspace{6.95cm} a_{2,1} \hspace{1.3cm}a_{3,1}
\hspace{1.6cm}
\eea
The ``+ \dots" in the EFT Lagrangian stands for operators of the form $\Tr D^{2k}F^n$ with $n \ge 4$. The rows below the  Lagrangian list the number of operators of the form $\Tr D^{2k}F^4$ that are independent under the use of equations of motion, the Bianchi identity, and integration by parts. In the first line, we assume no symmetries beyond Lorentz invariance, whereas in the second line the constraints of linearized $\mathcal{N}=4$ SUSY are imposed; for example, $ \Tr F^3$ is incompatible with any amount of SUSY, and there is only one linear combination of the three independent contractions of $\Tr F^4$ that is compatible with $\mathcal{N}=4$ SUSY. 
In the third row of \reef{gluonEFT}, we introduce the Wilson coefficients $a_{k,q}$ associated with each of the  $\mathcal{N}=4$ SUSY four-field operators $\Tr D^{2k}F^4$ (and all the other four-field operators in its SUSY orbit). 
Thus, $k$ tracks the derivative order and 
the number of  independent operators for given $k$ is labeled by $q=0,1,\dots,\lfloor k/2\rfloor$. 

The precise definition of the Wilson coefficients $a_{k,q}$ is through their contribution to the 4-point gluon amplitude. We work in the planar limit (i.e.~the large-$N$ limit, where $N$ is the rank of the gauge group), so the amplitudes can be decomposed into a sum over single traces of the generators. The coefficient of each single-trace structure is a color-ordered amplitude $A_n[1234\dots n]$. 
At 4 points, the SUSY Ward identities \cite{Grisaru:1976vm,Grisaru:1977px,Bianchi:2008pu,Elvang:2013cua} allow only the Maximally-Helicity-Violating (MHV) amplitudes with two positive and two negative helicity states,\footnote{Throughout the paper, all external momenta $p_i$ are outgoing and Mandelstam variables are defined as  $s_{i_1 \, \cdots \, i_k} \equiv (p_{i_1} + \dots + p_{i_k})^2$. For the 4-point scattering, we use the convention $s \equiv s_{12}=\<12\>[12]$, $t \equiv s_{13}=\<13\>[13]$ and $u \equiv s_{14}=\<14\>[14]$ with $s+t+u=0$.}
and they require the color-ordered single-trace MHV amplitude 
to take the form
\be
  \label{A4intro}
  A_4[--++] = \<12\>^2 [34]^2 F(s,u)\,,~~~~F(u,s) = F(s,u) \,.
\ee
The symmetry in $s \leftrightarrow u$ follows from cyclicity and maximal SUSY. The most general form of the low-energy expansion is
\be\label{4pamp}
  F(s,u) = -\frac{g^2}{su} + \sum_{0 \leq q \leq k} a_{k,q} \,s^{k-q} u^q\,,  ~~~~\text{with}~~~a_{k,k-q} = a_{k,q}\,,
\ee
where the crossing condition on the $a_{k,q}$'s is needed for $F(u,s) = F(s,u)$. The pole term $-\<12\>^2 [34]^2/(su)$ 
can be rewritten as the Parke–Taylor form of the gluon scattering amplitude in pure Yang–Mills theory; $g$ is the Yang–Mills coupling. The independent SUSY operators $\Tr D^{2k}F^4$ are in one-to-one correspondence with the matrix elements of the 4-point amplitude. This is how we count the operators in the second row of \reef{gluonEFT}. For example, $a_{2,0}$ and $a_{2,1}$ are the Wilson coefficients  of the two independent $\mathcal{N}=4$ SUSY-compatible operators of the form $\Tr D^4 F^4$.  
Apart from requiring $a_{k,k-q} = a_{k,q}$, the $\mathcal{N}=4$ SUSY Ward identities at 4 point do not restrict the values of the effective couplings $a_{k,q}$. 

The 5- and 6-point tree-level EFT amplitudes depend on the 4-point Wilson coefficients $a_{k,q}$ through tree-level factorization diagrams such as  
\be
\label{samplediagrams}
\raisebox{-4.9mm}{\includegraphics[width=0.6\textwidth]{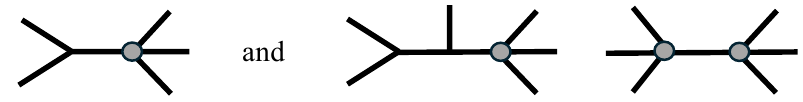}}
\ee
We construct the 5-point EFT amplitude using a manifestly $\mathcal{N}=4$ SUSY ansatz and impose the correct factorization on all the 2-particle poles (e.g.~as in the first diagram in  \reef{samplediagrams}) order-by-order in the Mandelstam expansion. The only parameters left in the ansatz are Wilson coefficients of local contact terms corresponding to operators $\Tr D^{2k}F^5$ that are independently solving the 5-point SUSY Ward identities. The same procedure can be applied to the 6-point MHV EFT amplitude, which has factorizations such as in the first 6-point diagram in \reef{samplediagrams}, but not of the second type. The 5- and 6-point MHV EFT amplitudes do not constrain the 4-point Wilson coefficients $a_{k,q}$, but both depend linearly on them. 

More non-trivial, and hence more interesting, is the 6-point NMHV sector. In this case, there is no manifestly SUSY ansatz available, so we need to construct the NMHV EFT amplitude from the bottom up by ensuring that it has the correct residue in every factorization channel and that all possible non-pole terms are parameterized with arbitrary coefficients. This process is significantly simpler to carry out using an amplitude with only external scalars and a small number of factorization channels. The amplitude we focus on is
\begin{equation}
  \label{Z1intro}
  \mathbb{Z}_1 
  \equiv 
  A_{6}\big[z_1 z_2 
   z_3 \bar{z}_3 
  \bar{z}_1 \bar{z}_2 \big]\,.
\end{equation}
The scalars $z_i$ and their conjugates $\bar{z}_i$ are the three complex scalars of the massless $\mathcal{N}=4$ supermultiplet; specifically, we choose
\be
  \label{defz1z2z3}
  \begin{split}
  &
  z_1 = z^{12}\,,\qquad
  z_2 = z^{13}\,,\qquad
  z_3 = z^{14}\,,\qquad
  \\
  &
  \bar{z}_1 = z^{34}\,,\qquad
  \bar{z}_2 = z^{24}\,,\qquad
  \bar{z}_3 = z^{23}\,
  .\qquad
  \end{split}
\ee
The scalars $z^{ab}$ transform in the 2-index antisymmetric representation of the $SU(4)$ R-symmetry, which we assume to be unbroken. 
The amplitude~\reef{Z1intro} has poles only in the 34-, 234-, and 345-channels, which makes it easy to construct directly from its factorization channels and a systematic parametrization of its local polynomial terms. 

\vspace{2mm}
\noindent {\bf Parity Condition.} At this point, it is useful to describe the parity condition in more detail. 
The pole terms 
of the amplitude $\mathbb{Z}_1$ depend only on Mandelstam variables, but the contact terms 
are polynomial in the Mandelstam variables and may or may not include contractions of the momenta with Levi–Civita symbols.

The operation of parity on an on-shell massless amplitude can be described in 4d spinor-helicity as the combination of a helicity flip (including $z \lra \bar{z}$) and the exchange of angle and square spinor brackets. Thus, under parity $\mathbb{Z}_1$ goes to 
\begin{equation}
  \overline{\mathbb{Z}}_1 
  \equiv 
  A_{6}\big[\bar{z}_1 
  \bar{z}_2 \bar{z}_3 z_3 z_1 
   z_2  \big]\,,
\end{equation}
and requiring that this leaves the amplitude unchanged is imposing that 
\be  
  \label{Z1isbarZ1cond}
\overline{\mathbb{Z}}_1
=\mathbb{Z}_1 \,.
\ee
Restoring 
the $SU(4)$ indices via \reef{defz1z2z3}, one finds that \reef{Z1isbarZ1cond} is not ensured by R-symmetry. 

It is in fact very simple to see what the condition \reef{Z1isbarZ1cond} imposes on $\mathbb{Z}_1$. 
Under the exchange of angle and square spinor brackets, the Mandelstams are invariant (``parity-even''), but the Levi–Civita contractions flip sign (``parity-odd''). Hence, the condition \reef{Z1isbarZ1cond} is exactly the requirement that the scalar amplitude is free of parity-odd terms, i.e.~that there are no Levi–Civita contractions in the ansatz for the local contact terms of $\mathbb{Z}_1$. 

It is in the above sense that we are imposing a certain parity condition on the 6-scalar amplitude $\mathbb{Z}_1$. As it turns out, once SUSY is imposed (see below), any other $\mathcal{N}=4$ SYM EFT 6-scalar amplitude is also parity-even. In fact, 
certain other identities then 
also hold for the scalar sector, as we now motivate. 

Since $SU(4) \sim SO(6)$, the scalars of the massless $\mathcal{N}=4$ supermultiplet can be described as complex scalars $z^{ab}$ transforming in the 2-index anti-symmetric representation of $SU(4)$, as above, or they can be equivalently regrouped into 6 real scalars $\phi_I$ transforming in the fundamental of $SO(6)$. The relation between the $z^{ab}$ and $\phi_I$ can be realized for example as
\be
  \begin{split}
  &z_1 
   = z^{12} 
   = \phi_1 + i \phi_2\,,
  ~~~~
  \bar{z}_1 
   = z^{34} 
   = \phi_1 - i \phi_2\,,\\
  &z_2 
   = z^{13} 
   = \phi_3 + i \phi_4\,,
  ~~~~
  \bar{z}_2
   = z^{24} 
   = \phi_3 - i \phi_4\,,\\
  &z_3 
   = z^{14} 
   = \phi_5 + i \phi_6\,,
  ~~~~
  \bar{z}_3
   = z^{23} 
   = \phi_5 - i \phi_6\,.
   \end{split}
\ee
When we require $\overline{\mathbb{Z}}_1
=\mathbb{Z}_1$, we are requiring invariance under 
$(z_1,z_2,z_3) \lra (\bar{z}_1,\bar{z}_2,\bar{z}_3)$, i.e.~
\be
(\phi_1,\phi_2,\phi_3,\phi_4,\phi_5,\phi_6 )
~~\to~~ 
(\phi_1,-\phi_2,\phi_3,-\phi_4,\phi_5,-\phi_6 )\,,
\ee  
which is clearly not an $SO(6)$ transformation, 
but rather a discrete transformation that extends it to $O(6)$ for the pure scalar sector.
For this to be true, 
our 6-scalar amplitude should also be invariant under any sign flip, e.g.~$\phi_2 \to -\phi_2$, which translates to invariance under $z_1 \lra \bar{z}_1$.
We have checked that with the parity-assumption and supersymmetry imposed, $\mathbb{Z}_1$ is indeed invariant under the exchange of any pair of conjugate scalars.

To summarize, we have given three different descriptions of the condition imposed on the ansatz for the local terms of the 6-scalar amplitude $\mathbb{Z}_1$: (1) absence of parity-odd terms, (2) 
imposing 
$\overline{\mathbb{Z}}_1
=\mathbb{Z}_1$, and (3) extending $SO(6)$ to $O(6)$ for the pure scalar amplitudes.

\vspace{2mm}
\noindent {\bf Imposing SUSY.} The construction of $\mathbb{Z}_1$ from its factorization structure does not ensure compatibility with SUSY; after all, we have simply constructed a particular component NMHV amplitude and SUSY relates it to other NMHV amplitudes. Thus, to impose the $\mathcal{N}=4$ SUSY Ward identities, we use the NMHV superamplitude  \cite{Elvang:2009wd,Elvang:2010xn} to derive a six-term NMHV SUSY Ward identity that relates $\mathbb{Z}_1$ to its five other cyclic permutations. The latter are obtained simply by momentum relabelings of $\mathbb{Z}_1$, so this does not introduce any new parameters. The 6-term SUSY Ward identity is imposed order-by-order in the low-energy expansion. This fixes a large number of the contact terms in the ansatz for the scalar amplitude. However, when we impose that there are no parity-odd terms in $\mathbb{Z}_1$ 
(or, equivalently, that $\mathbb{Z}_1=\overline{\mathbb{Z}}_1$), then 
starting at order $s^3$ in the EFT expansion, the NMHV SUSY Ward identity is only satisfied when the 4-point Wilson coefficients $a_{k,q}$ 
obey a strikingly restrictive set of {\em nonlinear} relationships.\footnote{Including the parity-odd terms in the ansatz, we find that SUSY Ward identities can  be solved without constraints among the 4-point Wilson coefficients.} The nonlinearities arise from the 3-particle factorizations such as the last diagram in  \reef{samplediagrams}, which give contributions to the 6-point amplitude that are quadratic in the $a_{k,q}$'s. 
At the lowest orders, these nonlinear relations are
\be 
  \label{nonlinIntro}
  g^2 a_{2,0} = \frac{2}{5} a_{0,0}^2\,,~~~~
  g^2 a_{2,1} = \frac{1}{10} a_{0,0}^2\,,~~~~
  g^2  a_{3,1} = 2g^2 a_{3,0} - a_{1,0} a_{0,0}\,, \ \dots\ 
\ee
with $g$ the Yang–Mills coupling. Continuing to higher order, up to and including $k=7$, we find that all $a_{k,q}$'s are fixed in terms of $a_{0,0}$ and $a_{2m-1,0}$'s.

We emphasize that these nonlinear constraints are derived in a purely bottom-up EFT approach where the only input is $\mathcal{N}=4$ SUSY in 4d, $SU(4)$ R-symmetry, tree-level factorization of the on-shell EFT scattering amplitudes (i.e.~we explicitly exclude cut contributions from loops, etc), and the assumption of parity for $\mathbb{Z}_1$. 
Contrasting standard EFT lore, the relations \reef{nonlinIntro} say that the Wilson coefficients $a_{2,0}$ and $a_{2,1}$ of the two local operators $\Tr D^4 F^4$ cannot be chosen independently from the coupling $a_{0,0}$ of $\Tr F^4$: $a_{2,0}$ and $a_{2,1}$ are exactly fixed in terms of $a_{0,0}^2$ and the YM coupling $g$. Thus, the nonlinear SUSY relations greatly reduce the number of free Wilson coefficients in the EFT; moreover, we also find that the coefficients of several of the lowest orders of the $\Tr D^{2k} F^5$ and $\Tr D^{2k} F^6$ operators are fixed in terms of the 4-point Wilson coefficients $a_{k,q}$. 

While SUSY is easy to impose at the level of the superamplitudes, the way the NMHV SUSY Ward identities relate individual component amplitudes is rather intricate. Imposing the Ward identity relating $\mathbb{Z}_1$ to its cyclic permutations is certainly necessary for $\mathcal{N}=4$ SUSY, but it is not obvious why it should be sufficient. However, for the 4-point Wilson coefficients $a_{k,q}$, the nonlinear relations we derive {\em are} a maximal set of algebraic constraints that can be derived from the basic assumptions of maximal SUSY, tree-factorization, and the parity condition \reef{Z1isbarZ1cond}. The reason for this is found by comparing to the open string, which in the low-energy expansion has tree-level amplitudes in the class of EFTs we are considering.  
Specifically, the $\alpha'$-expansion of the 4-gluon Veneziano amplitude~\cite{Veneziano:1968yb}
\be\label{eq:venezianointro}
A_4^{\text{str}}[--++] =
-g^{2}\alpha'^{2} \<12\>^2 [34]^2 \frac{
\Gamma(-\alpha' s)
\Gamma(-\alpha'u)}
{\Gamma\big(1-\alpha'(s+u)\big)} 
\ee
is of the form \reef{A4intro}-\reef{4pamp} with Wilson coefficients
\be
\label{eq:4pt_Wilson_coeffs_Veneziano}
  \begin{split}
  \text{Veneziano:}~~~
   &a_{0,0} = g^{2}\zeta_2 \,, 
   ~~~
   a_{1,0} = g^{2}\zeta_3 \,,
   ~~~a_{2,0} = g^{2}\zeta_4  
  \,, ~~~~
   a_{2,1} =\frac{g^{2}}{4} \zeta_4 
   \,, ~~~~
   \\[1.5mm]
   &
   a_{3,0} = g^{2}\zeta_5 \,,~~~ a_{3,1} = 2 g^{2}\zeta_5 - g^{2} \zeta_2 \zeta_3 \,, \ \dots .
  \end{split}
\ee
For simplicity, we have set $\alpha'=1$ above. The coefficients are written in terms of the Riemann zeta function $\zeta_b = \sum_{n=1}^\infty {1/n^b}$. Using that $\zeta_2 = \pi^2/6$ and $\zeta_4 = \pi^4/90$, one verifies straightforwardly 
that the coefficients in  \reef{eq:4pt_Wilson_coeffs_Veneziano} satisfy the EFT nonlinear relationships in  \reef{nonlinIntro}. 
The Wilson coefficients $a_{2m-1,0}$ that were left unfixed in the EFT analysis correspond exactly to the first appearances of $\zeta_\text{odd}$  (i.e.~$\zeta_{3}$, $\zeta_{5}$, \dots), since $a_{2m-1,0} = g^2 \zeta_{2m+1}$ for $m=1,2,\dots$. 
Because the $\zeta_\text{odd}$  (at least conjecturally) have no algebraic relationships with any of the other $\zeta_k$-values, it should not be possible  to fix $a_{2m-1,0}$ in the EFT analysis. Therefore, there cannot be additional constraints on the $a_{k,q}$'s from additional SUSY Ward identities or   from extending the analysis to higher points. In fact, it is quite remarkable that a purely bottom-up EFT analysis up to 6-points with maximal SUSY, tree-factorization, and the parity condition \reef{Z1isbarZ1cond} 
fixes every possible relation among the $a_{k,q}$'s.

The nonlinear constraints \reef{nonlinIntro} rule out certain 4-point amplitudes, such as the SUSY infinite spin tower amplitude \cite{Caron-Huot:2020cmc,Albert:2022oes,Berman:2023jys,Berman:2024eid,Berman:2025owb} and the amplitude with the tree-level exchange of a single massive supermultiplet. These amplitudes satisfy the 4-point SUSY Ward identities, but our analysis shows that they are not compatible with the combined constraints of maximal SUSY, tree-factorization, and the parity condition \reef{Z1isbarZ1cond}.

It is useful to note what makes $\mathcal{N}=4$ SYM special relative to theories with less SUSY. In general, supersymmetry is well-known to be extremely restrictive. For example, in non-gravitational theories with $\mathcal{N}=2$ SUSY, the off-shell superspace formulation is sufficient to reduce the space of $\mathcal{N}=2$ EFTs to the space of pre-potentials \cite{Salam:1974yz,Salam:1974jj,Salam:1974pp}, which are greatly constrained by powerful non-renormalization theorems \cite{Grisaru:1979wc,Affleck:1983rr}. However, because we do not have a finite {\em off-shell} superspace formalism for $\mathcal{N}=4$ SUSY, it is not known how to construct general effective actions with $\mathcal{N}=4$ SUSY. In our analysis, we instead use the {\em on-shell} $\,\mathcal{N}=4$ superspace \cite{Ferber:1977qx} to derive how maximal SUSY constrains Wilson coefficients. In particular, we exploit our computational control over the on-shell EFT amplitudes to restrict the couplings in the EFT Lagrangian.

\subsection{Part 2: S-matrix Bootstrap with Nonlinear Constraints}

In the first part of the analysis, everything is done in the low-energy EFT, without assumptions about the existence of any UV theory. In contrast, the S-matrix bootstrap relies on the assumption that the EFT is the low-energy expansion of a `sensible' UV theory. 
Broadly speaking, sensible here means that the full theory is unitary, that its 4-point amplitudes are analytic in the complex $s$-plane away from the real $s$-axis, and that the large $s$-behavior of $A_4/s^2$ goes to zero. In addition, we assume a non-zero mass gap. In the weak-coupling limit, unitarity simply becomes the requirement that the spectral density is positive. From these assumptions, one derives dispersive representations for the Wilson coefficients. These are then used to formulate an optimization problem that can be implemented numerically to obtain two-sided bounds on ratios of Wilson coefficients \cite{Caron-Huot:2020cmc}. 
Such S-matrix bootstrap bounds were studied for $\mathcal{N}=4$ SYM EFTs in Refs. \cite{Chiang:2023quf,Berman:2023jys,Berman:2024wyt,Berman:2024eid}. 

In the second part of this paper, we combine the nonlinear relations \reef{nonlinIntro} with the bootstrap bounds and find evidence that the resulting positivity bounds only allow a 1-parameter family of solutions. In fact, working with constraints up to $k=7$, i.e.~up to 18th-derivative order in the EFT Lagrangian, the bounds on $a_{1,0}/a_{0,0}$, $a_{2,0}/a_{0,0}$, and $a_{2,1}/a_{0,0}$ converge to a thin sliver that lies within $\sim 5 \cdot 10^{-5}$ of the open string Veneziano amplitude; see Figures \ref{fig:allowedRegs} and \ref{fig:kmax345}. The single free parameter is the arbitrary choice of bootstrap mass gap relative to the scale set by the string tension (i.e.~$\alpha'$).

Separately from the bootstrap analysis, it turns out that the nonlinear relations alone imply that the 4-point EFT amplitude obeys the string monodromy relations \cite{Plahte:1970wy,Stieberger:2009hq,Bjerrum-Bohr:2009ulz,Bjerrum-Bohr:2010mia,Bjerrum-Bohr:2010pnr} order-by-order in the low-energy expansion. It is surprising that maximal SUSY, tree-factorization, and the parity condition \reef{Z1isbarZ1cond} imply string monodromy. In earlier work \cite{Huang:2020nqy,Berman:2023jys,Chiang:2023quf}, it was shown that string monodromy together with the positivity bounds restrict the  values of the Wilson coefficients  $a_{k,q}$ with $k \le 8$ to be at very small intervals around the string values and that the size of the intervals shrink with the number of constraints in the bootstrap. With these past results in mind, it is not surprising that our new bootstrap results with nonlinear constraints from maximal SUSY and the parity condition single out the string values for the $a_{k,q}$'s with $k \le 2$. The SUSY nonlinear  constraints are actually more restrictive than the monodromy constraints on the EFT amplitude, so extending the bootstrap with nonlinear constraints to $a_{k,q}$'s with $k \ge 3$ can be expected to give bounds that are even more restrictive than those in \cite{Huang:2020nqy,Berman:2023jys,Chiang:2023quf}.

\subsection*{Outline of the Paper}

{\bf Part 1.} We start by reviewing the on-shell superspace formalism in Section~\ref{sec:Superamps_SYM}, where we also construct the most general 4- and 5-point EFT superamplitudes compatible with $\mathcal{N}=4$ SUSY.
Section~\ref{sec:SYM_6pt} is devoted to the analysis of the 6-point NMHV amplitudes. In Section \ref{sec:NMHV_SUSY_Ward_id}, we derive the SUSY Ward identity that relates the 6-scalar NMHV amplitude $\mathbb{Z}_1$ from \reef{Z1intro} to its cyclic permutations. Then, in Section \ref{sec:SYM_6pt_ansatz}, we perform a completely general tree-level bottom-up construction of the EFT amplitude $\mathbb{Z}_1$, and finally subject it in Section \ref{sec:SYM_nonlinear_constraints} to the SUSY Ward identity and the parity condition to derive the nonlinear constraints \reef{nonlinIntro} among the $a_{k,q}$'s. The full set of constraints for $k \le 7$ are given in \reef{eq:nonlinear_constraints_akq}. We discuss compatibility with the open string 4-point amplitude and how the nonlinear relations rule out examples of other 4-point amplitudes in Section \ref{sec:consequences}.

\vspace{2mm}
\noindent {\bf Part 2.} The bootstrap analysis is covered in Section  \ref{sec:pos}, which begins with a review of the positivity bounds for $\mathcal{N}=4$ SYM EFTs in Section \ref{sec:bootsetup}. How to incorporate the nonlinear relations from Part 1 into the bootstrap analysis is discussed in Section \ref{sec:nonlinbootstrap}, where the results of the bootstrap are also presented.  
In Section \ref{sec:mono}, we make the observation that the nonlinear relations (without unitarity assumptions) imply the string monodromy relations. This leads us to compare our analysis to a selection of other recent bootstrap results for EFTs with maximal SUSY in Section \ref{sec:compare}.

\vspace{2mm}
We conclude the paper in Section \ref{sec:conclusions} with a discussion of the results and a view toward future directions. We point out that the nonlinear constraints on the 4-point EFT amplitude allow it to be resummed to a compact form and we discuss examples of the particle spectrum obtained from the numerical bootstrap.

Some technical details are relegated to appendices. In Appendix \ref{app:BCFW_residues}, we show how we use BCFW shifts to enforce momentum conservation when computing pole residues. Details of the construction of the 6-point amplitude are discussed in Appendix \ref{app:6-point_ampls}: this includes both a brief discussion of the 6-point MHV amplitude and explicit calculations of all the pole residues needed for the ansatz of the 6-point NMHV amplitude \reef{Z1intro}. In addition to giving the nonlinear constraints among the $a_{k,q}$'s, the 6-point SUSY Ward identity also fixes many contact terms in the 5- and 6-point amplitudes. As consistency checks, we show in Appendix~\ref{app:comparison_string} that the resulting EFT amplitudes at 5- and 6-point level are indeed compatible with the open string amplitudes \cite{Stieberger:2006bh,Stieberger:2006te,Broedel:2009nsh,Mafra:2011nv,Mafra:2011nw,Broedel:2013aza}. 
Finally, in Appendix \ref{app:DBI}, we consider the Abelian version of the EFT analysis in Part 1 of the paper and show that maximal SUSY and tree-level factorization do not lead to nonlinear constraints on the Wilson coefficients of the higher-derivative corrections of  $\mathcal{N}=4$ Dirac-Born-Infeld theory.

\section{Superamplitudes in \texorpdfstring{$\mathcal{N}=4$}{N=4} SYM EFTs}
\label{sec:Superamps_SYM}

In this section, we review the $\mathcal{N}=4$ on-shell superspace and apply it to determine the EFT superamplitudes at 3, 4, and 5 points. The more involved 6-point analysis is presented in Section \ref{sec:SYM_6pt}. Because we work in the large-$N$ limit, we consider only single-trace higher-derivative operators and our scattering amplitudes are all single-trace. Specifically, we work with the color-ordered amplitudes throughout.  

\subsection{On-shell Superamplitudes}
\label{sec:On-shell_superamps}

The 4d massless $\mathcal{N}=4$ supermultiplet consists of the positive and negative helicity gluons, $g^+$ and $g^-=g^{1234}$, four pairs of positive and negative helicity gluinos $\lambda^a$ and $\lambda^{abc}$, and 3 complex scalars $z^{ab} = -z^{ba}$ with $\bar{z}_{ab}=(z^{ab})^* = \frac{1}{2} \eps_{abcd} z^{cd}$. The $SU(4)$ R-symmetry indices $a,b,\ldots$, take values $1,2,3,4$. Using the on-shell superspace formalism,\footnote{See Refs. \cite{Elvang:2008na,Elvang:2013cua,Elvang_Huang_2015} for a review.} the states in the massless supermultiplet can be arranged into a single on-shell superfield for each  external particle $i$:
\begin{equation}
\label{eq: superfield}
    \Phi_i = g_i^+ + \eta_{ia} \lambda_i^{a} - \frac{1}{2} \eta_{ia} \eta_{ib} z_i^{ab} - \frac{1}{6} \eta_{ia} \eta_{ib} \eta_{ic} \lambda_i^{abc} + \eta_{i1} \eta_{i2} \eta_{i3} \eta_{i4} g_i^-\,.
\end{equation}
The Grassmann variables $\eta_{ia}$ are very useful for organizing $n$-point amplitudes with different external states into a superamplitude $\mathcal{A}_n [1 2 3 \dots n] \equiv \mathcal{A}_n [\Phi_1 \Phi_2 \Phi_3 \dots \Phi_n]$.\footnote{We used Matthew Headrick's \texttt{Mathematica} package~\cite{Grassmann} to implement Grassmann variables.}
Component amplitudes with specific external states are extracted from $\mathcal{A}_n$ by taking Grassmann derivatives:
\begin{equation}
    g_i^+ \to 1\,, \ \quad \lambda_i^a \to \partial_i^a\,, \ \quad z_i^{ab} \to \partial_i^a \partial_i^b\,, \ \quad \lambda_i^{abc} \to \partial_i^a \partial_i^b \partial_i^c\,, \ \quad g_i^- \to \partial_i^1 \partial_i^2 \partial_i^3 \partial_i^4\,,
\end{equation}
where $\partial_i^a = \partial/\partial\eta_{ia}$. For example,  
\begin{align}
    A_n[z^{12} z^{34}  - +\,\!\dots\,\! +] =&\, (\partial_1^1 \partial_1^2) (\partial_2^3 \partial_2^4) (\partial_3^1 \partial_3^2 \partial_3^3 \partial_3^4) \mathcal{A}_n[1 2 3 \dots n]\Big|_{\eta_{ia}=0}  \,.
\end{align}
When the $SU(4)$ R-symmetry is unbroken, the superamplitude is a sum of polynomials of degree $4k$ in the $\eta_{ia}$'s.  
The $k$'th sector includes the pure gluon amplitude with $k$ negative and $n-k$ positive helicity states, so we label the sectors as N$^{k-2}$MHV and write
\begin{equation}
\label{eq: superamplitude_helicity_decomp}
    \mathcal{A}_n = \mathcal{A}_n^{\text{MHV}} +  \mathcal{A}_n^{\text{NMHV}} +  \mathcal{A}_n^{\text{N}^2\text{MHV}} + \dots + \mathcal{A}_n^{\text{anti-MHV}}\,.
\end{equation} 
In on-shell superspace, the $\mathcal{N}=4$ SUSY generators $Q^a$ and $ \widetilde{Q}_a$ are represented as 
\be   \label{tqq}
Q^a = \sum_{i=1}^n\,[i|\, \frac{\pa}{\pa\eta_{ia}}\,, \qquad\quad
\widetilde Q_a = \sum_{i=1}^n\, | i\>\, \eta_{ia} \,.
\ee

The SUSY Ward identities (first considered in \cite{Grisaru:1976vm,Grisaru:1977px} and explored systematically for $\mathcal{N}=4$ SYM in \cite{Bianchi:2008pu}) 
 relate amplitudes with SUSY-connected external states. They are encoded compactly in on-shell superspace as the condition that the superamplitude vanishes under the action of the supercharges:
\be \label{susyward}
Q^a {\cal A}_n =0\,, \qquad\quad \widetilde Q_a \, {\cal A}_n =0\,.
\ee
This makes it clear that the different N$^{k-2}$MHV helicity sectors do not mix under SUSY. 

It was shown in \cite{Elvang:2009wd,Elvang:2010xn} that superamplitudes explicitly solving \reef{susyward} can be written in terms of two basic building blocks, namely, the Grassmann delta function
\be
\label{eq:delta8Q}
  \delta^{(8)}\big( \widetilde{Q}\big)
  \equiv 
  \frac{1}{2^4} \prod_{a=1}^4 \sum_{i,j=1}^n \<ij\> \eta_{ia}\eta_{ja}\,,
\ee
and 
\begin{equation}
\label{mijks}
    m_{ijk,a} \equiv [ij] \eta_{ka} +[jk] \eta_{ia} +[ki] \eta_{ja}\,.
\end{equation} 
For instance, the 3-point anti-MHV superamplitude is a degree-4 Grassmann polynomial proportional to 
$\prod_{a=1}^4 m_{ijk,a}$. Superamplitudes with 
$n\geq4$ are all proportional to the Grassmann delta function. The MHV superamplitudes, being degree-8 Grassmann polynomials, are simply $\delta^{(8)}\big( \widetilde{Q}\big)$ times a kinematic function. The first nontrivial case is the NMHV superamplitude, which is a degree-12 polynomial given by $\delta^{(8)}\big( \widetilde{Q}\big)$ times a particular linear combination of products of $m_{ijk,a}$’s. We discuss the 6-point NMHV sector in Section~\ref{sec:NMHV_SUSY_Ward_id}. 
We now use this formalism to determine the most general expressions for 3-, 4-, and 5-point tree-level superamplitudes in an $\mathcal{N}=4$ SYM EFT.

\subsection{3- and 4-point EFT Amplitudes}
\label{sec:SYM_3_4pt_amps}

As discussed in Section~\ref{intro:New_constraints}, the operator $\Tr F^3$ is incompatible with SUSY, and higher-derivative 3-point operators of the form $\Tr D^{2k}F^3$ can be moved into higher-point terms in the action via field redefinitions. Consequently, 3-point amplitudes do not receive corrections in the SUSY EFT. The 3-point superamplitudes are therefore completely fixed by the requirement that they correctly produce the pure SYM gluon amplitudes, and hence they can be written as
\be
\mathcal{A}^{\text{MHV}}_3[123]
=\frac{g}{\<12\> \<23\> \<31\>}\, 
  \delta^{(8)}\big( \widetilde{Q}\big)\,,
  ~~~~~~
  \mathcal{A}^{\text{anti-MHV}}_3
  [123]
  = - \frac{g}{[12] [23] [31]}\, 
  \prod_{a=1}^4 m_{123,a} \,,
\ee
where $g$ is the Yang–Mills coupling. In general, conjugate component amplitudes are related (up to a sign) by exchanging angle and square brackets.

At 4 points, the anti-MHV sector is equivalent to the MHV, thus there is only one helicity sector in the superamplitude. Concretely, the tree-level MHV superamplitude is given by
\be
\mathcal{A}^{\text{MHV}}_4[1234]
   = \frac{A_4[--+\,+]}{\<12\>^4} \, 
  \delta^{(8)}\big( \widetilde{Q}\big) \,,
\ee
where, importantly, the 4-point amplitude $A_4[--+\,+]$ may include higher-derivative corrections from the EFT expansion. It was shown in \cite{Berman:2023jys} that the most general tree-level $\mathcal{N}=4$ superamplitude takes the form 
\be
  \label{A4super}
\mathcal{A}^{\text{MHV}}_4[1234]
   = \frac{[34]^2}{\<12\>^2} \, F(s,u) \,
  \delta^{(8)}\big( \widetilde{Q}\big)\,,
\ee
with $F(u,s) = F(s,u)$ being crossing-symmetric as required by SUSY and cyclicity. The function $F(s,u)$ has the low-energy expansion
\be
   \label{eq:def_F_SYM}
   F(s,u) = - \frac{g^2}{su}
   + f(s,u)\,,
\ee
with
\be
  \label{eq:def_f_SYM}
  f(s,u) = \sum_{0 \leq q \leq k} a_{k,q} s^{k-q} u^q = a_{0,0} + a_{1,0}(s+u) + a_{2,0} (s^2+u^2) + a_{2,1} su + \dots \,.
\ee
Since $F(u,s) = F(s,u)$, we must have $a_{k,k-q} = a_{k,q}$.
 The 4-point Wilson coefficients of operators $\Tr D^{2k} F^4$ compatible with linearized $\mathcal{N}=4$ SUSY are in one-to-one correspondence with the effective couplings $a_{k,q}$ with $q \le \lfloor k/2 \rfloor$; as in \eqref{gluonEFT}. 
The $-g^2/(su)$ term yields the leading Parke–Taylor amplitude, e.g.~
\begin{equation}
    A_4[--+\,+] = \<12\>^2 [34]^2 \, F(s,u) = g^2 \, \frac{\<12\>^4}{\<12\>\<23\>\<34\>\<41\>} + \<12\>^2 [34]^2 \, f(s,u) \,.
\end{equation}
Examples of other component amplitudes are 
\begin{equation}
\label{eq: scalar_ampls_SYM}
A_4[zz\bar{z}\bar{z}] = s^2 F(s,u)\,, \qquad A_4[z\bar{z}z\bar{z}] = t^2 F(s,u)\,,
\end{equation}
where we can choose $z=z^{12}$ and $\bar{z}=z^{34}$ as examples of conjugate scalars. 

\subsection{Construction of the 5-point EFT Amplitude}
\label{sec:SYM_5pt_ansatz}

At 5 points, the NMHV sector is conjugate to the MHV sector, so it suffices to study the MHV superamplitude, 
\be
\label{eq:superamplitude_A5}
  \mathcal{A}^{\text{MHV}}_5[12345]
   = \frac{A_5[--+++]}{\<12\>^4} \, 
  \delta^{(8)}\big( \widetilde{Q}\big) \,.
\ee
The goal now is to determine the most general expression for the EFT gluon amplitude $A_5[--+++]$.

The amplitude $A_5[--+++]$ has physical poles in four 2-particle channels: 
23, 34, 45, and 51. Analyzing the helicity configurations on the poles, one finds that the 3-point vertex is always of anti-MHV type; hence, the special 3-particle kinematics determine that the pole has to be located at the zero in the angle-bracket of the Mandelstam $s_{ij} = \<ij\>[ij]$.   
The poles at $\<23\>=0$, $\<34\>=0$, $\<45\>=0$, and $\<51\>=0$ can be made explicit by writing an ansatz for $A_5[--+++]$ that has an overall prefactor of the Parke–Taylor amplitude and the higher-derivative terms accounted for
by a function $G_5[12345]$ as follows:
\be
  \label{EFT5ptansatz}
  A_5[--+++] 
  = 
  \frac{\<12\>^4}{{\<12\> \<23\> \<34\> \<45\> \<51\>} } \, 
  G_5[12345]\,.
\ee
The benefit of the Parke–Taylor prefactor is that it fully accounts for the little group scaling and the poles, and it ensures that the MHV SUSY Ward identities are solved. 
Therefore, in the EFT expansion, $G_5[12345]$ must be polynomial in the external momenta and what remains is to ensure that the residues on each pole match the appropriate product of lower-point amplitudes.

To get a consistent ansatz for $G_5$, we have to include both parity-even and -odd terms:
\be
 \label{G5pt}
 G_5[12345] \equiv
  V_5[12345] - \frac{1}{2}\eps[1234] \, Q_5[12345]\,,
\ee
where $V_5$ and $Q_5$ are  polynomials in the 5-point Mandelstam variables $\{s_{12},s_{23},s_{34},s_{45},s_{15}\}$. 
The factor of $-1/2$ is a choice of convention and the Levi–Civita contraction $\eps[1234]$ can be written in  spinor-helicity formalism as 
\be
\label{eq:Levi_Civita_def}
 \eps[1234] = 4 i \eps_{\mu \nu \rho \sigma} p_1^\mu p_2^\nu  p_3^\rho  p_4^\sigma =
 [12]\<23\>[34]\<41\>
 -\<12\>[23]\<34\>[41] \,.
\ee 
Cyclicity of the superamplitude allows us to require $G_5[12345] = G_5[51234]$.\footnote{\label{footn}One may additionally impose invariance of the superamplitude under reversal of the external states, $\mathcal{A}_n[12\dots n] = (-1)^n \mathcal{A}_n[n\dots 21]$. We do not make this  assumption in order to avoid restricting the color structures of the local operators in the EFT. However, note that {\em if} reversal symmetry is imposed (e.g.~by requiring $G_5[12345] = G_5[54321]$ at 5 points) we find, up to and including $\mathcal{O}(s^8)$, that it does not modify the nonlinear constraints found in Section~\ref{sec:SYM_nonlinear_constraints}, but only reduces the number of remaining free contact terms in the 5- and 6-point amplitudes.} 
It can readily be seen from momentum conservation that the Levi–Civita contraction is invariant under cyclic permutations of the external states,
\be
  \eps[1234] ~\to~ 
  \eps[2345]
  = - \eps[2341]
  = \eps[1234]\,.
\ee
Thus, it follows that both $V_5$ and $Q_5$ should also be invariant under cyclic permutations of the external momenta. 
We then write an ansatz for 
$V_5$ and $Q_5$ as 
\be
  \label{V5Q5}
  \begin{split}
  V_5[12345] \,=& ~
   v_{0,1}
   +
   v_{1,1} \{s_{12}\}
  + v_{2,1}
    \{s_{12}^2\}
    + v_{2,2} 
    \{s_{12} s_{23}\}
    + v_{2,3} 
    \{s_{12}s_{34}\}
     + \ldots \,,  
    \\
    Q_5[12345] \,=& ~
   q_{0,1}
   +
   q_{1,1} \{s_{12}\}
  + q_{2,1}
    \{s_{12}^2\}
    + q_{2,2} 
    \{s_{12} s_{23}\}
    + q_{2,3} 
    \{s_{12}s_{34}\}
     + \ldots \,,  
    \end{split}    
\ee
where the curly brackets $\{ \ldots \}$ denote a sum over all inequivalent cyclic permutations of the five labels $12345$. The coefficients $v_{k,r}$ and $q_{k,r}$, where $k$ denotes the order $s^k$ and $r$ labels the different contributions for a given $k$, are --- for now ---  
arbitrary. A summary of the number of free coefficients can be found in Table \ref{tab:solution_ansatz_5pt}.

\vspace{2mm}
\noindent {\bf Factorization Constraints.} 
At this point, the ansatz \reef{EFT5ptansatz} with \reef{G5pt} and \reef{V5Q5} satisfies all constraints of 5-point SUSY, little group scaling, and cyclicity. The remaining requirement is that it factorizes on each pole to the correct product of lower-point amplitudes. 
Setting 
\be
   v_{0,1} = g^3
\ee
ensures that we match the leading 5-point Parke–Taylor amplitude. Every other pole in the EFT amplitude arises from a diagram with a cubic YM vertex and a 4-point local interaction; i.e.~Res${}_\text{pole}$($A_5$) $= - A_3 \times A_4$. Specifically, the $s_{15}$-pole  contribution to the amplitude has to be 
\begin{align}
\label{eq:A5gluon_factorization51}
  A_5[--+++] \Big|_{s_{15}\,\text{pole}}
  &=- \ \begin{tikzpicture}[baseline={([yshift=-0.1cm]current bounding box.center)}] 
	\node[] (a) at (0,0) {};
    \node[] (a1) at ($(a)+(-0.75,0.75)$) {};
    \node[] (a3) at ($(a)+(-0.75,-0.75)$) {};
	\node[] (b) at ($(a)+(3,0)$) {};
    \node[] (b1) at ($(b)+(0.75,0.75)$) {};
    \node[] (b2) at ($(b)+(1,0)$) {};
    \node[] (b3) at ($(b)+(0.75,-0.75)$) {};
	\draw[line width=0.2mm,decoration={coil, amplitude=1.25mm, segment length=1.75mm, aspect=0.75,post length=0.5em},decorate] (a1.center) -- (a.center);
    \node[] at ($(a1)+(-0.2,0.2)$) {$5^+$};
    \draw[line width=0.2mm,decoration={coil, amplitude=1.25mm, segment length=1.75mm, aspect=0.75,post length=0.5em},decorate] (a3.center) -- (a.center);
    \node[] at ($(a3)+(-0.3,-0.2)$) {$1^-$};
    \draw[line width=0.2mm,decoration={coil, amplitude=1.25mm, segment length=1.75mm, aspect=0.75,pre length=0.75em,post length=0.5em},decorate] ($(a)+(0,-0.05)$) -- ($(b)+(0,-0.05)$);
    \draw[line width=0.2mm,decoration={coil, amplitude=1.25mm, segment length=1.75mm, aspect=0.75,pre length=0.2em},decorate] (b.center) -- (b1.center);
    \node[] at ($(b1)+(0.3,0.2)$) {$4^+$};
    \draw[line width=0.2mm,decoration={coil, amplitude=1.25mm, segment length=1.75mm, aspect=0.75,pre length=0.58em},decorate] ($(b)+(0.05,-0.05)$) -- ($(b2)+(0.25,-0.05)$);
    \node[] at ($(b2)+(0.55,0)$) {$3^+$};
    \draw[line width=0.2mm,decoration={coil, amplitude=1.25mm, segment length=1.75mm, aspect=0.75,pre length=0.64em},decorate] ($(b)+(-0.15,-0.1)$) -- ($(b3)+(-0.15,-0.1)$);
    \node[] at ($(b3)+(0.2,-0.2)$) {$2^-$};
    \node[] at ($(a)+(0.95,-0.45)$) {$(-P)^{+}$};
    \node[] at ($(b)+(-0.7,-0.425)$) {$P^{-}$};
    \filldraw[gray!25] (a.center) circle (0.35cm);
    \draw[line width=0.2mm] (a.center) circle (0.35cm);
    \filldraw[gray!25] (b.center) circle (0.35cm);
    \draw[line width=0.2mm] (b.center) circle (0.35cm);
    \node[] at ($(a)+(0,0)$) {$A_3$};
    \node[] at ($(b)+(0,0)$) {$A_4$};
\end{tikzpicture} \nonumber \\
  &=
  -A_3\big[
  5^+ \, 1^- \, (-P)^{+}
  \big] \,
  \frac{1}{s_{15}} \,
  A_4\big[ P^{-}\, 2^- \, 3^+ \, 4^+ 
  \big] \nonumber \\
  &=
  -g \, \frac{[5P]^3}{[51][1P]}
  \,
  \frac{1}{s_{15}}
  \,
  {\<P2\>^2 [34]^2} \,
  f(s_{34},s_{23}) \nonumber \\
  &=
  g \,\frac{\<12\>^3 [34]^2}{\<15\> \<25\>} \,
  f(s_{34},s_{23}) \,.
\end{align}
Here we are using $P=P_{15}$ and the analytic continuation convention $|-p\> = -|p\>$, $|-p] = |p]$, and we have simplified the spinor brackets, for example
$[5P] \<P2\> = [51]\<12\>$. 
As anticipated, \reef{eq:A5gluon_factorization51} contains the physical pole at $\<15\>=0$, but  rather disturbingly there also appears to be an unphysical pole at $\<25\> = 0$. However, in the special kinematics of the 15-pole, we have $|1\> \propto |5\>$, hence $\<25\> \propto \<21\>$. This cancels against a factor of $\<12\>$ in the numerator of the residue and shows that there is in fact no pole at all. We can similarly compute pole contributions in the other channels. 

Now, we have to make sure that the ansatz \reef{EFT5ptansatz} factorizes correctly on every pole. Starting with the 15-channel, we match the $\<15\>=0$ residue of \reef{EFT5ptansatz} to the $- A_3 \times A_4$ expression obtained in  \reef{eq:A5gluon_factorization51}.\footnote{The calculation of the residues has to be done carefully by ensuring that momentum conservation holds in the special kinematics on the pole. A trick to do this is to use a BCFW-shift to go to the pole; we describe how in Appendix \ref{app:BCFW_residues}.} This is done order-by-order in the low-energy expansion and  it fixes the coefficients of the polynomials $V_5$ and $Q_5$ from \reef{V5Q5} in terms of the 4-point Wilson coefficients $a_{k,q}$; at the lowest orders, we find
\be
  \label{EFT5ptRes}
  \begin{split}
  \mathcal{O}(p^1):~~~&v_{1,1} = 0\,,\\
  \mathcal{O}(p^3):~~~&v_{2,1} = 0\,,~~~
  v_{2,2} = -\frac{g}{2} a_{0,0}\,,~~~
  v_{2,3} = 0\,,~~~
  q_{0,1} = g \,a_{0,0}\,,
  \\
  \mathcal{O}(p^5):~~~&
  v_{3,1} = v_{3,4} = v_{3,5} = v_{3,6} = 0 \,,~~~
  v_{3,2} = v_{3,3} = v_{3,7} = -\frac{g}{2} a_{1,0}\,,~~~
  q_{1,1} =  g\, a_{1,0}\,.
  \end{split}
\ee
Up to and including $\mathcal{O}(p^7)$ every coefficient in the ansatz is fixed. However, starting at $\mathcal{O}(p^9)$ some coefficients remain free. Table \ref{tab:solution_ansatz_5pt} summarizes the number of free coefficients before and after matching to the 15-residue. We have explicitly checked that once the 15-residue is matched, the other poles in the 23-, 34-, and 45-channels are automatically correct and no further constraints arise on the ansatz. 

\begin{table}[t]
\begin{center}
\begin{tabular}{|l|cccccccccc|}
\hline
Order in $A_5[--+++]$ & $p^{-1}$ & $p^1$ & $p^3$ & $p^5$ & $p^7$ & $p^9$ & $p^{11}$ & $p^{13}$ & $p^{15}$ & $p^{17}$ \\ \hline
Ansatz & 1 & 1 & 4 & 8 & 17 & 33 & 56 & 92 & 141 & 209 \\
$s_{15}$ pole residue matching & 0 & 0 & 0 & 0 & 0 & 2 & 5 & 14 & 28 & 52 \\
6-point SUSY Ward identity & 0 & 0 & 0 & 0 & 0 & 0 & 0 & 1 & 3 & 4 \\
\hline
\end{tabular}
\end{center}
\caption{Number of free parameters (other than the 4-point coefficients $a_{k,q}$) in the ansatz of the 5-point MHV amplitude $A_5[--+++]$ at each order up to $\mathcal{O}(p^{17})$, and the resulting simplification after matching the residue at the $s_{15}$ pole with the physical factorization channel, and solving the 6-point NMHV SUSY Ward identity (see Section~\ref{sec:SYM_nonlinear_constraints}).}
\label{tab:solution_ansatz_5pt}
\end{table}

\vspace{2mm}
\noindent {\bf Local Contact Terms.} 
The second row of Table \ref{tab:solution_ansatz_5pt} shows that starting at order $p^9$, there are unfixed coefficients in the 5-point ansatz \reef{V5Q5} after fixing the pole residues.  Since they are not determined by matching any of the poles, they must correspond to local contact terms in the 5-point amplitude, i.e.~we expect them to be associated with matrix elements of local operators $\Tr D^{2k}F^5$ that are independently compatible with the constraints of linear 
$\mathcal{N}=4$ SUSY. To characterize such matrix elements, we use that they have to satisfy the 5-point SUSY Ward identities. This includes the constraint that any pure gluon 5-point MHV matrix element with negative helicity states at positions $i$ and $j$ has to be proportional to the pure gluon MHV matrix elements $\big\<-- +++\big\>$,
\be
 \big\<\dots i^- \dots j^-\dots\big\>
  = 
 \frac{\<ij\>^4}{\<12\>^4}\big\<-- +++\big\> \,.
\ee
Since the matrix elements on both sides have to be local, i.e.~cannot have any poles, the only option is that 
$\big\<-- +++\big\> \propto \<12\>^4 B[+++++]$, where $B[+++++]$ has to be polynomial in the spinor brackets and have the same little group scaling of a positive helicity gluon for each external line. In addition, $B[+++++]$ has to be cyclic invariant.

Now, in order to satisfy the little group scaling, each spinor $|i]$ has to appear twice in $B[+++++]$. This means that the lowest possible momentum order where such local terms can exist will be 4+5 = 9.  We can directly write down two linearly independent candidates at $\mathcal{O}(p^9)$, namely
\be
 \begin{split}
 \big\<-- +++\big\> =&
 \<12\>^4 [12][23]
 [34][45][51]\,,
 \\
 \big\<-- +++\big\> =&
 \<12\>^4 \big( [12]^2 [34][45][53] + \text{cyclic} \big)\,.
  \end{split}
\ee
Indeed,  the kinematic prefactors of the two unfixed coefficients at order $\mathcal{O}(p^9)$ in the ansatz \reef{EFT5ptansatz} directly match these two local matrix elements. 
Likewise, at higher orders, we have found a one-to-one correspondence between each term in the ansatz with unfixed coefficient and the possible local matrix elements compatible with $\mathcal{N}=4$ SUSY. 
Translating it to the Lagrangian, this implies that $\Tr D^4 F^5$ is the lowest-dimension 5-field operator that is {\em independently}\footnote{Not taking into account constraints from 6 points.} compatible with linearized $\mathcal{N}=4$ SUSY and that there are two such  operators.

This analysis of local contact terms verifies that our ansatz is fully general, and we have therefore solved the problem of a bottom-up construction of the EFT 5-point MHV superamplitude. The anti-MHV sector is simply obtained by conjugation. Importantly, the construction of the 5-point superamplitude placed no constraints on the 4-point Wilson coefficients. Next, we calculate the 6-point EFT amplitudes using the 4- and 5-point EFT amplitudes as an input.

\section{6-point EFT Amplitudes with \texorpdfstring{$\mathcal{N}=4$}{N=4} SUSY}
\label{sec:SYM_6pt}

There are two helicity sectors of the 6-point amplitude to study. The MHV sector is very similar to the 5-point MHV sector and yields no constraints on the 4-point Wilson coefficients; we present the analysis in Appendix \ref{app:6-point_ampls}. The NMHV sector is much more nontrivial and is the subject of this section. 

We begin in Section \ref{sec:NMHV_SUSY_Ward_id} with the derivation of the $\mathcal{N}=4$ SUSY constraint for a particular 6-scalar NMHV amplitude and its cyclic permutations. Next, in Section \ref{sec:SYM_6pt_ansatz}, we derive the most general bottom-up EFT ansatz for the 6-scalar amplitude, and in Section \ref{sec:SYM_nonlinear_constraints} we combine the results of the previous two subsections to derive nonlinear constraints on the 4-point Wilson coefficients assuming the absence of Levi–Civita contractions in the local contact terms. In Section \ref{sec:consequences} we discuss 
some immediate consequences of the nonlinear constraints in the context of the open string amplitude, and give examples of other 4-point amplitudes that are ruled out.

\subsection{SUSY Constraints in the 6-point NMHV Sector}
\label{sec:NMHV_SUSY_Ward_id}

Unlike MHV amplitudes that are simply proportional to each other by maximal SUSY, the constraints of the SUSY Ward identities for NMHV amplitudes are more involved. Nonetheless, by solving the $\mathcal{N}=4$ SUSY Ward identities \reef{susyward} in on-shell superspace, it was shown in \cite{Elvang:2009wd,Elvang:2010xn} that, with $SU(4)$ R-symmetry, the 6-point NMHV superamplitude can be written as
\begin{equation}
   {\ca}_6^\text{NMHV}[123456] =
    \mathbb{A}_1\, X_{(1112)}
   +\mathbb{A}_2\,
   X_{(1122)}
   + \mathbb{A}_3\,X_{(1222)}
   + \mathbb{A}_4\, X_{2222}
   + \mathbb{A}_5\, X_{1111}
    \, .
    \label{nmhv6B}
\end{equation}
The five ``basis amplitudes'' $\mathbb{A}_i$
are
\be 
  \label{defAi}
  \begin{split}
   &\mathbb{A}_1  = A_6[  \lambda^{123} \lambda^{4}++ --]\,,~~~
   \mathbb{A}_2 = A_6[ z^{12}\, z^{34} ++--]\,,~~~
   \mathbb{A}_3 = A_6[ \lambda^{1} \lambda^{234} ++--]\,, \\[2mm]
   &\mathbb{A}_4 = A_6[ +-++-\,-] \,,\hspace{0.8cm}
   \mathbb{A}_5 = A_6[ -+++-\,-] \,,
  \end{split}
\ee
and $X_{(ijkl)}$ denotes the sum over the inequivalent permutations of the $ijkl$ indices of the degree-4 Grassmann polynomials  
\be\label{defXs}
    X_{ijkl} ~\equiv~ \delta^{(8)}\big(\widetilde{Q}\big)
  ~ \frac{m_{i34,1}\,m_{j34,2}\,m_{k34,3}\,m_{l34,4}}{[34]^4\<56\>^4}   \,.
\ee
These are defined in terms of the manifestly SUSY building blocks $\delta^{(8)}\big( \widetilde{Q}\big)$ and $m_{ijk,a}$ introduced in~\eqref{eq:delta8Q}--\eqref{mijks}.
For instance, we have $X_{(1112)} = X_{1112}+ X_{1121}+ X_{1211}+ X_{2111}$ and 
$X_{(1122)} = X_{1122}+ X_{1212}+ X_{1221}+ X_{2112}+ X_{2121}+ X_{2211}$.

In our case, the basis amplitudes would be the EFT amplitudes. Since the five amplitudes in \reef{defAi} have different external states that are not simply related by cyclicity and momentum relabeling, this basis requires five independent bottom-up constructions of EFT amplitudes with different parametrizations of contact terms. That is rather involved and it is instead desirable to use five amplitudes related by cyclic symmetry. One option would be the pure gluon amplitudes, e.g.~$A_6[---+++]$ and its cyclic permutations; however, it is a lot simpler to work with external scalars rather than vectors. 

Picking a scalar amplitude with only few factorization channels makes the bottom-up construction even simpler. After considering various options, our choice fell on 
\be 
  \label{ourZ1}
  \mathbb{Z}_1 = 
   A_6[z_1 z_2 z_3 \bar{z}_3 \bar{z}_1 \bar{z}_2] \,,
\ee
where we have used the shorthand notation
\be
  \begin{split}
  &
  z_1 = z^{12}\,,\qquad
  z_2 = z^{13}\,,\qquad
  z_3 = z^{14}\,,\qquad
  \\
  &
  \bar{z}_1 = z^{34}\,,\qquad
  \bar{z}_2 = z^{24}\,,\qquad
  \bar{z}_3 = z^{23}\,,\qquad
  \end{split}
\ee
for the complex $SU(4)$ scalars. The amplitude \reef{ourZ1} has poles only in the 34-, 234-, and 345-channels; for a detailed discussion of the factorization structure we refer to Section \ref{sec:SYM_6pt_ansatz}, where we construct $\mathbb{Z}_1$. 

Let us now proceed to explain how we impose SUSY. 
Starting with $\mathbb{Z}_1$, 
we cyclically permute the external states to get 
\be 
 \label{defZi}
  \begin{split}
   &\mathbb{Z}_1 = 
   A_6[z_1 z_2 z_3 \bar{z}_3 \bar{z}_1 \bar{z}_2]
   \,,~~~~~~
   \mathbb{Z}_2 = 
   A_6[z_2 z_3  \bar{z}_3 \bar{z}_1 \bar{z}_2 z_1]
   \,,~~~~~~
   \mathbb{Z}_3 = 
   A_6[z_3  \bar{z}_3 \bar{z}_1 \bar{z}_2 z_1 z_2]   
   \,,\\[2mm]
   &\mathbb{Z}_4 = 
    A_6[\bar{z}_3 \bar{z}_1 \bar{z}_2 z_1 z_2 z_3]      
   \,,~~~~~~
   \mathbb{Z}_5 = 
   A_6[\bar{z}_1 \bar{z}_2 z_1 z_2 z_3 \bar{z}_3]   
   \,,~~~~~~
   \mathbb{Z}_6 =
   A_6[\bar{z}_2 z_1 z_2 z_3 \bar{z}_3 \bar{z}_1]  
   \,.
  \end{split}
\ee
These six amplitudes are related by simple cyclic momentum relabeling, but they have the 2-particle pole in different locations, so as kinematic functions they are distinct. Now, the key idea is to use the five amplitudes $\mathbb{Z}_1$, \ldots, $\mathbb{Z}_5$ as the basis amplitudes for the 6-point NMHV superamplitude instead of the $\mathbb{A}_i$'s in \reef{defAi}. We then project out the sixth amplitude $\mathbb{Z}_6$ from the superamplitude to obtain an expression for it as a linear combination of the five other $\mathbb{Z}_1$, \ldots, $\mathbb{Z}_5$. The resulting NMHV SUSY Ward identity turns out to be a very powerful consistency condition on the EFT construction.

To change basis in the superamplitude, we first project out 
$\mathbb{Z}_1$, \ldots ,$\mathbb{Z}_5$; for example, 
\begin{equation}
\label{exproj}
  \mathbb{Z}_1 = (\partial_1^1 \partial_1^2) (\partial_2^1 \partial_2^3) (\partial_3^1 \partial_3^4) (\partial_4^2 \partial_4^3) (\partial_5^3 \partial_5^4) (\partial_6^2 \partial_6^4)\mathcal{A}_6^{\text{NMHV}}[1 2 3 4 5 6]\,.
\end{equation}
With $\mathcal{A}_6^{\text{NMHV}}[1 2 3 4 5 6]$ given by \reef{nmhv6B}, this gives 
each $\mathbb{Z}_1$, \ldots ,$\mathbb{Z}_5$ as a linear combination of the amplitudes $\mathbb{A}_1$, \ldots ,$\mathbb{A}_5$. We can write the linear system as 
\be
   \begin{pmatrix}
      \mathbb{Z}_1 \\
      \mathbb{Z}_2 \\
      \mathbb{Z}_3 \\
      \mathbb{Z}_4 \\
      \mathbb{Z}_5 \\
   \end{pmatrix}
    = C_{5 \times 5}
       \begin{pmatrix}
      \mathbb{A}_1 \\
      \mathbb{A}_2 \\
      \mathbb{A}_3 \\
      \mathbb{A}_4 \\
      \mathbb{A}_5 
     \end{pmatrix},
\ee
where $C_{5 \times 5}$ is a $5 \times 5$ matrix consisting of spinor-helicity products arising from the projections. Inverting the matrix $C_{5 \times 5}$ is straightforward, and it yields a solution ``$\text{sol}({\mathbb{A}_i)}$'' for the $\mathbb{A}_i$ in terms of linear combinations of the $\mathbb{Z}_i$. This is then plugged back into equation \reef{nmhv6B}, allowing us to change the basis for the superamplitude,
\begin{align}
\label{nmhv6C}
   {\ca}_6^\text{NMHV}[123456] =&\,
   \bigg( \mathbb{A}_1\, X_{(1112)}
   +\mathbb{A}_2\,
   X_{(1122)}
   + \mathbb{A}_3\,X_{(1222)}
   + \mathbb{A}_4\, X_{2222}
   + \mathbb{A}_5\, X_{1111}
   \bigg) \bigg|_{\text{sol}({\mathbb{A}_i})} \nonumber \\
   =&\,  
   \mathbb{Z}_1 \, C_1+
   \mathbb{Z}_2 \, C_2+
   \mathbb{Z}_3 \, C_3+
   \mathbb{Z}_4 \, C_4+
   \mathbb{Z}_5 \, C_5   
    \, .
\end{align}
The $C_i$ are degree-12 Grassmann polynomials that are rather complicated spinor-bracket-dependent linear combinations of the $X_{(ijkl)}$'s from \reef{nmhv6B}. From this expression, we now project out the sixth scalar amplitude $\mathbb{Z}_6$, and thereby we obtain a nontrivial 6-point NMHV SUSY Ward identity 
\be  
  \label{consistency}
  \mathbb{Z}_6
  = 
   c_1\, \mathbb{Z}_1
  +c_2\, \mathbb{Z}_2
  +c_3\, \mathbb{Z}_3
  +c_4\, \mathbb{Z}_4
  +c_5\, \mathbb{Z}_5 \,,
\ee
where the $c_i$ are complicated (but explicitly computed) spinor-helicity expressions. 

The Ward identity \reef{consistency} is a necessary condition for the amplitudes $\mathbb{Z}_i$ to be compatible with $\mathcal{N}=4$ SUSY and $SU(4)$ R-symmetry. 
Importantly, the coefficients $c_i$ do not receive corrections: 
for the {\em same} coefficients $c_i$, 
the Ward identity \reef{consistency} has to hold at any order in perturbation theory and, as is relevant for us in this analysis, at any order in the EFT expansion. 

As a basic test of the Ward identity \eqref{consistency}, we compute $\mathbb{Z}_1$ in  pure SYM.  This can be done straightforwardly using   BCFW recursion~\cite{Britto:2004ap,Britto:2005fq} and the result simplifies to 
\be
\label{LO}
   \mathbb{Z}_1^{\text{(SYM)}} = A_6^{\text{(SYM)}}[z_1 z_2 z_3 
   \bar{z}_3 \bar{z}_1 \bar{z}_2]
   = g^4 \bigg(
    \frac{1}{s_{34}}
    - \frac{s_{24}}{s_{34}s_{234}}
    - \frac{s_{35}}{s_{34}s_{345}}
   \bigg) \,.
\ee
The other $\mathbb{Z}_i$ are obtained from cyclic momentum relabelings and together it is easy to verify that they solve \reef{consistency}.\footnote{In practice, this is done numerically using arbitrary rational kinematics for the spinor brackets.} 

The next step is to construct the EFT corrections to $\mathbb{Z}_1$; then, to use them to solve \reef{consistency} order-by-order in the derivative expansion. 
While \reef{consistency} is necessary for compatibility with maximal SUSY, it is not at all obvious that it is sufficient to fix the EFT ansatz. In the coming sections and based on the solutions we find, we are going to argue that for the purpose of constraining the 4-point Wilson coefficients $a_{k,q}$, the single Ward identity \reef{consistency} and the parity condition \reef{Z1isbarZ1cond} are indeed sufficient.

\subsection{EFT Construction of the 6-point NMHV Basis Amplitudes}
\label{sec:SYM_6pt_ansatz}

The bottom-up construction of the 6-point EFT amplitude $\mathbb{Z}_1=A_6[z_1 z_2 z_3 \bar{z}_3 \bar{z}_1 \bar{z}_2]$ is in principle straightforward. As a tree-level amplitude, it is given by a sum of its pole terms, whose residues are products of lower-point amplitudes, and a general ansatz for polynomial terms in the external momenta. The main difficulty is dealing with internal gluon lines, but we explain below how this is done.  

The pole diagrams for $\mathbb{Z}_1$ are computed from their lower-point factorizations; full details are in Appendix \ref{app:6-point_ampls}. Here we simply summarize the contributions of the three different types of pole diagrams:
\begin{enumerate}
\item {\bf Two 3-particle channel diagrams} with poles at  $s_{234}=0$ and $s_{345}=0$, respectively, and factorization $A_4 \times  A_4$. These are the easiest to compute since they only involve scalar internal lines. They contribute to terms in the 6-point amplitude that are quadratic in the 4-point Wilson coefficients, i.e.~$a_{k,q} a_{k',q'}$, starting at order $\mathcal{O}(s^3)$, as well as to linear terms of the form $g^2 a_{k,q}$ starting at order $\mathcal{O}(s^1)$.
\item {\bf Two double-pole diagrams} with $s_{34}=s_{234}=0$ or
$s_{34}=s_{345}=0$, respectively, and factorization structure $A_3 \times A_3 \times A_4$ that simplifies nicely. Starting at order $\mathcal{O}(s^1)$, these diagrams produce contributions to $\mathbb{Z}_1$ of the form $g^2 a_{k,q}$.
\item {\bf One 2-particle channel diagram} with $s_{34}=0$ and factorization structure $A_3 \times A_5$. For the purpose of the analysis that follows, it is useful to give the explicit result for the 34-pole term. For kinematics with $[34]=0$, it is 
\be
  \label{eq:Ressq34MT}
  \mathbb{Z}_1 \Big|_{[34]=0}
  =
  g \, \bigg(
  \frac{1}{s_{34}} - \frac{s_{24}}{s_{34}s_{234}} 
   - 
  \frac{s_{35}}{s_{34}s_{345}}
  \bigg) \,
\bigg( V_5[5612P] - \frac{1}{2}\eps[1256] \, Q_5[5612P]\bigg),~
\ee
where $P=P_{34}$. 
For $\<34\>=0$ kinematics, one simply conjugates $\<{...}\> \lra [...]$ in the expression \reef{eq:Ressq34MT}.
The result of the 5-point analysis (equations \reef{V5Q5} and \reef{EFT5ptRes} for $V_5$ and $Q_5$) shows that
the EFT contributions of this factorization diagram are of the form $g^2 a_{k,q}$ as well as  $g v_{k,r}$ and $g q_{k,r}$, where $v_{k,r}$ and $q_{k,r}$
are the Wilson coefficients of the 5-point contact terms left unfixed by the 5-point SUSY analysis in Section \ref{sec:SYM_5pt_ansatz}. 
Note that the Mandelstam prefactor in the residue \reef{eq:Ressq34MT} is nothing but that of the 6-point pure SYM amplitude \reef{LO}, which is therefore reproduced by the leading term $g^3$ in $V_5[1234P]$.

\end{enumerate}

The fact that the 34-residue \reef{eq:Ressq34MT} has poles also in the 234- and 345-channels complicates the bottom-up construction of the 6-point amplitude slightly; this is due to the gluon exchange in the 34-channel. However, we can write an ansatz for 6-point EFT amplitude $\mathbb{Z}_1$ that gets all the 3-particle poles and double poles (types 1 and 2 above)
correct and parametrizes the 34-pole fully generally:
\begin{align}
\label{scalarZ1ans}
   \mathbb{Z}_1 =& \, g^4 \bigg(
    \frac{1}{s_{34}}
    - \frac{s_{24}}{s_{34}s_{234}}
    - \frac{s_{35}}{s_{34}s_{345}}
   \bigg)  +
   g^2 \frac{s_{16} s_{24} s_{56}}{s_{34} s_{234}}
  f(s_{16},s_{56})
  +g^2 \frac{s_{12} s_{16} s_{35}}{s_{34} s_{345}}
  f(s_{12},s_{16})
  \nonumber \\
  & \,
  + g^2 \frac{s_{23} s_{24}}{s_{234}}f(s_{23},s_{34}) - \frac{s_{16} s_{23} s_{24} s_{56}}{s_{234}}
  f(s_{23},s_{34})
  f(s_{16},s_{56})  + g^2 \frac{s_{35} s_{45}}{s_{345}}f(s_{34},s_{45})
  \nonumber \\
  & \,
  - \frac{s_{12} s_{16} s_{35} s_{45}}{s_{345}}
  f(s_{34},s_{45})
  f(s_{12},s_{16}) +\frac{G_6[123456]}{s_{34}}
  + \text{CT}_6[123456]\,.
\end{align}
The $g^4$-term is simply the leading-order SYM result from \reef{LO}. By construction, the pole terms involving $f$ ensure that the residues of the 3-particle poles and the double poles are correct. As an ansatz to capture the $s_{34} = 0$ pole, we include the function $G_6$, soon to be determined. 
Afterwards, we turn to the polynomial terms $\text{CT}_6$, which parametrize the 6-point local contact terms.

\vspace{2mm} 
\noindent {\bf Fixing $G_6$ and the 34-pole Terms.} 
Naively, the most general ansatz for $G_6$ consists of all independent polynomials in a choice of nine independent  Mandelstam variables, e.g.~
\begin{equation}
\label{eq:6pt_Mandelstams}
    \big\{ s_{12},\, s_{23},\, s_{34},\, s_{45},\, s_{56},\, s_{16},\, s_{123},\, s_{234},\, s_{345} \big\}\,,
\end{equation}
and a choice of the independent Levi–Civita contractions 
\begin{equation}
\label{eq:6pt_LC}
    \big\{ \epsilon[1234],\, \epsilon[1235],\, \epsilon[1245],\, \epsilon[1345],\, \epsilon[2345] \big\} \,.
\end{equation}
However, by inspecting the residue \reef{eq:Ressq34MT} that the 34-pole of \reef{scalarZ1ans} has to reproduce, it is clear that the parity-odd contributions from $G_6$ can only be of the form $\eps[1256]=\eps[1235]+\eps[1245]$. Therefore, we can write 
\begin{equation}
    G_6[123456] = V_6[123456] - \frac{1}{2} \epsilon[1256] \,Q_6[123456]\,,
\end{equation}
where $V_6$ and $Q_6$ are polynomials in the Mandelstams.

Next, rather than working with polynomials in nine independent Mandelstams, we can take advantage of the special kinematics on the $s_{34}=0$ pole. For a 6-point scalar factorization, the residue can only depend on five 5-point Mandelstams, namely
\begin{equation}\label{eq:5pointeffective}
\{ s_{12},\, s_{16},\, s_{56},\, s_{234},\, s_{345} \} \, .
\end{equation}
However, in our case, the internal particle is a gluon, so there can also be a single contraction of $p_3$ or $p_4$ with momenta from the 5-point factor. 
This allows for the three other nonzero Mandelstam variables $\{ s_{123}, s_{23}, s_{45} \}$ to appear at most linearly in $G_6$. Consequently, a refined version of the ansatz is given by
\begin{equation}
\label{eq:ansatz6pt_refined}
\begin{split}
V_{6}[123456]&=V_{6,1}[123456]+s_{123}V_{6,2}[123456]+s_{23}V_{6,3}[123456]+s_{45}V_{6,4}[123456]\,, \\
Q_{6}[123456]&=Q_{6,1}[123456]+s_{123}Q_{6,2}[123456]+s_{23}Q_{6,3}[123456]+s_{45}Q_{6,4}[123456]\,, \\
\end{split}
\end{equation}
where the $V_{6,i}[123456]$ and $Q_{6,j}[123456]$ are polynomials in the effective 5-point planar Mandelstam variables \eqref{eq:5pointeffective}. This ansatz has far fewer free parameters than the most general polynomial in 6-particle variables, and that allows us to go to higher orders. For instance, at low orders, we have
\begin{equation}
\label{V61ansatz}
    V_{6,1}[123456] = w_{0,1}
   + w_{1,1} s_{16} + w_{1,2} s_{56} + w_{1,3} s_{345} + w_{1,4} s_{234} + w_{1,5} s_{12} + \ldots \,,
\end{equation}
while the counting at higher orders can be found in the first row of Table \ref{tab:solution_ansatz_G6}. For example, at order $\mathcal{O}(s^1)$, the 8 parameters correspond to 3 parameters from the constant terms in $V_{6,2}$, $V_{6,3}$ and $V_{6,4}$ in \reef{eq:ansatz6pt_refined} and the five $\mathcal{O}(s^1)$ coefficients from $V_{6,1}$ in \reef{V61ansatz}. The $Q_{6,i}$'s do not contribute until order $\mathcal{O}(s^2)$ because of the four powers of momentum in the Levi–Civita contraction.

\begin{table}[t]
\begin{center}
\begin{tabular}{|l|cccccccccc|}
\hline
Order in $G_6[123456]$ & $s^0$ & $s^1$ & $s^2$ & $s^3$ & $s^4$ & $s^5$ & $s^6$ & $s^7$ & $s^8$ & $s^9$ \\ \hline
Ansatz & 1 & 8 & 31 & 88 & 205 & 416 & 763 & 1296 & 2073 & 3160 \\
$[34]$ pole residue matching & 0 & 0 & 0 & 0 & 3 & 14 & 40 & 90 & 175 & 308 \\
$\<34\>$ pole residue matching & 0 & 0 & 0 & 0 & 0 & 0 & 0 & 0 & 0 & 0 \\
\hline
\end{tabular}
\end{center}
\caption{Number of free parameters (not including the 4-point and 5-point coefficients) in the ansatz of $G_6[123456]$ at each order up to $\mathcal{O}(s^9)$. The last two rows show the number of free parameters after first matching to the $[34]=0$ residue of \reef{eq:Ressq34MT} and next matching to the $\<34\>=0$ residue. The combined matching fixes all parameters in the ansatz of $G_6[123456]$ in terms of the 4- and 5-point coefficients.}
\label{tab:solution_ansatz_G6}
\end{table}

Now that we have determined an efficient ansatz for $G_6$, we simply compute the full $[34]=0$ residue of \reef{scalarZ1ans} and compare it to that of \reef{eq:Ressq34MT}.\footnote{As in the 5-point analysis, this is done using the BCFW trick from Appendix \ref{app:BCFW_residues} to ensure that the residue satisfies the appropriate momentum conservation. In practice, we solve the systems numerically using 
rational kinematics and, for large linear systems, finite field methods \cite{Mangan:2023eeb}.} This fixes a vast number of parameters in $G_6$ in terms of the coupling $g$ and the 4- and 5-point EFT couplings $a_{k,q}$,  $v_{k,r}$, and $q_{k,r}$. Next, matching to the $\<34\>=0$ residue  fixes the remaining free parameters in $G_6$, showing that our ansatz for $G_6$ was not only sufficient but also minimal; see Table \ref{tab:solution_ansatz_G6} for the parameter count at each step. 

We find that the parity-odd contributions included in the $G_6$-ansatz are required to be zero in the final answer. This means that all the pole terms of  $\mathbb{Z}_1$ in \reef{scalarZ1ans} are parity-even. The pole terms also turn out to be invariant under reversal of the external states, i.e.~under relabeling the momenta $\{123456\} \lra \{654321\}$. Note that this is not a symmetry the $\mathbb{Z}_1$ amplitude is expected to have.

\vspace{2mm}
\noindent {\bf Contact Terms $\text{CT}_6$.} 
With the pole terms now fixed in $\mathbb{Z}_1$, we turn to the local terms $\text{CT}_6[123456]$ in  \reef{scalarZ1ans}.
$\text{CT}_6$ consists of the most general parity-even and -odd polynomial terms, i.e.~going order-by-order in the derivative expansion, we can write all
polynomials in a choice of the nine independent Mandelstams (e.g.~\reef{eq:6pt_Mandelstams}), and five Levi–Civita's (e.g.~\reef{eq:6pt_LC}). At lowest order, there is simply a constant and at $\mathcal{O}(s^1)$ there are nine independent terms: 
\begin{align}
\label{eq: 6pt_CTs}
    \text{CT}_6[123456] = &\ 
   y_{0,1}
   +
   y_{1,1} s_{345} 
   + y_{1,2} s_{234} + y_{1,3} s_{123} + y_{1,4} s_{16} + y_{1,5} s_{56} + y_{1,6} s_{45} \nonumber \\
   &\, 
   + y_{1,7} s_{34} + y_{1,8} s_{23} + y_{1,9} s_{12}
     + \ldots \,.
\end{align}
The counting at higher orders can be found in Table~\ref{tab:6pt_CTs}.

\begin{table}[t]
\begin{center}
\begin{tabular}{|l|ccccccccc|}
\hline
Order in $\text{CT}_6[123456]$ & $s^0$ & $s^1$ & $s^2$ & $s^3$ & $s^4$ & $s^5$ & $s^6$ & $s^7$ & $s^8$  \\ \hline
Ansatz with Levi–Civita's & 1 & 9 & 50 & 210 & 720 & 2112 & 5478 & 12870 & 27885 \\
6-point SUSY Ward identity & 0 & 0 & 0 & 0 & 9 & 83 & -- & -- & -- 
\\ \hline \hline
Ansatz without Levi–Civita's & 1 & 9 & 45 & 165 & 495 & 1287 & 3003 & 6435 & 12870 \\
6-point SUSY Ward identity & 0 & 0 & 0 & 0 & 0 & 0 & 2 & 9 & 13 \\
\hline
\end{tabular}
\end{center}
\caption{Number of free coefficients in the ansatz of the 6-point local contact terms $\text{CT}_6[123456]$ for $\mathbb{Z}_1$ at each order up to $\mathcal{O}(s^8)$, with and without parity-odd terms proportional to the 4d Levi–Civita's. 
In each case, the table also shows the number of free coefficients after imposing SUSY, as described in  Section~\ref{sec:SYM_nonlinear_constraints}. 
With 4d Levi–Civita's in the local ansatz, we find no constraints on the 4-point Wilson coefficients $a_{k,q}$; this was checked to $\mathcal{O}(s^5)$. However, excluding the Levi–Civita terms in $\mathbb{Z}_1$ we find nonlinear constraints among various $a_{k,q}$, as described in Section~\ref{sec:SYM_nonlinear_constraints}.}
\label{tab:6pt_CTs}
\end{table}

\vspace{2mm}
At this point, we have completed the bottom-up construction of the 6-point EFT amplitude $\mathbb{Z}_1$. The next step is to subject it to the SUSY constraint from Section \ref{sec:NMHV_SUSY_Ward_id}.

\subsection{Nonlinear Constraints from \texorpdfstring{$\mathcal{N}=4$}{N=4} SUSY}
\label{sec:SYM_nonlinear_constraints}

At leading order, $\mathcal{O}(s^{-1})$, the pure SYM amplitude \reef{LO} was already verified to solve the SUSY Ward identity \reef{consistency}. At $\mathcal{O}(s^{0})$, none of the pole terms contribute, so the only input from the $\mathbb{Z}_{i}$ is the constant term $y_{0,1}$ in CT${}_6$ and it is set to zero by the SUSY Ward identity. Next, at  $\mathcal{O}(s^1)$ and  $\mathcal{O}(s^2)$, all local parameters in CT${}_6$ become fixed, but otherwise the SUSY Ward identity is solved without constraints on the lower-point Wilson coefficients.

When Levi–Civita terms are {\em included} in the ansatz for the local terms $\text{CT}_6[123456]$ in $\mathbb{Z}_{1}$, we find no constraints on the 4-point Wilson coefficients $a_{k,q}$, but a large number of coefficients of the local terms are fixed, as shown in Table \ref{tab:6pt_CTs}. We have carried out this analysis to $\mathcal{O}(s^5)$.

However, this changes dramatically when the Levi–Civita terms are {\em excluded}.\footnote{We find that upon imposing the constraints from the SUSY Ward identity, the absence of Levi–Civita symbols in $\mathbb{Z}_1$ implies that any other 6-scalar amplitude in the $\mathcal{N}=4$ SYM EFT is also free of them.} 
Starting at order  $\mathcal{O}(s^3)$,  the $\mathcal{N}=4$ SUSY Ward identity \reef{consistency} can in this case no longer be solved for generic values of the lower-point Wilson coefficients. In fact, we find that \reef{consistency} requires the following {\em nonlinear} constraints among the 4-point Wilson coefficients $a_{k,q}$:
{\allowdisplaybreaks
\begin{align}
\mathcal{O}(s^3)\!: \quad & g^2 a_{2,0}= \frac{2}{5}a_{0,0}^2 \,, \qquad \ \, g^2 a_{2,1}= \frac{1}{10} a_{0,0}^2\,, \nonumber \\[1.5mm]
\mathcal{O}(s^4):\! \quad & g^2a_{3,1}= 2 g^2 a_{3,0}-a_{1,0} a_{0,0} \,, \nonumber \\[1.5mm]
\mathcal{O}(s^5)\!: \quad &g^4 a_{4,0}= \frac{8}{35} a_{0,0}^3\,, \qquad  g^4 a_{4,1}= -\frac{g^2}{2} a_{1,0}^2 + \frac{6}{35} a_{0,0}^3 \,, \qquad g^4 a_{4,2}= -g^2 a_{1,0}^2 + \frac{23}{70} a_{0,0}^3\,, \nonumber \\[1.5mm]
\mathcal{O}(s^6)\!: \quad &g^4 a_{5,1}=3g^4 a_{5,0} - g^2 a_{3,0} a_{0,0} - \frac{2}{5} a_{1,0} a_{0,0}^2 \,, \nonumber \\
&  g^4 a_{5,2}= 5 g^4 a_{5,0} - 2 g^2 a_{3,0} a_{0,0} - \frac{1}{2} a_{1,0} a_{0,0}^2 \,, \nonumber \\[1.5mm]
\mathcal{O}(s^7)\!: \quad &g^6 a_{6,0}= \frac{24}{175} a_{0,0}^4 \,, \quad \ \, g^6 a_{6,1} = - g^4 a_{3,0} a_{1,0} + \frac{6}{35} a_{0,0}^4\,, \nonumber \\
& g^6 a_{6,2} = - 3 g^4 a_{3,0} a_{1,0} + \frac{1}{2} g^2 a_{1,0}^2 a_{0,0} + \frac{61}{175} a_{0,0}^4\,, \nonumber \\
& g^6 a_{6,3} = - 4 g^4 a_{3,0} a_{1,0} + g^2 a_{1,0}^2 a_{0,0} + \frac{499}{1400} a_{0,0}^4\,, \nonumber \\[1.5mm]
\mathcal{O}(s^8)\!: \quad &g^6 a_{7,1}= 4 g^6 a_{7,0} - g^4 a_{5,0} a_{0,0} - \frac{2}{5} g^2 a_{3,0} a_{0,0}^2 - \frac{8}{35} a_{1,0} a_{0,0}^3 \,, \nonumber \\
&g^6 a_{7,2}= \frac{28}{3} g^6 a_{7,0} - 3 g^4 a_{5,0} a_{0,0} - \frac{9}{10} g^2 a_{3,0} a_{0,0}^2 + \frac{1}{6}g^2 a_{1,0}^3 - \frac{2}{5} a_{1,0} a_{0,0}^3 \,, \nonumber \\
&g^6 a_{7,3}= 14 g^6 a_{7,0} - 5 g^4 a_{5,0} a_{0,0} - \frac{7}{5} g^2 a_{3,0} a_{0,0}^2 + \frac{1}{2}g^2 a_{1,0}^3 - \frac{1}{2} a_{1,0} a_{0,0}^3 \,. \nonumber \\[-0.3cm]
\label{eq:nonlinear_constraints_akq}
\end{align}
}%
The first of these nonlinear constraints imply that the effective couplings $a_{2,0}$ and $a_{2,1}$ of the two SUSY-compatible $\Tr(D^4 F^4)$ interactions are completely fixed by the value of the coupling $a_{0,0}$ of the $\Tr(F^4)$ operator and the YM coupling $g$. Thus, there is no consistent {\em tree-level} $\mathcal{N}=4$ SYM  EFT with a $\Tr(F^4)$ interaction but without the two $\Tr(D^4 F^4)$ interactions! Similarly, all of the coefficients $a_{2k,0}$ are required if $a_{0,0}\ne 0$; so at least one SUSY-compatible $\Tr(D^{4k} F^4)$ operator is required too. In contrast, in the presence of $\Tr(F^4)$, SUSY does not require a $\Tr(D^2 F^4)$ operator, as can be seen from the coefficient $a_{1,0}$ being left unfixed.

From the SUSY results above, we infer a general pattern: the only 4-point Wilson coefficients that remain free after imposing maximal SUSY, tree-factorization, and the parity condition \reef{Z1isbarZ1cond} are $a_{2m-1,0}$ for $m=1,2,\dots$ and $a_{0,0}$. All other coefficients $a_{k,q}$ are fixed in terms of the previous ones. We expect this pattern to continue at higher orders. 
In fact, as we discuss in Section~\ref{sec:openstring}, we expect the nonlinear constraints to be complete when it comes to relations among the 4-point coefficients $a_{k,q}$. In other words, extending the analysis to higher points will not further constrain the $a_{k,q}$'s; it may, however, constrain some of the 5- and higher-point local operator coefficients further. 

As for the Wilson coefficients at 5 point (the effective couplings of operators $\Tr D^{2k}F^5$ compatible with linearized $\mathcal{N}=4$ SUSY), we summarized their count in  
Table \ref{tab:solution_ansatz_5pt}. In the last row of that table we show how many of the 5-point Wilson coefficients remain free after imposing the 6-point SUSY Ward identity \reef{consistency}. The lowest-dimension 5-point operators, the two at $\mathcal{O}(p^{9})$ and five at order $\mathcal{O}(p^{11})$, have all their coefficients completely fixed in terms of the 4-point Wilson coefficients $a_{k,q}$ and the YM coupling $g$. Of the 14 operators at $\mathcal{O}(p^{13})$, 13 are fixed by the 6-point SUSY Ward identity \reef{consistency} while one coefficient remains free, etc. Going to higher points in the SUSY analysis may well fix more, however, not all; see Section~\ref{sec:openstring}.

At 6-point, the coefficients of the local contact terms are almost all fixed (most vanish, the rest are expressed in terms of lower-point Wilson coefficients), as shown in the last row of Table \ref{tab:6pt_CTs}. For example, \reef{eq: 6pt_CTs} becomes 
\be
    \text{CT}_6[123456] = 
    a_{0,0} \big(
      s_{16} + s_{34} 
      -s_{234} - s_{345}
    \big)
         + \ldots \,.
\ee
The first freedom in CT${}_6$ is at order $\mathcal{O}(s^6)$, but it is quite possible that a higher-point analysis would fix more of these parameters. 

We emphasize that the nonlinear constraints \reef{eq:nonlinear_constraints_akq} have been derived using a bottom-up EFT approach; consequently, they must be satisfied by any UV completion of $\mathcal{N}=4$ SYM theory so long as the low-energy limit has EFT amplitudes with tree-level factorization and absence of Levi–Civita's in the scalar amplitudes. 
One such example is the massless open string amplitude. In the following, we examine examples of amplitudes from different UV completions and consider the consequences of the nonlinear relations among the $a_{k,q}$'s.

\subsection{Consequences: Compatibility vs.~Ruling Out}
\label{sec:consequences}

In this section, we consider explicit examples of 4-point amplitudes that are compatible with linearized maximal SUSY at 4 points and test if they are compatible with the nonlinear constraints \reef{eq:nonlinear_constraints_akq}. As discussed in Section~\ref{sec:SYM_3_4pt_amps}, the 4-point $\mathcal{N}=4$ SUSY Ward identities allow any amplitude of the form
\be
 \label{4ptprototype}
 A_4[--++] = \<12\>^2 [34]^2 F(s,u)\,,
 ~~~\text{with}~~~F(u,s) = F(s,u).
\ee 
For examples where the low-energy expansion of $F(s,u)$ reproduces the Parke–Taylor amplitude at leading order and otherwise only has local terms, we test if the Wilson coefficients obey the nonlinear constraints \reef{eq:nonlinear_constraints_akq}. This way, we assess whether there exists for these amplitudes an underlying EFT compatible with the joint constraints of 6-point tree-factorization, maximal SUSY, and the parity condition \reef{Z1isbarZ1cond}.

\subsubsection{Open Superstring Tree-Level Amplitudes}
\label{sec:openstring}
The 4-point type IIB open superstring amplitude is the gluon Veneziano amplitude 
\begin{equation}
\label{eq:Veneziano_ampl_4pt}
 A^\text{str}_4[--++] = 
-g^{2}\alpha'^{2}\<12\>^2 [34]^2\,\frac{
\Gamma(-\alpha' s)
\Gamma(-\alpha'u)}
{\Gamma\big(1-\alpha'(s+u)\big)}\,.
\end{equation}
Expanding it at low-energy $s \alpha', u\alpha' \ll 1$,
we identify the first Wilson coefficients $a_{k,q}$ to be 
\begin{alignat}{3}
\label{eq:Wilson_coeffs_Veneziano}
  & a_{0,0} = g^2 \alpha'^2 \zeta_2 = g^2 \alpha'^2\frac{\pi^2}{6}\,, \quad && a_{1,0} = g^2 \alpha'^3 \zeta_3 \,, \quad & & a_{2,0} = g^2 \alpha'^4 \zeta_4  = g^2 \alpha'^4 \frac{\pi^4}{90}\,, \nonumber \\
   & a_{2,1} =g^2 \alpha'^4 \frac{\zeta_4}{4}\,, \quad && a_{3,0} = g^2 \alpha'^5 \zeta_5\,, \quad & & a_{3,1} = g^2 \alpha'^5 \Big( 2 \zeta_5 - \zeta_2 \zeta_3 \Big), \\
   & a_{4,0} = g^2 \alpha'^6 \zeta_6 = g^2 \alpha'^6 \frac{\pi^6}{945}\,, \quad && a_{4,1} = g^2 \alpha'^6 \Big( \frac{3}{4} \zeta_6 - \frac{1}{2} \zeta_3^2 \Big), \quad & & a_{4,2} = g^2 \alpha'^6 \Big( \frac{23}{16} \zeta_6 - \zeta_3^2 \Big), \nonumber
\end{alignat}
where $\zeta_b = \sum_{n=1}^\infty \frac{1}{n^b}$ are Riemann zeta values with integer $b$. 
For given $k$, all $a_{k,q}$ have transcendentality $k+2$. 
It is straightforward to check that all of the nonlinear constraints \eqref{eq:nonlinear_constraints_akq} are satisfied. Indeed, the 6-scalar open string tree amplitudes do not have any parity-odd terms.

Importantly, for the Veneziano amplitude, as we go up in Mandelstam order, the first appearance of each odd-integer zeta value $\zeta_\text{2m+1}$ is in the $a_{2m-1,0}$ coefficient, specifically
\be
 a_{2m-1,0}=g^2 (\alpha')^{2m+1} \zeta_{2m+1} \,,
 ~~~~m=1,2,3,\dots\,.
\ee
These are precisely the coefficients that maximal SUSY, tree-factorization, and the parity condition \reef{Z1isbarZ1cond} left unfixed (in addition to $a_{0,0}$).
Conjecturally, the odd zeta values $\zeta_{2m+1}$ do not have algebraic relations to any other Riemann zeta values; hence, we do not expect that extending the SUSY analysis to higher points would fix the values of $a_{2m-1,0}$. Consequently, the nonlinear constraints obtained in Section \ref{sec:SYM_nonlinear_constraints} are expected to be complete at 4-point order. 

We noted in Section \ref{sec:SYM_nonlinear_constraints} that the 6-point analysis completely fixes the low-energy expansion of the 5-point EFT amplitude up to and including $\mathcal{O}(p^{11})$ and the 6-point EFT amplitude to $\mathcal{O}(s^{5})$ in terms of the $a_{k,q}$'s and $g$, in addition to fixing a large set of higher-order coefficients for the local terms. The resulting 5- and 6-point EFT amplitudes can be compared with low-energy expansions of the open superstring 5- and 6-point  amplitudes using the results in 
Refs.~\cite{Mafra:2011nv,Mafra:2011nw,Stieberger:2006bh,Stieberger:2006te,Broedel:2009nsh,Broedel:2013aza}. We present details of this comparison in Appendix \ref{app:comparison_string}, where we find a complete match of every fixed coefficient. 

The low-energy expansion of the 4-point superstring tree-level amplitude contains only Riemann zeta values $\zeta_k$; however, multi-zeta values appear in higher-point amplitudes. Specifically, $\zeta_{3,5}$ appears first at  $\mathcal{O}(p^{15})$ at 5 points and at order $\mathcal{O}(s^7)$ at 6-points. Since $\zeta_{3,5}$ is algebraically independent from the $\zeta_k$'s, the corresponding coefficient in the low-energy expansion should not be fixed by maximal SUSY, tree-factorization, and the parity condition \reef{Z1isbarZ1cond}. Yet, we do expect that the other lower-order coefficients in the 5-point amplitude left free by maximal SUSY, tree-factorization, and the parity condition \reef{Z1isbarZ1cond} would be fixed in terms of the $a_{k,q}$'s if the analysis were extended to 7-point or higher.

\subsubsection{Ruled Out: Infinite Spin Tower}
\label{sec:IST}
An amplitude that frequently shows up in the tree-level S-matrix bootstrap is the Infinite Spin Tower (IST) amplitude, which describes the exchange of an infinite set of same-mass massive states with spin $J=0,1,2,3,\dots$ \cite{Caron-Huot:2020cmc,Albert:2022oes,Berman:2023jys,Berman:2024eid,Berman:2025owb}. In the context of $\mathcal{N}=4$ SUSY, the IST amplitude takes the form
\begin{equation}
\label{A4IST}
     A_4[--++] = 
\<12\>^2 [34]^2
  \bigg(    
  -\frac{g^2}{su}
  +
\frac{\lambda^{2}}
{(m^2-s)(m^2-u)}
 \bigg) \,,
\end{equation}
where $\lambda$ sets the scale of the 3-point coupling constant. An infinite tower of spins with the same mass is considered unphysical from the point of view of both string theory and particle phenomenology, but is not ruled out by positivity at 4-point.\footnote{Arguments to exclude infinite tower amplitudes are discussed in Refs.~\cite{Cheung:2024uhn,Caron-Huot:2016icg,Cheung:2023uwn}.} Closed-form expressions for the higher-point versions of such an amplitude are unknown.

We find that the Wilson coefficients 
\begin{equation}
  a_{k,q} = \frac{\lambda^2}{m^{2k+4}}\,, \quad \text{ for all }k,q\,,
\end{equation}
 of the IST amplitude \reef{A4IST} 
are {\em incompatible} with the nonlinear constraints. Specifically, there is no choice of $\lambda$ and $g$ such that  \eqref{eq:nonlinear_constraints_akq} holds. Thus, if there exist higher-point Infinite Spin Tower amplitudes compatible with maximal SUSY and tree-level factorization, their 6-scalar 
amplitudes must have parity-odd terms.

\subsubsection{Ruled Out: Tree-Level Exchange of a Massive $\mathcal{N}=4$ Supermultiplet}
\label{s:massiveN4tree}
A very simple example of a 4-point amplitude of the form \reef{4ptprototype} is
\begin{equation}
    A_4[--++] = 
\<12\>^2 [34]^2
  \bigg(    
  -\frac{g^2}{su}
  + \frac{\lambda^{2}}
{m^2-s} 
+ \frac{\lambda^{2}}
{m^2-u}
\bigg)
\,.
\end{equation}
This describes the exchange of a massive $\mathcal{N}=4$ supermultiplet with mass $m$ and coupling $\lambda$ to the massless external states. A spin-0 state of the massive supermultiplet is exchanged in the $s$-channel and a spin-$2$ state in the $u$-channel. 
Expanding the amplitude at low-energy, the Wilson coefficients are found to be
\begin{equation}
  a_{0,0}= \frac{2\lambda^2}{m^2}\,, \quad \qquad a_{k,0} = \frac{\lambda^2}{m^{2k+2}}\,, \quad \text{ for all }k\,,
\end{equation}
while all other Wilson coefficients vanish. In this case, the nonlinear constraints from \eqref{eq:nonlinear_constraints_akq} cannot be satisfied for any nonzero choice of couplings $g$ and $\lambda$.  Thus, for such a theory to be consistent with maximal SUSY and tree-level factorization, the 6-scalar amplitudes must include parity-odd terms. 

This result reflects that  $\mathcal{N}=4$ SUSY is significantly more restrictive than $\mathcal{N}=1$ or $\mathcal{N}=2$ SUSY. For example, in   $\mathcal{N}=1$ SYM, we are free to add massive supermultiplets with interactions such as $\Tr (\Phi W_\alpha W^\alpha)$, where $\Phi$ is the superfield of a massive chiral supermultiplet. This contributes   
$ \frac{\<12\>^2 [34]^2}
{m^2-s}$ to the helicity amplitude
$A_4[--++]$, but does not include a $u$-channel exchange.

It is useful to compare this with the string tree-level amplitude. 
In the 4d open string Veneziano amplitude \reef{eq:Veneziano_ampl_4pt}, the lowest massive state exchanged has mass $m^2 = 1/\alpha'$ and is exactly a massive $\mathcal{N}=4$ supermultiplet of the form discussed here. However, the full massive tower of states, $m_n^2 = n/\alpha'$ for $n=1,2,3,\dots$, restores the compatibility with maximal SUSY, tree-factorization, and the parity condition \reef{Z1isbarZ1cond} at higher points, as we have seen in Section \ref{sec:openstring}.

\subsubsection{Ruled Out: Loop of a Massive $\mathcal{N}=4$ Supermultiplet (Coulomb Branch)}
\label{sec:Coulomb_branch}

Instead of coupling a massive supermultiplet linearly to the gauge supermultiplet as discussed above, we could attempt to couple it quadratically. The massive supermultiplet would then contribute to $A_4[--++]$ starting at 1-loop order. Using  $\mathcal{N}=4$ SUSY, the leading contribution to the $A_4[--++]$ comes from the 1-loop box diagram with a massive $\mathcal{N}=4$ supermultiplet running in the loop and the result is~\cite{Davydychev:1993ut,Berman:2023jys}
\begin{equation}
\label{Coulomb4pt}
    A_4[--++] = 
\<12\>^2 [34]^2
  \bigg(  - \frac{g^2}{su} + \frac{g'^4}{m^4} \Gamma(5/2) \sum_{i,j=0}^\infty \frac{\Gamma(1+i) \Gamma(1+j)}{\Gamma(5/2+i+j)} \Big( \frac{s}{4m^2} \Big)^i \Big( \frac{u}{4m^2} \Big)^j \bigg)  
  \,,
\end{equation}
where $g'$ is the coupling of the massive supermultiplet to the massless SYM external states. 

In the low-energy expansion, taking the mass $m$ as the UV scale, the Wilson coefficients obtained from \reef{Coulomb4pt} are at the lowest orders
\begin{equation}
\label{akqCoulomb}
  a_{0,0} = \frac{g'^4}{m^4}\,, \quad a_{1,0} = \frac{g'^4}{10 m^6}\,, \quad a_{2,0} = \frac{g'^4}{70m^8}\,, \quad a_{2,1} = \frac{g'^4}{140m^8}\,.
\end{equation}
Since $a_{2,1} \ne a_{2,0}/4$, the Wilson coefficients \reef{akqCoulomb} are {\em not} compatible with the nonlinear  constraint \eqref{eq:nonlinear_constraints_akq}. 
This should in a sense not be surprising because the nonlinear terms in the 6-point amplitude arise from the 3-particle channel, which contains two 4-point EFT interactions that effectively count as a 2-loop contribution. Working consistently at one loop, one can then remove these 3-particle-channel diagrams, but the effect is then that $a_{2,0}$ and $a_{2,1}$ are set to zero by our SUSY Ward identity. Clearly that is in contradiction with \reef{akqCoulomb}. We thus conclude that the corresponding 1-loop 6-scalar amplitudes must include Levi–Civita's in the local terms.  

A realization of the quadratically-coupled massive $\mathcal{N}=4$ supermultiplet to the massless SYM states is the Coulomb branch of $\mathcal{N}=4$ SYM. 
In that scenario, the massive states arise as the $W$-strings extended between two stacks of D3-branes. 
Consider the Coulomb branch with the SYM gauge group $SU(N)$ being spontaneously broken to $SU(N_1) \times SU(N_2)$, with $N_1 + N_2 = N$. The tree-level amplitudes with mixed massless and massive external states were studied in \cite{Craig:2011ws,Kiermaier:2011cr,Herderschee:2019dmc}. 
Here, we are interested in the scattering of only massless external states transforming in the adjoint of $SU(N_1)$. At leading order in the low-energy expansion, these are just pure $\mathcal{N}=4$ SYM amplitudes. However, they receive loop corrections from massive $W$-supermultiplets. At one-loop order, the only contribution is from a box diagram and the resulting amplitude is exactly \reef{akqCoulomb} with $g'=g$; this does not obey the nonlinear constraints \eqref{eq:nonlinear_constraints_akq}.

While a priori one may think that the Coulomb branch breaks the $SU(4)$ R-symmetry, there exist more complicated breaking patterns that can restore this symmetry. Similarly, one can also consider whether the diagrams with massless states at 1-loop interfere with the assumption of tree-level factorization, but scaling $N_2 \gg N_1$ handles that.\footnote{We are grateful to Simon Caron-Huot for pointing out these two loopholes in our original argument.}
Thus, the 6-scalar amplitudes of the Coulomb branch must include Levi–Civita's in the local terms.\footnote{An alternative way to avoid the massless loops is to consider the case $N_1 =1$, i.e. the EFT of a massless Abelian $\mathcal{N}=4$ supermultiplet. There is no Parke–Taylor scattering, no cubic interactions, and the leading interaction is  $\Tr F^4$.  We have analyzed this case (see Appendix \ref{app:DBI} for details) and find that {\em no nonlinear constraints} arise on the 4-point Wilson coefficients from the combination of tree-level factorization and maximal SUSY of the 6-point amplitude. So, in the Abelian case the Coulomb branch amplitudes survive.}

\section{Positivity Bounds and Finding String Theory}
\label{sec:pos}

In this section, we combine the nonlinear constraints from maximal SUSY, tree-factorization, and the parity condition \reef{Z1isbarZ1cond} obtained in Section~\ref{sec:SYM_nonlinear_constraints} with positivity constraints of the S-matrix bootstrap.

\subsection{Dispersion Relations, Positivity, and Convex Allowed Regions}\label{sec:bootsetup}

In our analysis so far, we have assumed maximal SUSY, tree-factorization, and the parity condition \reef{Z1isbarZ1cond} in the EFT. We now additionally assume that the EFT arises as the low-energy limit of a {\em unitary} theory whose perturbative 4-point amplitudes are {\em analytic} in the complex $s$-plane away from the real $s$-axis and obey a Froissart–Martin-like bound \cite{Froissart:1961ux,Martin:1962rt}\footnote{Our assumptions are generally considered to be theorems in axiomatic quantum field theory \cite{PhysRev.135.B1375,Martin:1965jj,Correia:2020xtr,Caron-Huot:2020cmc}. See Ref.~\cite{Arkani-Hamed:2020blm} for discussion on the Froissart–Martin bound.}  at large $s$ for constant $u<0$ or $t<0$, respectively:
\be
\label{A4regge}
  \lim_{|s|\to \infty}\frac{A(s,u)}{s^2} = 0
  \,,
  ~~~~~
  \lim_{|s|\to \infty}\frac{A(s,-s-t)}{s^2} = 0\,.
\ee
Since $\mathcal{N}=4$ SUSY requires all 4-point amplitudes to be proportional, we focus on the scalar amplitude
\be
  \label{bootA4}
   A_4[zz\bar{z}\bar{z}] = s^2 F(s,u)
\ee
from \reef{eq: scalar_ampls_SYM}. Thanks to the overall $s^2$ factor in  \reef{bootA4}, we obtain unsubtracted dispersive representations for each of the Wilson coefficients in the low-energy expansion of $F(s,u)$ in equations \reef{eq:def_F_SYM}-\reef{eq:def_f_SYM}. Assuming the spectrum to have a mass gap $M_\text{gap} > 0$, the result is
\begin{equation}
\label{eq:dispersiveSUSY}
a_{k,q}= \sum_{\ell=0}^\infty\int_{1}^{\infty} dy \,\rho_{\ell}(y) \,y^{-k-1}
v_{\ell,q}\,, 
\qquad P_{\ell}(1+2\delta)=\sum_{q=0}^{\ell} v_{\ell,q}\delta^{q} \, ,
\end{equation}
where $y= s'/M_\text{gap}^{2}$ and all $a_{k,q}$ have been rescaled to be dimensionless in units of $M_\text{gap}$. The integral over the center-of-mass energy squared $s'$ from $M_\text{gap}^{2}$ to $\infty$ links the UV spectrum above $M_\text{gap}$ to the IR behavior encoded in the 4-point Wilson coefficients $a_{k,q}$. 
The numbers $v_{\ell,q} \ge 0$ are the coefficients of the Legendre polynomials;\footnote{Not to be confused with the variables $v_{k,q}$ appearing in the 5-point amplitude from Section~\ref{sec:SYM_5pt_ansatz}.} it is highly relevant for the analysis that they are non-negative numbers. 
The full derivation of \reef{eq:dispersiveSUSY}, along with a more detailed statement of the assumptions, can be found in \cite{Berman:2023jys,Berman:2024wyt}. 
The assumption of unitarity is imposed in a rather weak form, namely as simply the requirement of positivity of the spectral density $\rho_{\ell}(M^{2}) > 0$. 

An immediate consequence of \reef{eq:dispersiveSUSY} is that $a_{k,q} \ge 0$ and $a_{k,q} \ge a_{k',q}$ for $k\le k'$. Analyzing the consequences of the dispersive representation \reef{eq:dispersiveSUSY} leads to universal two-sided bounds on ratios of the Wilson coefficients \cite{Berman:2023jys}. In a few cases, such bounds can be determined analytically, for example 
\be
  \label{unibounds}
  0\le \Big(\frac{a_{1,0}}{a_{0,0}}\Big)^2 \le \frac{a_{2,0}}{a_{0,0}}  \le \frac{a_{1,0}}{a_{0,0}} \le 1\,.
\ee
This restricts the values of  $a_{1,0}/a_{0,0}$ and $a_{2,0}/a_{0,0}$ to the allowed region defined as the convex wedge between the parabola and the straight line with slope 1 through the origin, as shown in the plot on the left of Figure \ref{fig:allowedRegs}.

More generally, the positivity bounds have to be computed numerically. The crossing conditions $s \lra u$ are imposed as null constraints up to some maximal order in the derivative expansion, i.e.~including crossing for the $a_{k,q}$ up to some maximal $k \le k_\text{max}$.\footnote{See \cite{Berman:2023jys} for details of the constant $u$- and constant $t$-constraints and the formulation of the dispersive representation as an optimization problem.} We use the semi-definite optimizer SDPB \cite{Simmons-Duffin:2015qma} to obtain the numerical bounds. The plot on the right in Figure \ref{fig:allowedRegs} shows the $k_\text{max}=7$ numerical bounds (dark gray) for the allowed parameters space in the $(a_{2,0}/a_{0,0},a_{2,1}/a_{0,0})$-plane.

In general, any positive linear combination of amplitudes with positive spectral density also obeys positivity, hence any allowed region resulting from \reef{eq:dispersiveSUSY} is necessarily convex. 
In contrast, the nonlinear constraints \reef{eq:nonlinear_constraints_akq} required by maximal SUSY, tree-factorization, and the parity condition \reef{Z1isbarZ1cond} are generally not respected by linear combinations of amplitudes. Hence, when they are incorporated into the bootstrap, we should expect more restrictive bounds that do not necessarily have to yield convex allowed regions. This is the subject of the next section.

\subsection{Bootstrap with Nonlinear Constraints: Non-Convexity}
\label{sec:nonlinbootstrap}

\begin{figure}[t]
\centering
\includegraphics[width=0.44\textwidth]{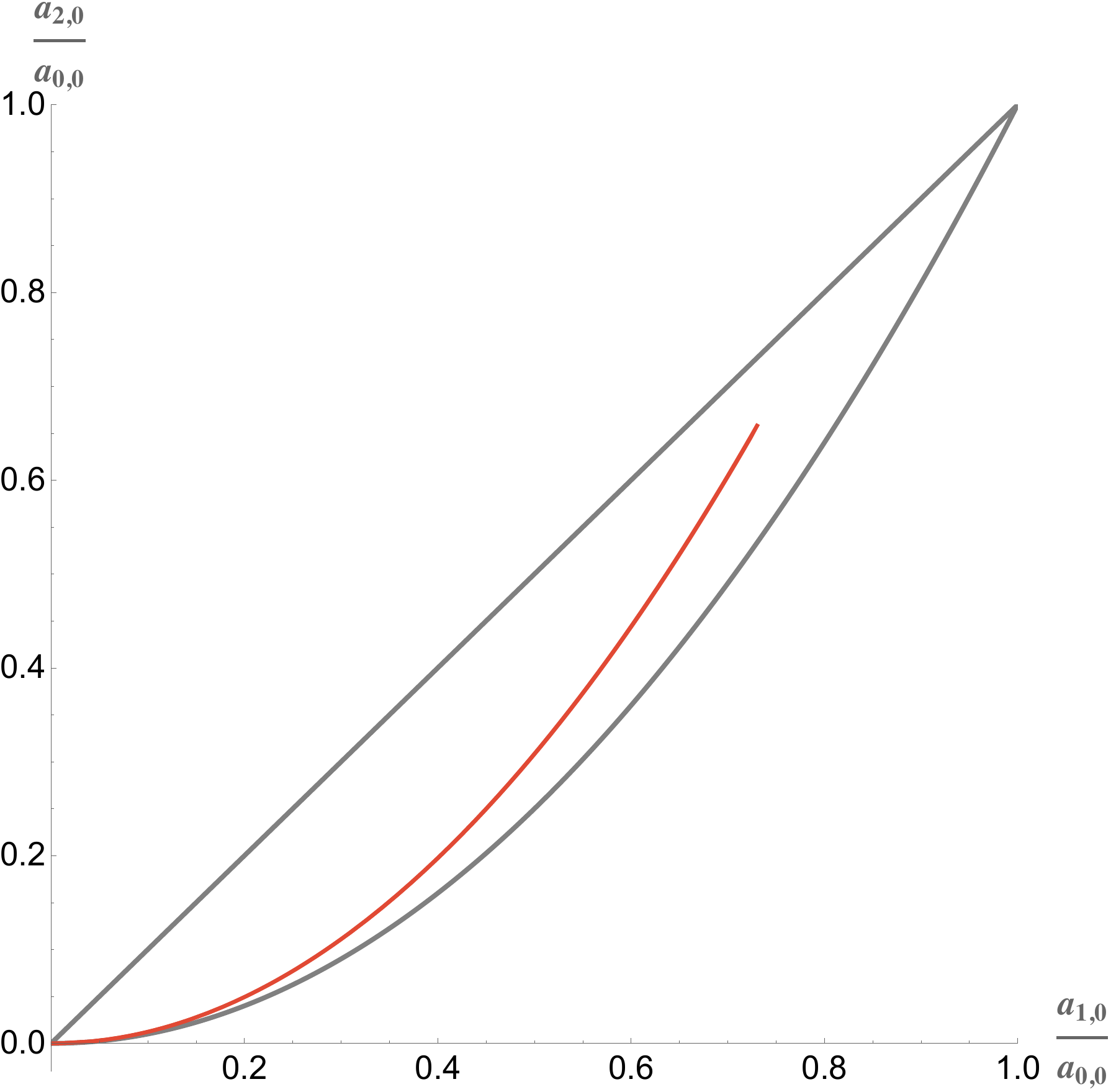}
~~~~~
\includegraphics[width=0.44\textwidth]{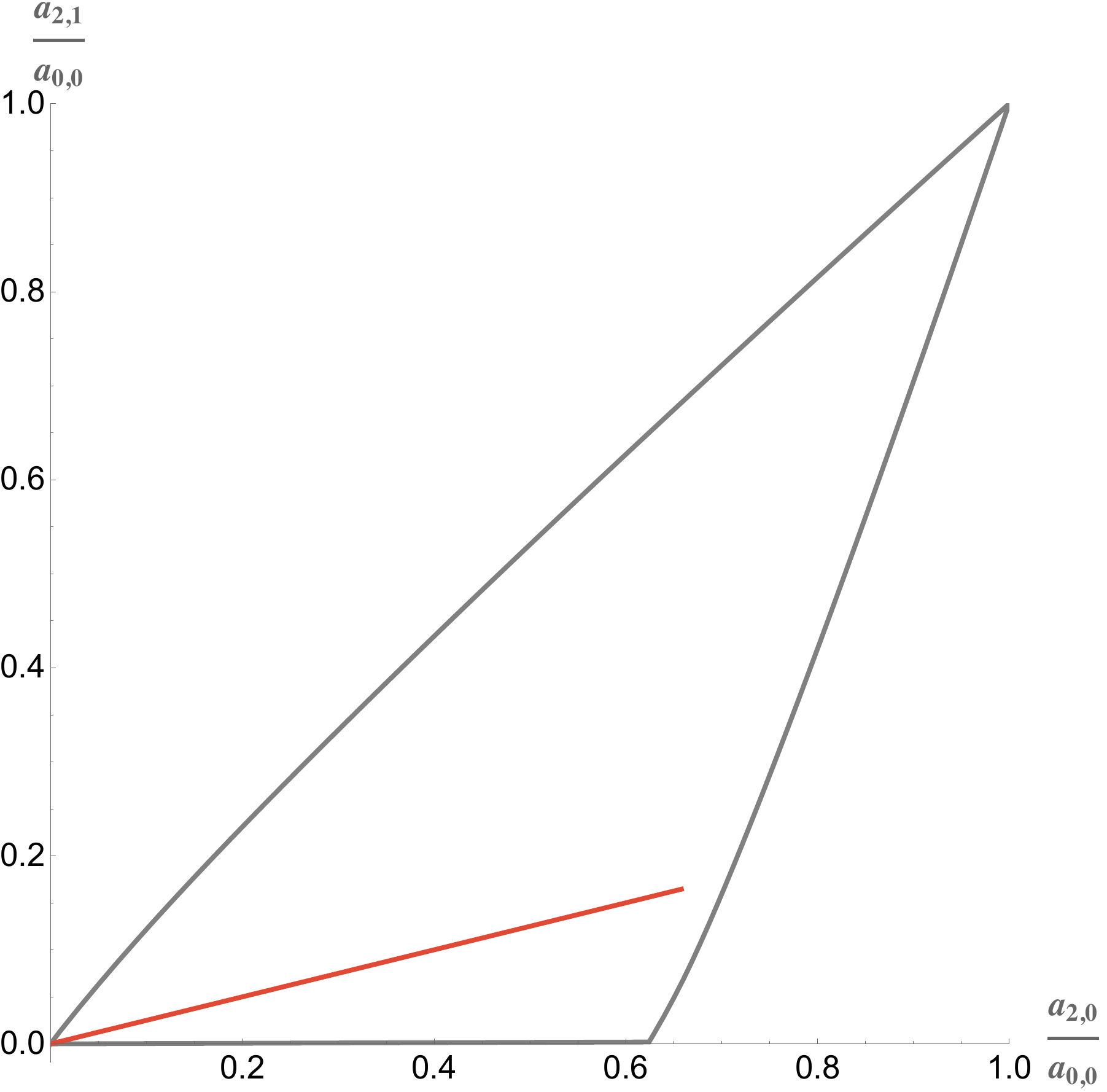}
\caption{4d positivity bounds. {\bf Dark Gray:} The universal positivity constraints restrict the lowest-order Wilson coefficients to lie in the convex regions bounded by the  dark gray curves. In the plot on the left, the curves are the  exact bounds \reef{unibounds}; on the right, the universal bounds are obtained numerically with SDPB for $k_\text{max}=7$. 
{\bf Red:}
When the nonlinear constraints \reef{eq:nonlinear_constraints_akq}  arising from SUSY and the parity condition \reef{Z1isbarZ1cond} with higher-point tree-level factorization in the EFT are imposed, the positivity bounds reduce the allowed regions to those shown in red. On the right, the constraints \reef{eq:nonlinear_constraints_akq} require $a_{2,1}/a_{0,0}=(1/4)a_{2,0}/a_{0,0}$, which directly restricts the allowed space to a line. However, it also imposes an upper bound on $a_{2,0}/a_{0,0}$: at $k_\text{max}=7$ it is $0.65802$, which can be compared with the string value $\zeta_4/\zeta_{2} \approx 0.657974\dots$. On the left, the red bounds are obtained numerically with SDPB at $k_\text{max}=7$: the upper and lower bound on $a_{1,0}/a_{0,0}$ for given  $a_{2,0}/a_{0,0}$ differ by less than $5 \times 10^{-4}$. This difference is not visible in the plot, hence the allowed region on the red appears as a curve. Figure \ref{fig:kmax345} illustrates how the bounds for lower values of $k_\text{max}$ converge to $k_\text{max}=7$ bounds shown here. 
}
\label{fig:allowedRegs}
\end{figure}

The constraints \reef{eq:nonlinear_constraints_akq} from maximal SUSY, the parity condition, and higher-point tree-level factorization are nonlinear in the $a_{k,q}$'s and also involve the YM coupling $g$. To incorporate them into SDPB, we eliminate $g$ and rewrite the relations such that they can be imposed as linear null constraints. Note, though, that a subset of the nonlinear constraints can directly be incorporated; for example the two $\mathcal{O}(s^3)$ constraints in \reef{eq:nonlinear_constraints_akq} imply the single linear condition $a_{2,1} = \tfrac{1}{4}a_{2,0}$. While this immediately reduces the entire 2d allowed region in the $(a_{2,0}/a_{0,0},a_{2,1}/a_{0,0})$-plane to a single line, as shown by the red line on the right of Figure \ref{fig:allowedRegs}, obtaining a strong bound for the maximal allowed value in that line requires incorporation of  additional nonlinear constraints.  

To rewrite the nonlinear constraints, we first solve for $g$ using the $a_{2,0}$ nonlinear constraint, giving $g^2 = 2 a_{0,0}^2/(5a_{2,0})$, and we use that to eliminate $g$ in all other relations. Second, we use that for any given numerical value of $\bar{a}_{2,0} = a_{2,0}/a_{0,0}$, many of the nonlinear constraints can be linearized and thereby directly implemented in SDPB. For example, for the lowest orders in $k$, we have
\be
  \label{linSUSYcond}
  \begin{split}
  4a_{2,1} - \bar{a}_{2,0} a_{0,0} &= 0\,,
  \\
  2 a_{3,1}- 4 a_{3,0}+5 \bar{a}_{2,0} a_{1,0} &= 0\,,
  \\
  14 a_{4,0}-20 \bar{a}_{2,0}^2 a_{0,0} 
  &=0\,,
  \\
  16 a_{4,2}- 32 a_{4,1} 
  +a_{4,0}&=0\,, ~~\text{etc}.
  \end{split}
\ee 
We incorporate these constraints into SDPB for $k \le k_\text{max}$ along with the regular crossing null constraints also up to $k_\text{max}$.
With that, we first determine the maximal allowed value of $\bar{a}_{2,0}$. For $k_\text{max}=7$, we find max$(\bar{a}_{2,0})$= max$(a_{2,0}/a_{0,0}) = 0.65802$.\ 
This sets the upper bound on the red line on the right in Figure \ref{fig:allowedRegs}, which is the strictly 1d allowed region in the $(a_{2,0}/a_{0,0},a_{2,1}/a_{0,0})$-plane when nonlinear constraints are included. 

\begin{figure}[t]
\centering
\includegraphics[width=0.44\textwidth]{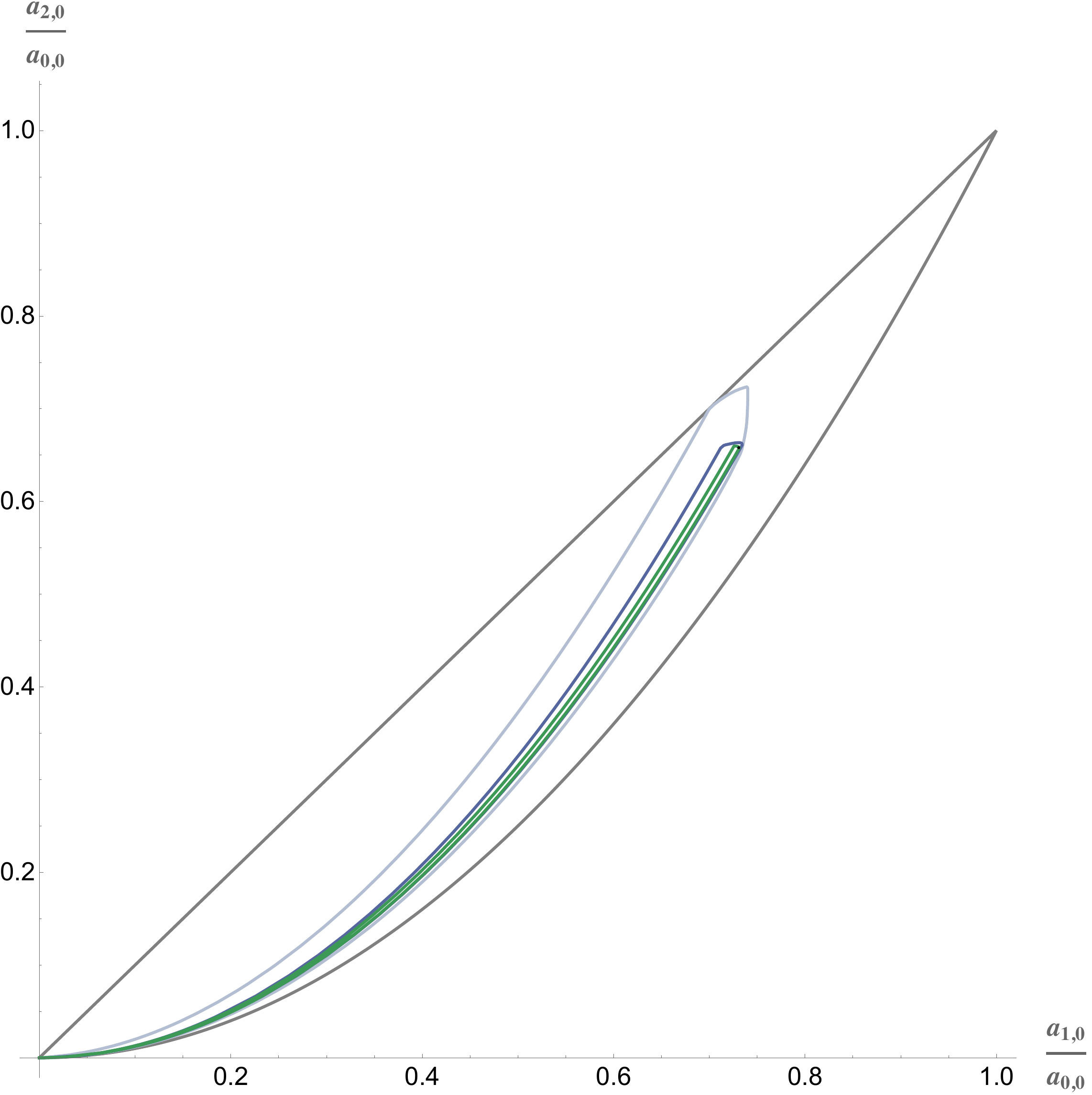}
~~~~~
\includegraphics[width=0.44\textwidth]{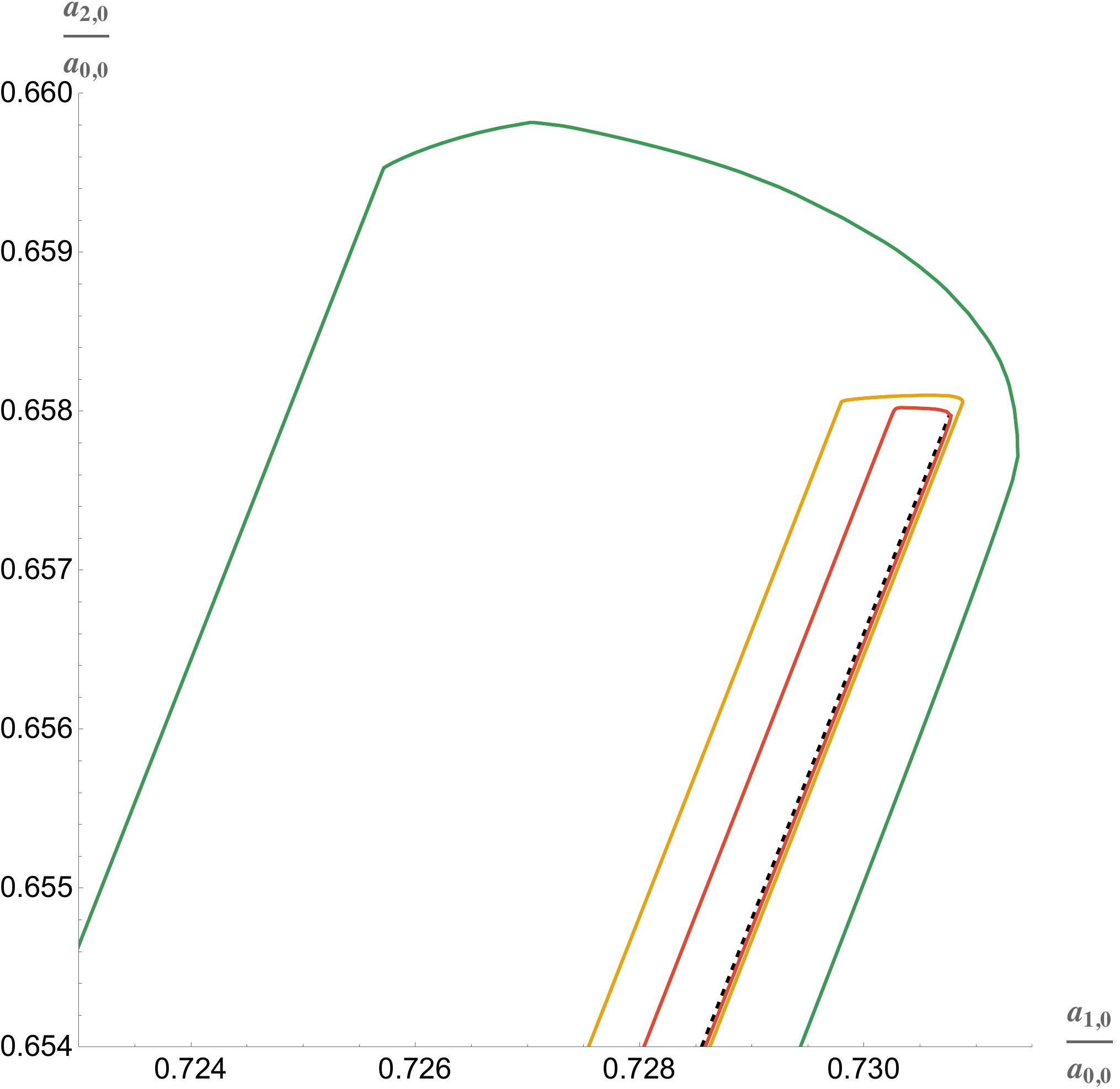}
\caption{The nonlinear constraints reduce the convex universal region bounded by the dark gray curves to the smaller non-convex regions shown on the left for $k_\text{max}=3$ (light blue) $4$ (dark blue), and $5$ (green). The small black dot indicates the Veneziano values of the Wilson coefficients for $\alpha' M_\text{gap}^2 = 1$. On the right, we zoom in on the region near the top of the $k_\text{max}=5$ region on the left and show the bounds for $k_\text{max}=5$ (green), $k_\text{max}=6$ (orange), and $k_\text{max}=7$ (red). The black dashed curve is the open string with $\alpha' M_\text{gap}^2 \le 1$. The bootstrap is done in 4d.}
\label{fig:kmax345}
\end{figure}

Turning to the $(a_{1,0}/a_{0,0},a_{2,0}/a_{0,0})$-plane, we compute for the given $\bar{a}_{2,0}=a_{2,0}/a_{0,0}$ in the range from $0$ to $0.65802$,  the minimum and maximum allowed values of $a_{1,0}/a_{0,0}$. For $k_\text{max} = 7$, the resulting bounds are shown in red on the left in Figure \ref{fig:allowedRegs}. Since we find that
\be
k_\text{max}=7\!:~~~~
\text{max}(a_{1,0}/a_{0,0})-\text{min}(a_{1,0}/a_{0,0}) 
< 5 \cdot 10^{-4}\,,
\ee
 the allowed region appears as a curve in Figure \ref{fig:allowedRegs}, though it actually has a small thickness; note also the non-convexity. To illustrate how  the allowed region converges as  $k_\text{max}$ increases, the plot on the left of Figure \ref{fig:kmax345} shows the bootstrap bounds in the $(a_{1,0}/a_{0,0},a_{2,0}/a_{0,0})$-plane for $k_\text{max} = 3$ (light blue), $4$ (dark blue), and $5$ (green). Zooming in on the tip of the allowed region, the plot on the right of Figure \ref{fig:kmax345} shows the bounds for $k_\text{max} = 5$ (green), $6$ (orange), and $7$ (red).

Extrapolating from these results, 
it looks like the allowed region
in the $(a_{1,0}/a_{0,0},a_{2,0}/a_{0,0})$-plane 
converges with increasing $k_\text{max}$ to a parabola of the form
\be
   \label{parabolic}
   \frac{a_{2,0}}{a_{0,0}} = \gamma 
   \Big(\frac{a_{1,0}}{a_{0,0}}\Big)^2.
\ee
The maximal values of $a_{1,0}/a_{0,0}$ for given $a_{2,0}/a_{0,0}$ appear to converge faster than the minimum values of $a_{1,0}/a_{0,0}$ (see zoomed plot in Figure \ref{fig:kmax345}). Therefore, fitting the values of max$(a_{1,0}/a_{0,0})$ to a curve of the form \reef{parabolic}, we find 
\be  \label{gammafit}
  k_\text{max} = 7\,:~~~~
  \gamma = 1.232
  \dots
\ee
The precise value depends on exactly how many points along the bound are included in the fit, but changing it within reason does not affect the digits shown. 

\vspace{2mm}
\noindent {\bf Comparison with the Open String.}   
Let us compare our results so far with string theory. Recall that for the derivation of the dispersion relations, we assume a mass gap $M_\text{gap}$, i.e.~there cannot be massive states contributing to the 4-point amplitude for $s < M_\text{gap}^2$. We express all $a_{k,q}$'s in units of $M_\text{gap}$, so the positivity bounds are with respect to $M_\text{gap}$. Including the $M_\text{gap}$ rescalings of the $a_{k,q}$'s, we find from \reef{eq:Wilson_coeffs_Veneziano} that
\be
 \frac{a_{1,0}}{a_{0,0}} 
 = \alpha' M_\text{gap}^2\frac{\zeta_3}{\zeta_2}\,,    
 ~~~~~ 
 \frac{a_{2,0}}{a_{0,0}} 
 = \alpha'^2M_\text{gap}^4\frac{\zeta_4}{\zeta_2}   
 ~~~~\implies~~~~
 \frac{a_{2,0}}{a_{0,0}}
 = \frac{\zeta_4 \zeta_2}{\zeta_3^2}
 \Big(\frac{a_{1,0}}{a_{0,0}} \Big)^2.
\ee
Thus, for the string, the $\gamma$ coefficient in \reef{parabolic} is 
\be
  \label{gammastr}
  \gamma^\text{string} = 
  \frac{\zeta_4 \zeta_2}{\zeta_3^2}
  \approx 
  1.23213\dots
\ee
The fitted value in \reef{gammafit} is indeed  close to the string value.
The parabolic curve with $\gamma = \gamma^\text{string}$ is shown as the black dashed curve on the right plot in Figure \ref{fig:kmax345}. What varies along the curve is the dimensionless quantity $M_\text{gap}^2 \alpha'$. To understand this better, note that the massless 4-point open string tree-level amplitude \reef{eq:Veneziano_ampl_4pt} has a spectrum of equally-spaced massive states with masses $m^2 = n/\alpha'$ for $n=1,2,3,\dots$. The lowest-mass state should be above our assumed mass gap, and that requires 
$1/\alpha' \ge M_\text{gap}^2$. Since the positivity bounds allow the first massive string state $1/\alpha'$ to be anywhere above $M_\text{gap}^2$, the freedom in choosing $1/\alpha'$ above $M_\text{gap}^2$ gives the one-parameter family of open string amplitudes, parameterized by $\alpha' M_\text{gap}^2 \le 1$. 
In the $(a_{1,0}/a_{0,0},a_{2,0}/a_{0,0})$-plane this is the parabola \reef{parabolic} with $\gamma = \gamma^\text{string}$ shown as the black dashed curve on the right plot in Figure \ref{fig:kmax345} with end point $(\zeta_3/\zeta_2,\zeta_4/\zeta_2)$. In the 
$(a_{2,0}/a_{0,0},a_{2,1}/a_{0,0})$-plane, this is the straight line 
$a_{2,1}/a_{0,0} = \tfrac{1}{4} a_{2,0}/a_{0,0}$ through the origin with endpoint at $(\zeta_4/\zeta_2,\tfrac{1}{4}\zeta_4/\zeta_2)$. Note that the value of $a_{2,0}/a_{0,0}$ at the endpoint of the string curves $\zeta_4/\zeta_2 \approx 0.65797\dots$ only differs from the $k_\text{max}=7$ bootstrap value max($a_{2,0}/a_{0,0})=0.65802$   by $4.6 \times 10^{-5}$.

In summary, the bootstrap results for the regions in the planes $(a_{1,0}/a_{0,0},a_{2,0}/a_{0,0})$
and
$(a_{2,0}/a_{0,0},a_{2,1}/a_{0,0})$ in  Figures \ref{fig:allowedRegs} and \ref{fig:kmax345}, respectively, show bounds that converge towards the string curves parameterized by $\alpha' M_\text{gap}^2 \le 1$. 
This is numerical evidence that the massless open string tree-level amplitude is the only weakly-coupled unitary completion of an $\mathcal{N}=4$ SYM EFT with tree-level factorization, given the parity condition on the 6-scalar amplitude. Numerical bounds for Wilson coefficients $a_{k,q}$ with $k>2$ are discussed in Section \ref{sec:unique}.

\subsection{Tree-Level \texorpdfstring{$\mathcal{N}=4$}{N=4} SUSY \texorpdfstring{$\implies$}{Implies} String Monodromy}
\label{sec:mono}
In the previous section, we made the connection to tree-level string theory via the assumption of positivity. However, our nonlinear constraints on the 4-point Wilson coefficients imply an interesting connection to the open string tree-level amplitude even without unitarity. 

The open string disk amplitudes obey a set of linear relations known as the ``string monodromy relations". They follow from the fact that the amplitudes resulting from different orderings of the vertex operators on the rim of the disk are related by contour deformations that pick up monodromy factors 
\cite{Plahte:1970wy,Stieberger:2009hq,Bjerrum-Bohr:2009ulz,Bjerrum-Bohr:2010mia,Bjerrum-Bohr:2010pnr}. For the 4-point disk amplitude, one form of the monodromy relation is
\be 
  \label{mono1}
  A_4[1324] 
  + e^{i\pi\alpha'u} A_4[1234] 
  + e^{-i\pi\alpha't} A_4[1342] = 0\,.
\ee
This is simply a linear relation among the different color-orderings of the planar amplitude. 
For the 4-point superstring amplitude, we can use the $\mathcal{N}=4$ SUSY Ward identities to write the amplitude in terms of a polarization factor times a symmetric function of $s$ and $u$ as in~\reef{4ptprototype}. The real and imaginary parts of \reef{mono1} then become equations for the function $F$:
\begin{equation}
\label{eq:monodromy}
\begin{split}
F(t,u)+\cos(\pi\alpha' u)\,F(s,u)+\cos(\pi\alpha't)\,F(s,t) & = 0\,,\\
\sin(\pi\alpha'u)\,F(s,u)-\sin(\pi\alpha't)\,F(s,t)& = 0\,.
\end{split}
\end{equation}
The function $F$ for the string amplitude in \eqref{eq:Veneziano_ampl_4pt} obeys these relations. In the leading low-energy limit where the Veneziano amplitude becomes the pure SYM amplitude, i.e.~$F(s,u) = -g^2/(su)$, the first monodromy relation in  \reef{eq:monodromy} is satisfied for $\alpha'=0$ because it simply 
becomes the $U(1)$ decoupling identity, and the $\alpha'$-expansion of the second monodromy relation in  \reef{eq:monodromy} gives at order $\mathcal{O}(\alpha')$ the BCJ relation~\cite{Bern:2008qj}, which is also satisfied by pure YM.

Including higher-derivative corrections to the SYM amplitude, there is no reason a priori to expect the EFT amplitude to satisfy the monodromy relations. In fact, plugging the low-energy expansion of $F$ from \reef{eq:def_F_SYM}-\reef{eq:def_f_SYM} into \reef{eq:monodromy} and solving order-by-order in the low-energy expansion, we find non-trivial linear relations among the Wilson coefficients $a_{k,q}$.
At the lowest order after pure YM theory, the monodromy relations \reef{eq:monodromy} require
\be 
 \label{a00Mono}
 a_{0,0} = g^2 \alpha'^2 \frac{\pi^2}{6} \,.
\ee
Because $a_{0,0}$ and $\alpha'$ are both dimensionful, this can be viewed as a choice of the EFT scale. 
Proceeding to the next orders assuming \reef{a00Mono}, the constraints are
\be 
 \label{akqMono}
 \begin{split}
 &a_{2,0}  = g^2 \alpha'^4 \frac{\pi^4}{90} \,,\,~~~~~~
 a_{2,1}  = g^2 \alpha'^4 \frac{\pi^4}{360}\,,\\
  &2a_{3,0}- a_{3,1}  = \alpha'^2 \frac{\pi^2}{6} a_{1,0}\,,\\
  &a_{4,0}  = g^2 \alpha'^6\frac{\pi^6}{945} \,,
  ~~~~~
  2a_{4,1} - a_{4,2}  =g^2 \alpha'^6 \frac{\pi^6}{15120}\,,
 \end{split}
\ee
as well as constraints for $k =5$ and beyond. 

Now, the connection to the work in this paper is that once $\alpha'$ is relabeled in terms of the leading EFT coefficients $a_{0,0}$ by the relation \reef{a00Mono}, the nonlinear relations \reef{eq:nonlinear_constraints_akq}
{\em imply} that the monodromy conditions \reef{akqMono} hold. Thus, maximal SUSY in conjunction with tree-level factorization and scalar-amplitude parity imply that the EFT amplitude obeys the string monodromy relations order by order in the low-energy expansion! Note that no properties of the UV theory were assumed for this; thus, it is purely a low-energy EFT analysis. Surprisingly, it results in a property among the color-ordered EFT amplitudes only known to arise from the worldsheet description of the open string.\footnote{In earlier work \cite{Chen:2022shl}, the string monodromy relations emerged similarly from an EFT analysis, but in the context of bi-adjoint scalar theory. In that case, it was found that if KK \cite{Kleiss:1988ne} and BCJ \cite{Bern:2008qj} relations were imposed for one of the two color-orderings of the bi-adjoint scalar tree-level EFT amplitudes, the amplitudes had to obey the string monodromy relations with respect to the other color ordering. As in this work, the analysis used tree-level factorization for higher-point amplitudes in the EFT and the scale of the EFT was fixed by a relation similar to \reef{a00Mono}. See \cite{Chen:2023dcx} for implications on generalizations of the KLT double-copy \cite{Kawai:1985xq} for EFTs.} 

Note that the tree-level SUSY constraints \reef{eq:nonlinear_constraints_akq} are stronger than the monodromy conditions \reef{a00Mono}-\reef{akqMono}. For example, for $k=4$, monodromy only requires $a_{4,0}$ and the linear combination $2a_{4,1}- a_{4,2}$ to be fixed whereas SUSY plus tree-level factorization and the parity condition \reef{Z1isbarZ1cond} fixes all three coefficients $a_{4,0}$, $a_{4,1}$, and $a_{4,2}$ completely in terms of $a_{0,0}^3$ and $a_{1,0}^2$.

\subsection{Comparison with Other Bootstraps of the Open String}
\label{sec:compare}
There have in recent years been various papers on the bootstrap of the open string tree-level amplitudes, including \cite{Cheung:2022mkw,Cheung:2023adk,Cheung:2023uwn,Huang:2020nqy,Chiang:2023quf,Berman:2023jys,Haring:2023zwu,Arkani-Hamed:2023jwn,Cheung:2024uhn,Albert:2024yap,Cheung:2024obl,Cheung:2025tbr}. The purpose of this section is to compare our analysis with some of these approaches, so that the similarities, differences and assumptions are clear. 

\subsubsection{Positivity and Monodromy}
\label{sec:unique}

It was proposed in \cite{Huang:2020nqy}, and further tested using numerical bounds computed with SDPB in \cite{Chiang:2023quf,Berman:2023jys}, that the Veneziano amplitude is the only solution to the positivity bounds if the EFT expansion of the 4-point amplitude is assumed to satisfy the monodromy relations \reef{a00Mono}-\reef{akqMono}. Specifically, it was shown in \cite{Berman:2023jys,Chiang:2023quf} 
that each $a_{k,q}$ with $k \le 8$ that was left unfixed by monodromy, was restricted by the positivity bounds to a very small interval around the Veneziano value with $\alpha' M_\text{gap}^2=1$. See for example equation~(1.3) as well as Figure 10 in \cite{Berman:2023jys} for how the bounds tighten when $k_\text{max}$ increases. 
Given that the bootstrap in \cite{Huang:2020nqy,Berman:2023jys,Chiang:2023quf} assumes the stringy property of monodromy relations, it is perhaps not so surprising that the output is the open string, but it was a useful test of the S-matrix bootstrap ideas. 

In contrast, in the present paper, we only assumed field theory properties. In particular, we have shown that the tree-level factorization in the EFT together with $\mathcal{N}=4$ SUSY and the parity condition implies (but is strictly stronger than) the string monodromies at EFT level. Since \cite{Huang:2020nqy,Berman:2023jys,Chiang:2023quf} already established that monodromy plus positivity restricts the Wilson coefficients to very small islands around the string values, we have not repeated the analysis here for the bootstrap of $a_{k,q}$ with $k>2$. However, the bootstrap in Section \ref{sec:nonlinbootstrap} does illustrate the important point that  from purely positivity and SUSY, the scale of $\alpha'$ is not fixed, but we can freely vary it via the dimensionless quantity $\alpha' M_\text{gap}^2 \le 1$. Hence the bootstrapped regions in Section \ref{sec:nonlinbootstrap} are curves rather than islands around points. The scale that is varied relative to 
$M_\text{gap}^2$ is equivalent to the choice of how $\alpha'$ is fixed relative to the Wilson coefficient $a_{0,0}$ in the monodromy relation \reef{a00Mono}.

\subsubsection{Positivity and Spectrum Input}
\label{sec:spectrum}

Inspired by the massless pion bootstrap results  \cite{Albert:2022oes,Albert:2023jtd,Albert:2023seb}, the $\mathcal{N}=4$ SYM EFT 4-point amplitude was bootstrapped with mild field theoretic assumptions about the lowest massive states in \cite{Berman:2024wyt,Albert:2024yap}. It was shown that absent of infinite towers of spin, the lowest state exchanged in the $s$-channel for the amplitude \reef{bootA4} had to be a scalar. At the next mass level a scalar and vector would be allowed, etc., with the maximal allowed spin growing by at most one unit at each mass level.\footnote{See \cite{Berman:2024kdh} for an analytic derivation of this consecutive spin property for color-ordered scalar theories and \cite{Berman:2024owc} for the analogue statement without color-ordering.}
Allowing for a scalar with mass-squared $M_\text{gap}^2$ to be exchanged at tree-level and then assuming a gap $\mu$ to the next massive state, the authors of \cite{Berman:2024wyt,Albert:2024yap} showed that the positivity bounds developed a sharp corner. This gave a 1-parameter family of corner theories parameterized by the value gap $\mu$. When the gap was chosen to be $\mu^2=2M_\text{gap}^2$, the corner was very close to the value of the Wilson coefficients for the Veneziano amplitude. 

Further, it was shown in \cite{Berman:2024wyt} that if the coupling of the scalar at $M_\text{gap}$ was chosen to be the value known from the Veneziano amplitude, the maximum gap allowed was found to be $2M_\text{gap}^2$ and the resulting bounds on the ratios of Wilson coefficients became tiny islands shrinking around the Veneziano amplitude with $\alpha' M_\text{gap}^2=1$.

In the paper \cite{Albert:2024yap}, the authors exploited the dispersion relations further to express the bounds on the $a_{k,q}$'s relative to the YM coupling $g$ rather than relative to $a_{0,0}$ as was done here and in \cite{Berman:2024wyt}. This required a smearing procedure. It could be interesting to implement it together with the nonlinear bounds \reef{eq:nonlinear_constraints_akq} so that one does not need to eliminate $g$ as we did in Section \ref{sec:nonlinbootstrap}.

Finally, in an ambitious setting, the authors of \cite{Alday:2025pmg} seek to bootstrap the full open string 4-point amplitude, even at finite coupling that interpolates between the extremes of the Coulomb branch amplitude and the Veneziano amplitude. The approach does not use higher point information, but instead involves a number of sophisticated ideas and techniques, including integrability and the AdS/CFT correspondence. 

\subsubsection{Positivity and Hidden Zeros/Split}
\label{sec:HZS}

The first time nonlinear constraints arising from the analysis of higher-point amplitudes were implemented into the positivity bootstrap for 4-point amplitudes was in \cite{Berman:2025owb}. In that case, no SUSY was assumed. Rather, what was studied was an adjoint scalar EFT with a leading $\Tr\, \phi^3$ interaction and single-trace higher-derivative corrections. It was known from \cite{Arkani-Hamed:2023swr,Arkani-Hamed:2024fyd} that both the pure $\Tr \,\phi^3$ theory as well as its string theory completion had tree-level  amplitudes with hidden zeros (the amplitudes vanish in certain kinematic regions) and hidden splits (the amplitudes split into products of lower-point amplitudes in subregions of the kinematic space). Assuming hidden zeros and splits to hold also for the tree-level amplitudes in the scalar EFT, it was shown in \cite{Berman:2025owb}  that the 4-point Wilson coefficients (similar to the $a_{k,q}$'s in this paper) had to satisfy nonlinear relationships involving the coupling of $\Tr \,\phi^3$ (playing the role of the Yang–Mills coupling in this paper). Indeed, the nonlinear relations found in \cite{Berman:2025owb} turn out to be a subset\footnote{For example, requiring the hidden splits for the 5-point tree-level EFT amplitude was found in \cite{Berman:2025owb} to fix a linear combination of $a_{2,0}$ and $a_{2,1}$ in terms of $g^{-2}a_{0,0}^2$ rather than each of them as in \reef{eq:nonlinear_constraints_akq}. The 6-point hidden zeros and split conditions did not restrict these two Wilson coefficients further.} of the ones we derived from maximal SUSY! That particular subset of course allowed the string amplitude but in addition they did not rule out the hidden zero version of the 4-point infinite spin tower amplitude. 

When the nonlinear relations found in \cite{Berman:2025owb} were implemented into the positivity bootstrap, the bounds of the simplest regions were shown to be non-convex with two sharp corners: the open string beta function amplitude and the hidden zero infinite spin tower. Additionally assuming the absence of the infinite spin tower at the mass gap and a small gap to the next state, the non-convex region was shown to bifurcate. The nontrivial part of the resulting allowed region was a small island shrinking with increasing $k_\text{max}$ around the values of the Wilson coefficients for the open string amplitude.

In contrast, the reason we did not have to make additional assumptions about the spectrum in the SUSY bootstrap in Section \ref{sec:nonlinbootstrap} was that our more restrictive nonlinear relations already disallowed the infinite spin tower amplitude (cf.~Section \ref{sec:IST}).

\subsubsection{Multi-Positivity Bounds}
\label{sec:deform}

Working in the context of a scalar EFT, the authors of \cite{Cheung:2025nhw} studied the relation between Wilson coefficients at higher and lower points via tree-level factorization in the amplitudes. Invoking positivity constraints, they found inequalities for a certain subset of Wilson coefficients. For example, one of their ``multi-positivity bounds" says that a product of a particular choice of a 4-point and a 6-point Wilson coefficient is bounded from below by the square of a certain 5-point coefficient. They use such bounds to rule out some proposed deformations of the Veneziano amplitude. 
However, the inequalities of \cite{Cheung:2025nhw} do not directly constrain the 4-point amplitude unless there is explicit knowledge about the Wilson coefficients of the higher-point amplitude. In contrast, in our work here, we use tree-level factorization at higher point to directly derive constraints on the 4-point Wilson coefficients from SUSY and the parity condition.

\section{Conclusions and Outlook}
\label{sec:conclusions}
In this paper, we have considered the constraints of $\mathcal{N}=4$ supersymmetry on the space of  weakly-coupled 4d gluon effective field theories in the planar limit. 
We have assumed that the leading-order low-energy interactions are those of $\mathcal{N}=4$ super Yang–Mills and that the color-ordered amplitudes of the EFT have standard tree-level factorization. The analysis is done at the level of the on-shell scattering amplitudes whose polynomial terms in the low-energy expansion are in one-to-one correspondence with the independent local higher-derivative operators in the EFT Lagrangian. 

Supersymmetry is imposed on the amplitudes via the SUSY Ward identities. Because these are derived from the action of SUSY on the asymptotic states, the SUSY Ward identities are linear relations among 
scattering amplitudes with SUSY-related external states. When the $\mathcal{N}=4$ SUSY Ward identities are combined with tree-level factorization for the 5- and 6-point EFT amplitudes, and assuming the 6-scalar amplitudes to be free of Levi–Civita symbols 
(or equivalently that $\mathbb{Z}_1=\overline{\mathbb{Z}}_1$), we find that highly nontrivial nonlinear relations \reef{eq:nonlinear_constraints_akq} are required among the Wilson coefficients contributing to the 4-point amplitude. 

The nonlinear constraints significantly reduce the number of free parameters of the EFT. Based on 4-point SUSY Ward identities alone, there are $\lfloor (k+2)/2 \rfloor$ independent maximally-SUSY 
operators of the form $\Tr D^{2k} F^4$; we denote their Wilson coefficients by $a_{k,q}$ for $q=0,1,2,\ldots,\lfloor k/2 \rfloor$. However, when the nonlinear constraints from imposing SUSY on the 5- and 6-point tree-level amplitudes are taken into account, all $a_{k,q}$'s, except $a_{0,0}$ and $a_{2k-1,0}$, are fixed in terms of Wilson coefficients of operators with fewer derivatives. Specifically, for  $k \le 7$  we have found that the only coefficients that remain free are $a_{0,0}$, the Wilson coefficient of $\Tr F^4$, and $a_{2k-1,0}$ for $k=1,2,3,\dots$. 

The nonlinear constraints are satisfied by the Wilson coefficients of the massless open string 4-point amplitude and indeed this example illustrates that no additional algebraic constraints among the free $a_{2k-1,0}$ can be expected. In contrast, the nonlinear constraints also rule out higher-point tree-level completions of certain 4-point amplitudes that from the 4-point perspective alone appear to be fine. This exclusion is demonstrated for the infinite spin tower, single massive supermultiplet exchange, and the  massive 1-loop amplitude. Also, the SUSY Ward identities at 4 points are trivially solved by a sum of two Veneziano amplitudes with different values of $\alpha'$, but our nonlinear constraints show that this is not compatible with tree-level factorization from higher point. 

From the perspective of the low-energy EFT, that is, without requiring specific properties of a UV completion or even its existence, the Wilson coefficients $a_{0,0}$ and $a_{2k-1,0}$ can take arbitrary values. We now seek to express the most general supersymmetric 4-point amplitude in closed form as a function of $a_{0,0}$ and $a_{2k-1,0}$. Surprisingly, the dependence on $a_{0,0}$ factorizes from that of the $a_{2k-1,0}$. Moreover, the contribution from the $a_{2k-1,0}$ exponentiates. Specifically, upon enforcing the nonlinear relations \reef{eq:nonlinear_constraints_akq}, we can rewrite the low-energy expansion in \reef{eq:def_F_SYM} and \reef{eq:def_f_SYM} as (setting $g=1$  for simplicity)
\be
\label{A4wU}
\begin{split}
  A_4[--++] =& \<12\>^2 [34]^2 F(s,u) 
  \\
  =& \<12\>^2 [34]^2 \,
  F_0(s,u) \,
  \exp\left ( \sum_{k=1}^{\infty}\frac{a_{2k-1,0}}{2k+1}\big(s^{2k+1}+t^{2k+1}+u^{2k+1}\big)\right ).
  \end{split}
\ee 
The $s \lra u$ symmetric function $F_0$ depends only on $a_{0,0}$. To the order we have computed, we  verified that $F_0$ can be resummed into the compact expression
\be
  \label{theF0}
  F_0(s,u) =\frac{-\alpha}{s u}\sqrt{\frac{\pi  s u \sin (\pi  \alpha (s+u))}{\alpha (s+u) \sin (\pi  \alpha s) \sin (\pi  \alpha u)}}\,,
\ee 
where we have introduced $\alpha$ such that  
$a_{0,0} =\alpha^2 \pi^2/ 6$.
These expressions arise purely from the low-energy EFT analysis, but were inspired by a factorized form of the open string tree-level amplitude \cite{Schlotterer:2012ny,Chen:2022shl}. Indeed, the massless open string Veneziano amplitude is reproduced if we make the choice  $a_{2k-1,0} \to \alpha^{2k+1}\zeta_{2k+1}$ for $k=1,2,\dots,$ and relabel $\alpha \to \alpha'$. With this choice, the apparent violation of tree-level simple-pole factorization in \reef{theF0} is eliminated in a simplification between the square-root expression in $F_0$ and a resummation of the sum in the exponential factor of \reef{A4wU} that reduces the product to the familiar gamma-function expression \reef{eq:Veneziano_ampl_4pt} for the Veneziano amplitude.

\vspace{2mm}
Without assumptions of a UV completion, any choice  of  $a_{0,0}$ and $a_{2k-1,0}$ gives a fully fixed 4-point EFT amplitude via the nonlinear relations \reef{eq:nonlinear_constraints_akq} and their expected extensions to higher-$k$. 
To constrain the Wilson coefficients further, we assumed in Section \ref{sec:pos} that the EFT arises as the low-energy limit of a unitary EFT with analytic perturbative 4-point amplitudes with the standard Regge behavior \reef{A4regge}. Dispersive representations then  link the low-energy Wilson coefficients to integrals over the spectrum above an assumed mass gap. When combined with the nonlinear relations  \reef{eq:nonlinear_constraints_akq}, we found that the numerical S-matrix bootstrap bounds appear to converge towards a unique solution, namely the open string. This strongly suggests that the only UV completion of a unitary strictly tree-level $\mathcal{N}=4$ SYM EFT is the open string.

\begin{figure}[t]
\centering
\includegraphics[width=0.94\textwidth]{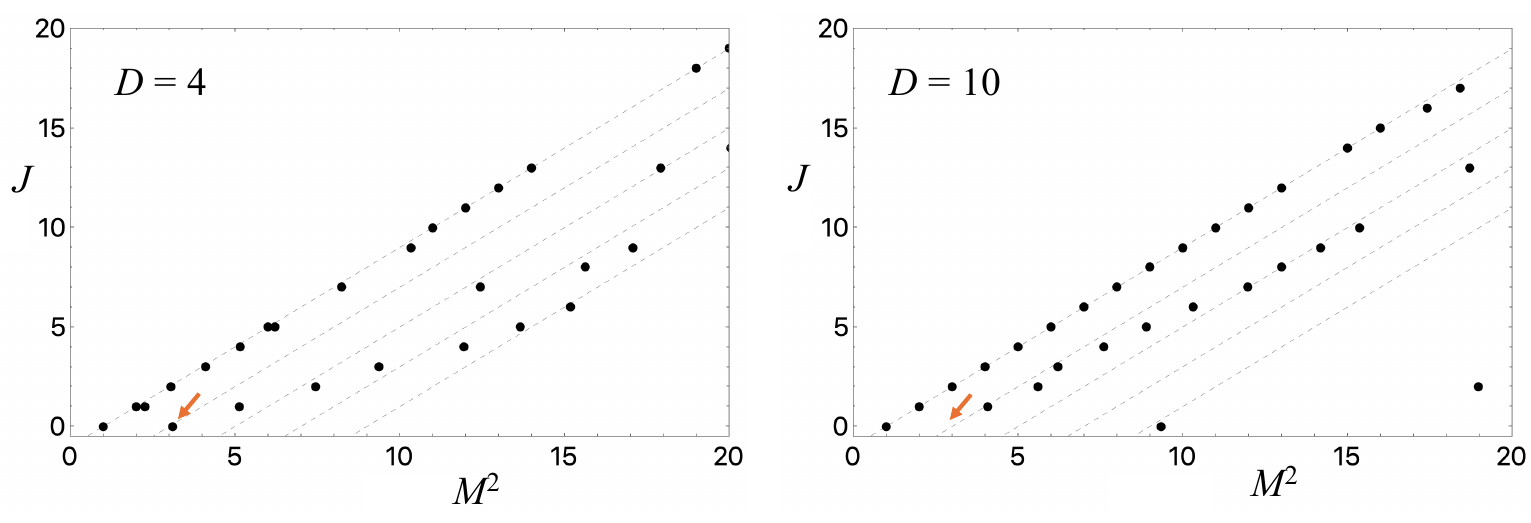}
\caption{Chew–Frautschi plots generated by SDPB when $a_{1,0}/a_{0,0}$ is maximized for $a_{2,0}/a_{0,0}$ fixed to the value of the string, $\zeta_4/\zeta_2$, assuming the leading Regge trajectory to have slope 1. The nonlinear SUSY relations are imposed for  $k\le 7$ while the regular crossing null constraints are imposed to $k_\text{max}=12$. The plot on the left shows the spectrum resulting from having 4d Legendre polynomials in the partial wave expansion used to the derive dispersion relations \reef{eq:dispersiveSUSY}. On the right, the $D=4$ Legendre polynomials are replaced by the $D=10$ Gegenbauer polynomials; this corresponds to bootstrapping the $D=10$ amplitude with the external states restricted to a $D=4$ subspace, but the massive exchanged states having spins in 10d representations. 
For $D=4$ and $D=10$, respectively, the maximal value of $a_{1,0}/a_{0,0}$ is within $9 \cdot 10^{-6}$ and $4 \cdot 10^{-7}$, respectively, of the string value. 
In both cases, the bootstrap finds multiple states along the leading Regge trajectory and a few along the daughter trajectories (indicated with dashed lines), but much less convincingly so, especially in $D=4$. One feature worth noting is that the $D=4$ spectrum includes the spin-0 state at mass-squared $3M_\text{gap}^2 = 3/\alpha'$, but it is absent in the $D=10$ spectrum (see indication by the orange arrow in both plots). This is exactly the situation for the string spectrum: the string is critical in $D=10$, where the coupling of the $3M_\text{gap}^2$ scalar is exactly zero, but in sub-critical dimensions it is non-vanishing (and positive). It is encouraging that this key feature of the string spectrum is captured by the SDPB spectrum.}
\label{fig:spectrum}
\end{figure}

\vspace{2mm}
When the optimizer SDPB is used to determine the numerical bounds, one can also extract the spectrum of states. Doing so with the nonlinear constraints included, SDPB finds several states along the leading slope-1 Regge trajectory, but it also has a tendency to include states with high spin at low masses. These states are known  analytically to be disallowed \cite{Berman:2024kdh,Berman:2024owc}, but  SDPB allows them with very small couplings at finite $k_\text{max}$. To eliminate such states, we make the additional assumption that the spectrum does not contain them; effectively, this means that  we assume the leading Regge trajectory to have slope one. (Note we make this additional assumption only for the purpose of making the spectrum plots, not for the bounds shown in Section \ref{sec:pos}.) Examples of the extracted spectra are shown in Figure~\ref{fig:spectrum}. A description of the spectra and how they were generated is provided in the caption. 

The string amplitude is critical in $D=10$: one way to see this is that, in the Veneziano amplitude, the scalar string state with mass-squared $3/\alpha'$ has coupling-squared which for $3< D < 10$ is positive, vanishes in $D=10$, and is negative for $D>10$. Thus, $D=10$ is the critical dimension of the string: the highest dimension for which the Veneziano amplitude is unitary. The vanishing of this scalar coupling is captured by the SDPB spectrum in $D=10$; cf.~Figure~\ref{fig:spectrum}. The $D=10$ spectrum also captures the leading state on the first daughter trajectory, namely the spin-1 state with mass-squared  $4/\alpha'$. A lot more can be done to analyze and clean up the spectra from SDPB, here we only  present a sample of what such plots look like. Compared to other perturbative bootstraps of the Veneziano amplitude (e.g.~\cite{Berman:2024wyt,Berman:2025owb}), we find that the nonlinear constraints make SDPB slightly better at finding states on the daughter trajectories, but to the orders computed here, it is still not overly convincing.

\vspace{2mm}
Looking back at the pure SUSY analysis of the EFT, we found that a large number of local terms in the ansatz for the 5-point and 6-point EFT amplitudes were fixed by the SUSY Ward identities. Extending the analysis to higher point may fix more of them; however, it would not further restrict the Wilson coefficients $a_{2k-1,0}$ of the 4-point EFT amplitude. The reason is that for the particular example of the string, they are given by odd zeta values,  $a_{2k-1,0} = (\alpha')^{2k+1}\zeta_{2k+1}$, which are (conjecturally) algebraically unrelated to any other single-zeta values. Hence, we cannot expect the $a_{2k-1,0}$ to satisfy any algebraic relations that would fix them in terms of $a_{0,0}$ or relate them to each other. The main motivation for extending the analysis to higher points would then be to restrict the higher-point coefficients further and isolate the first appearances of multi-zeta values, such as $\zeta_{3,5}$ in the 5-point amplitude. Note that while our analysis leads to a unique answer for the 4-point amplitude, the higher-point amplitudes may differ depending for example on whether full $SU(4)$ R-symmetry is imposed or not. 

\vspace{2mm}
The analysis in this paper was done for a non-gravitational EFT. An obvious question is whether an analogous analysis would constrain the Wilson coefficients in a low-energy supergravity EFT. We are currently investigating this for the case of $\mathcal{N}=8$ supergravity. In the $\mathcal{N}=4$ SYM EFT, the color-ordering was essential for making the connection between the nonlinear constraints among the Wilson coefficients and the string monodromy relations (see Section \ref{sec:mono}), but in gravity there is no color-ordering. In the Abelian limit of the $\mathcal{N}=4$ SYM EFT, i.e.~$\mathcal{N}=4$ super DBI, no restrictions were found on the Wilson coefficients from the combined constraints of maximal SUSY and tree-factorization (see Appendix \ref{app:DBI}). However, supergravity is different because of the non-vanishing 3-point gravitational interaction and we should expect Newton's constant $G_N$ to play the role of the Yang–Mills coupling $g$ in the supergravity analysis. Thus, we find it plausible that the $\mathcal{N}=8$ supergravity EFT analysis is going to reveal nonlinear constraints among the Wilson coefficients.\footnote{One could of course double-copy the $\mathcal{N}=4$ SYM EFT amplitudes to obtain $\mathcal{N}=8$ EFT amplitudes, but that procedure would not necessarily generate the most general supergravity EFT, not even with EFT versions of the double copy \cite{Broedel:2012rc,Carrasco:2019yyn,Chi:2021mio,Carrasco:2021ptp,Adamo:2022dcm,Chen:2023dcx}. In fact, a completely bottom-up supergravity analysis should not presume the double copy.}

\vspace{2mm}

Using the 4d spinor-helicity formalism and 4d on-shell superspace tied our analysis to four dimensions, but in earlier work \cite{Wang:2015jna} a quite similar analysis was done in 10d. There, tree-level factorization in conjunction with $\mathcal{N}=1$ on-shell superspace was used to restrict the $\alpha'$-corrections to the 10d open string amplitude. The focus in \cite{Wang:2015jna} was on using string-input at low-derivative order to restrict the higher-derivative operators, and --- just like in our analysis --- the signs of the nonlinear constraints are there. They could  have been extracted from the 10d analysis if the focus had been on the general EFT setting. 

\vspace{2mm}

A promising avenue for future work involves exploring whether the methods developed here could be generalized to loop-level calculations, and whether fundamental physical requirements like factorization and unitarity cuts might provide useful constraints for the S-matrix bootstrap program beyond weak coupling. At loop level, KK modes from compactified dimensions begin to contribute, introducing dependence on the compactification manifold. While this means there is no longer a unique solution as at tree level in less than ten dimensions, it opens an intriguing possibility: bootstrap constraints combined with maximal SUSY could potentially restrict the space of allowed compactifications. Alternatively, restricting to ten dimensions would eliminate KK-mode ambiguities entirely.

\vspace{2mm}

Another interesting direction is to move beyond flat space and ask whether our results extend to the conformal bootstrap and holographic settings. In close analogy to flat-space scattering amplitudes, boundary correlators of SUSY theories in anti–de Sitter (AdS) space should obey linearized SUSY Ward identities. Moreover, our analysis relies only on locality and factorization properties of the low-energy sector, which should hold in some form for AdS boundary correlators. It is therefore natural to anticipate that our results admit an AdS generalization. Furthermore, by the AdS/CFT correspondence \cite{Maldacena:1997re,Gubser:1998bc,Witten:1998qj}, such boundary correlators of gravitational theories in AdS are dual to correlators in particular CFTs. The canonical example is $\mathcal{N}=4$ SYM at strong coupling, which is dual to type IIB supergravity on AdS$_5\times S^5$. As expected, correlation functions in $\mathcal{N}=4$ SYM are known to obey SUSY Ward identities \cite{Belitsky:2014zha} that can be interpreted as the AdS generalization of the flat space SUSY Ward identities studied in this paper. Consequently, if our flat-space results can be promoted to gravitational theories in AdS, our computational strategy could provide a sharp tool for constraining, and perhaps even classifying, the space of maximally SUSY CFTs.

\section*{Acknowledgments}

The authors  
thank Jan Albert, Justin Berman, Paolo Di Vecchia, Loki L.~Lin, Juan Maldacena, Pedro Vieira and Edward Witten for useful discussions, as well as Justin Berman, Aditi Chandra, Nicholas Geiser and Loki L.~Lin for comments on the manuscript. The authors are indebted to Simon Caron-Huot for pointing out the importance of the local Levi–Civita terms in the 6-scalar amplitudes and for discussions related to the Coulomb branch 1-loop amplitudes.
The work of HE and RM is supported in part by Department
of Energy grant DE-SC0007859. RM is also funded by the Leinweber Postdoctoral Fellowship from the University
of Michigan. AH is grateful to the Simons Foundation as well as the Edward and Kiyomi Baird Founders’ Circle Member Recognition for their support.  HE would like to thank the Niels Bohr Institute and Niels Bohr International Academy for hospitality during the early parts of this work. RM would moreover like to thank the University of Wisconsin–Milwaukee for the hospitality received during part of this work.

\appendix

\section{BCFW Shifts and Residues}
\label{app:BCFW_residues}

When we evaluate residues on massless poles numerically, it is important to do so with proper momentum conservation in the restricted kinematics on the poles. This can be done by explicitly coding numerics, but a useful trick is to use BCFW shifts \cite{Britto:2004ap,Britto:2005fq}. We review this approach in this appendix.

Let us begin with a 2-particle pole $s_{ij}=0$. The aim is to find a way to set $s_{ij}=\<ij\>[ij]$ to zero while preserving momentum conservation in the rest of the diagram. Focusing first on the $\<ij\>$ pole, perform an $[X,i\>$-shift
\be
  \label{2partshift}
  |\hat{i}\> = |i\> - z |X\> \,,\qquad |\hat{X}] = |X] + z |i]\,, \quad \text{for \ } X \neq i,j\,.
\ee
No other helicity spinors are shifted. Then
\be
   \hat{s}_{ij} = \< \hat{i} j\> [ij] =  (\<ij\>-z\<Xj\>) [ij] 
   ~~~\text{vanishes when }~~ z = z_{ij} \equiv \frac{\<ij\>}{\<Xj\>}\,.
\ee
For $n \ge 5$, no other $s_{Xk}$ vanish under this shift when $z = z_{ij}$. In particular, for $k=i$, $\hat{s}_{Xi} = s_{Xi}\neq0$, whereas for $k \ne i$, we have
\be
 \hat{s}_{Xk} =  
 \<Xk\> \bigg( [Xk] +  \frac{\<ij\>}{\<Xj\>} [ik]\bigg) = -\frac{\<Xk\>}{\<Xj\>} \<j|X+i|k]\,,
\ee 
which is nonzero for generic momenta when $n \ge 5$.

Redefining $z = z_{ij} + \delta$, we therefore have
\be
\label{eq: Residue_2_pole}
 \text{Res}_{s_{ij}=0} \hat{A}_n=  \lim_{\delta\to 0} \Big( \hat{s}_{ij} \, \hat{A}_n \Big)\,.
\ee
In the case of an $[ij]$ pole, one simply switches to an $[i,X\>$-shift, for which $z_{i,j}=-[ij]/[Xj]$.

For a 3-particle pole $s_{ijk}$, we  perform the same $[X,i\>$-shift as in~\reef{2partshift}, and find that
\be  
 \hat{s}_{ijk} = s_{ijk} + z [i| j+ k|X\>
 ~~~\text{vanishes when }~~ 
 z = z_{ijk} \equiv -\frac{s_{ijk}}{[i| j+ k|X\>}\,.
\ee
It is easy to check that no other Mandelstams vanish in this limit. Thus, we proceed analogously as before, and redefine $z = z_{ijk} + \delta$, so that 
\be
 \text{Res}_{s_{ijk}=0} \hat{A}_n=  \lim_{\delta\to 0} \Big( \hat{s}_{ijk} \hat{A}_n \Big)\,.
\ee
In each case, the shifted momenta single out the desired residue and ensure momentum conservation in standard spinor helicity numerics.

\section{Further Details on the 6-point Amplitudes}
\label{app:6-point_ampls}

In this appendix, we provide details of both the MHV and NMHV sectors of the 6-point amplitudes in the EFT expansion of $\mathcal{N}=4$ SYM theory. Concretely, in Section~\ref{app:MHVsec} we outline the computation and ansatz required for the 6-point MHV amplitude, and in Section~\ref{app:details_SYM} we calculate the factorization channels of the NMHV amplitude needed in Section \ref{sec:SYM_6pt_ansatz}.

\subsection{MHV Amplitude}
\label{app:MHVsec}

In this section, we consider the 6-point MHV sector. As in the lower-point cases, the entire 6-point MHV superamplitude is uniquely fixed by a single component amplitude:
\be
  \mathcal{A}^{\text{MHV}}_6[123456]
   = \frac{A_6[--+++\,+]}{\<12\>^4} \, 
  \delta^{(8)}\big( \widetilde{Q}\big) \, .
\ee
The gluon amplitude $A_6[--+++\,+]$ has physical poles on five different channels, given by $s_{16}$, $s_{23}$, $s_{34}$, $s_{45}$, and $s_{56}$. Now, we can follow the same approach as for the 5-point MHV amplitude from Section~\ref{sec:SYM_5pt_ansatz} and consider an ansatz
\begin{align}
\label{MHVamplitudeansa}
A_{6}[--+++\,+]=& \ A^{\text{(PT)}}_6[--+++\,+] \, \Big( V_{6}[123456]-\frac{1}{2}\epsilon[1234]Q_{6,1}[123456] \nonumber \\
&-\frac{1}{2}\epsilon[1235]Q_{6,2}[123456]-\frac{1}{2}\epsilon[1245]Q_{6,3}[123456]-\frac{1}{2}\epsilon[1345]Q_{6,4}[123456] \nonumber \\
&-\frac{1}{2}\epsilon[2345]Q_{6,5}[123456] \Big) \, ,
\end{align}
where $A^{\text{(PT)}}_6[--+++\,+]$ denotes the Parke–Taylor amplitude. Notice that at 6 points we need five different Levi–Civita symbols to account for all possible odd-parity factors. In addition, $V_{6}[123456]$ and the five 
$Q_{6,i}[123456]$ denote generic polynomials in the nine 6-point planar Mandelstam variables $s_{12},\, s_{23},\, s_{34},\, s_{45},\, s_{56},\, s_{16},\, s_{123},\, s_{234},\, s_{345}$.
In general, each polynomial involves $\binom{N+8}{8}$ monomials at order $s^N$, with arbitrary coefficients. As in the 5-point case, we can fix these coefficients by matching to the residues in the physical factorization channels. However, the lack of factorization channels involving two 4-point amplitudes in the MHV sector implies an absence of nonlinear constraints on the 4-point Wilson coefficients $a_{k,q}$.

\subsection{Factorization Channels of the NMHV Amplitude}
\label{app:details_SYM}

In this section, we construct the residues of each factorization channel of the 6-point NMHV amplitude
\be
   \mathbb{Z}_1 \equiv A_6[z_1 z_2 z_3 
   \bar{z}_3 \bar{z}_1 \bar{z}_2] = A_6[1^{12} \, 2^{13} \, 3^{14} \, 4^{23} \, 5^{34} \, 6^{24}]\,.
\ee
There are three different factorization channels in this amplitude, with poles in $s_{34}$, $s_{234}$ and $s_{345}$, respectively. However, since the latter two contain the $s_{34}$ channel themselves, we need to compute both single and double factorizations to lower-point amplitudes. 

Recall that the functions $F(s,u)$ and $f(s,u)$, which capture the higher-derivative corrections to the 4-point amplitude, are defined in~\eqref{eq:def_F_SYM}-\eqref{eq:def_f_SYM}. Additionally, it is useful to note that we can project out any specific component amplitude by taking Grassmann derivatives of the superamplitude,\footnote{Alternatively, they can be obtained by doing Wick contractions of the lines with matching R-indices, resulting in a factor $\<ij\>$ when the states $i$ and $j$ have the same R-index.} as explained in Section~\ref{sec:On-shell_superamps}. Notice, however, that this might introduce relative minus signs between different factorization channels, due to the different order of the Grassmann derivatives. In the following, we have fixed all relative minus signs from the beginning, and in practice it can be simply done by comparing to numerics.

\subsubsection{\texorpdfstring{$s_{234}$}{s234} Pole Diagram}

The $s_{234}$ pole diagram is straightforward to compute because, by SUSY, the exchanged particle has to be a scalar. Introducing $P=P_{234}$ as the momentum exchange, and using that $|-p\> = -|p\>$ and $|-p] = |p]$ with all momenta taken to be outgoing, we have
\begin{align}
  \mathbb{Z}_1 \Big|_{s_{234} = 0}
  &=- \ \begin{tikzpicture}[baseline={([yshift=-0.1cm]current bounding box.center)}] 
	\node[] (a) at (0,0) {};
    \node[] (a1) at ($(a)+(-0.75,0.75)$) {};
    \node[] (a2) at ($(a)+(-1,0)$) {};
    \node[] (a3) at ($(a)+(-0.75,-0.75)$) {};
	\node[] (b) at ($(a)+(3,0)$) {};
    \node[] (b1) at ($(b)+(0.75,0.75)$) {};
    \node[] (b2) at ($(b)+(1,0)$) {};
    \node[] (b3) at ($(b)+(0.75,-0.75)$) {};
	\draw[line width=0.2mm] (a.center) -- (a1.center);
    \node[] at ($(a1)+(-0.2,0.2)$) {$2^{13}$};
    \draw[line width=0.2mm] (a.center) -- (a2.center);
    \node[] at ($(a2)+(-0.35,0)$) {$3^{14}$};
    \draw[line width=0.2mm] (a.center) -- (a3.center);
    \node[] at ($(a3)+(-0.3,-0.2)$) {$4^{23}$};
    \draw[line width=0.2mm] (a.center) -- (b.center);
    \draw[line width=0.2mm] (b.center) -- (b1.center);
    \node[] at ($(b1)+(0.3,0.2)$) {$1^{12}$};
    \draw[line width=0.2mm] (b.center) -- (b2.center);
    \node[] at ($(b2)+(0.35,0)$) {$6^{24}$};
    \draw[line width=0.2mm] (b.center) -- (b3.center);
    \node[] at ($(b3)+(0.3,-0.2)$) {$5^{34}$};
    \node[] at ($(a)+(0.95,-0.3)$) {$(-P)^{24}$};
    \node[] at ($(b)+(-0.7,-0.275)$) {$P^{13}$};
    \filldraw[gray!25] (a.center) circle (0.35cm);
    \draw[line width=0.2mm] (a.center) circle (0.35cm);
    \filldraw[gray!25] (b.center) circle (0.35cm);
    \draw[line width=0.2mm] (b.center) circle (0.35cm);
    \node[] at ($(a)+(0,0)$) {$A_4$};
    \node[] at ($(b)+(0,0)$) {$A_4$};
\end{tikzpicture} \nonumber \\
  &=
  -A_4\big[
  2^{13} \, 3^{14} \, 4^{23} \, (-P)^{24}
  \big] \,
  \frac{1}{s_{234}} \,
  A_4\big[
  5^{34} \, 6^{24} \, 1^{12} \, P^{13}
  \big] \nonumber \\
  &=
  \frac{s_{23} [23] \<24\> \<3P\>}{\<4P\>} \,
  F(s_{23},s_{34})
  \,
  \frac{1}{s_{234}}
  \,
  \frac{-s_{56} \<16\> [56] \<5P\>}{\<1P\>} \,
  F(s_{16},s_{56}) \nonumber \\
  &=
  - \frac{s_{16} s_{23} s_{24} s_{56}}{s_{234}} \,
  F(s_{23},s_{34})
  \, 
  F(s_{16},s_{56}) \,,
\end{align}
where we multiply and divide by $[P2]$ and $[P6]$ to obtain the fourth equality. Since the functions $F(s,u)$ have a $1/(su)$ pole, one might worry that they could create spurious poles in the $s_{16}$, $s_{23}$ and $s_{56}$ channels. However, we nicely see that the prefactor $s_{16} s_{23} s_{56}$ precisely cancels these poles. Using~\reef{eq:def_F_SYM} to express $F(s,u)$ in terms of $f(s,u)$, we can then write the result as
\be
\boxed{
  \label{eq:Res234}
  \begin{aligned}
  \mathbb{Z}_1 \Big|_{s_{234} = 0}
  =& \,
  -g^4 \frac{s_{24}}{s_{34} s_{234}} 
  + g^2 \frac{s_{16} s_{24} s_{56}}{s_{34} s_{234}} \,
  f(s_{16},s_{56})
  + g^2 \frac{s_{23} s_{24}}{s_{234}} \, f(s_{23},s_{34})
  \\
  & \,
  - \frac{s_{16} s_{23} s_{24} s_{56}}{s_{234}} \,
  f(s_{23},s_{34}) \,
  f(s_{16},s_{56}) \,.
 \end{aligned}
 }
\ee
The $\mathcal{O}(s^{-1})$ term is the leading-order contribution from pure SYM. Then, nonlinear terms in the $a_{k,q}$'s arise from the last term, which only begins to contribute at order $\mathcal{O}(s^3)$.

\subsubsection{\texorpdfstring{$s_{345}$}{s345} Pole Diagram}

The $s_{345}$ pole residue can be computed similarly as the $s_{234}$ pole diagram. The result is
\be
\boxed{
  \label{eq:Res345}
  \begin{aligned}
  \mathbb{Z}_1 \Big|_{s_{345} = 0}
  =& \,
  -g^4 \frac{s_{35}}{s_{34} s_{345}} 
   +g^2 \frac{s_{12} s_{16} s_{35}}{s_{34} s_{345}}\,
  f(s_{12},s_{16})
  + g^2 \frac{s_{35} s_{45}}{s_{345}}\,f(s_{34},s_{45})
  \\
  & \,
  - \frac{s_{12} s_{16} s_{35} s_{45}}{s_{345}}\,
  f(s_{34},s_{45})\,
  f(s_{12},s_{16})\,.
 \end{aligned}
 }
\ee

\subsubsection{\texorpdfstring{$s_{34}s_{234}$}{s34s234} Pole Diagram}

Let us consider the diagram with both an $s_{34}$ and an $s_{234}$ pole. With $P=P_{34}$ and $\widetilde{P}=P_{234}$, we obtain
\begin{align}
  \mathbb{Z}_1 \Big|_{s_{34}=s_{234} = 0}
  &= \begin{tikzpicture}[baseline={([yshift=-0.3cm]current bounding box.center)}] 
	\node[] (a) at (0,0) {};
    \node[] (a1) at ($(a)+(-0.75,0.75)$) {};
    \node[] (a3) at ($(a)+(-0.75,-0.75)$) {};
	\node[] (b) at ($(a)+(2.75,0)$) {};
    \node[] (b1) at ($(b)+(0,1)$) {};
    \node[] (c) at ($(b)+(2.75,0)$) {};
    \node[] (c1) at ($(c)+(0.75,0.75)$) {};
    \node[] (c2) at ($(c)+(1,0)$) {};
    \node[] (c3) at ($(c)+(0.75,-0.75)$) {};
	\draw[line width=0.2mm] (a.center) -- (a1.center);
    \node[] at ($(a1)+(-0.2,0.2)$) {$3^{14}$};
    \draw[line width=0.2mm] (a.center) -- (a3.center);
    \node[] at ($(a3)+(-0.3,-0.2)$) {$4^{23}$};
    \draw[line width=0.2mm] (b.center) -- (b1.center);
    \node[] at ($(b1)+(0.1,0.3)$) {$2^{13}$};
    \draw[line width=0.2mm,decoration={coil, amplitude=1.25mm, segment length=1.75mm, aspect=0.75,pre length=0.75em,post length=0.5em},decorate] ($(a)+(0,-0.05)$) -- ($(b)+(0,-0.05)$);
    \draw[line width=0.2mm] (b.center) -- (c.center);
    \draw[line width=0.2mm] (c.center) -- (c1.center);
    \node[] at ($(c1)+(0.3,0.2)$) {$1^{12}$};
    \draw[line width=0.2mm] (c.center) -- (c2.center);
    \node[] at ($(c2)+(0.35,0)$) {$6^{24}$};
    \draw[line width=0.2mm] (c.center) -- (c3.center);
    \node[] at ($(c3)+(0.3,-0.2)$) {$5^{34}$};
    \node[] at ($(a)+(0.95,-0.45)$) {$(-P)^{\pm}$};
    \node[] at ($(b)+(-0.7,-0.425)$) {$P^{\mp}$};
    \node[] at ($(b)+(0.95,-0.3)$) {$(-\widetilde{P})^{24}$};
    \node[] at ($(c)+(-0.7,-0.275)$) {$\widetilde{P}^{13}$};
    \filldraw[gray!25] (a.center) circle (0.35cm);
    \draw[line width=0.2mm] (a.center) circle (0.35cm);
    \filldraw[gray!25] (b.center) circle (0.35cm);
    \draw[line width=0.2mm] (b.center) circle (0.35cm);
    \filldraw[gray!25] (c.center) circle (0.35cm);
    \draw[line width=0.2mm] (c.center) circle (0.35cm);
    \node[] at ($(a)+(0,0)$) {$A_3$};
    \node[] at ($(b)+(0,0)$) {$A_3$};
    \node[] at ($(c)+(0,0)$) {$A_4$};
\end{tikzpicture} \nonumber \\
  &=
  A_{3}\big[3^{14} \, 4^{23} \, (-P)^{\pm}\big]  \,
  \frac{1}{s_{34}}  \,
  A_{3}\big[2^{13} \, P^{\mp} \, (-\widetilde{P})^{24}\big]  \,
  \frac{1}{s_{234}}  \,
  A_{4}\big[5^{34} \, 6^{24} \, 1^{12} \, \widetilde{P}^{13} \big] \,. 
\end{align}
The two helicity options $(-P)^\pm$ for the internal gluon should be considered separately. They simply differ by whether $s_{34} = 0$ is obtained by setting $\<34\> = 0$ or $[34] = 0$, respectively. Let us consider first the case of $[34] = 0$:
{\allowdisplaybreaks
\begin{align}
  \mathbb{Z}_1 \Big|_{[34]=s_{234} = 0}
  &=
  A_{3}\big[3^{14} \, 4^{23} \, (-P)^-\big]  \,
  \frac{1}{s_{34}}  \,
  A_{3}\big[2^{13} \, P^+ \, (-\widetilde{P})^{24}\big]  \,
  \frac{1}{s_{234}}  \,
  A_{4}\big[5^{34} \, 6^{24} \, 1^{12} \, \widetilde{P}^{13} \big]
  \nonumber \\
  &=
  \frac{-g \, \<P3\>\<P4\>}{\<34\>} \,
  \frac{1}{s_{34}} \,
  \frac{-g \, [2P][\widetilde{P}P]}{[\widetilde{P}2]} \,
  \frac{1}{s_{234}} \,
  \frac{-s_{56} \<16\> [56] \<5 \widetilde{P}\>}{\<1 \widetilde{P}\>} \,
  F(s_{16}, s_{56})
  \nonumber \\
  &= g^2\frac{[2|P|3\> [\widetilde{P}|P|4\>}{\<34\> [\widetilde{P}2]} \, \frac{1}{s_{34} s_{234}} \,
  \frac{-s_{56} \<16\> [56] [6| \widetilde{P}|5\>}{[6| \widetilde{P}|1\>}\,
  F(s_{16}, s_{56})
  \nonumber \\
  &= g^2\frac{[24]\<43\> [\widetilde{P}3]}{[\widetilde{P}2]} \, \frac{1}{s_{34} s_{234}} \,
  s_{16} s_{56}\,
  F(s_{16}, s_{56})
  \nonumber \\
  &=
  g^2 \frac{s_{16} s_{24} s_{56}}{s_{34} s_{234}} \,
  F(s_{16}, s_{56}) \,,
\end{align}
}
\!\!where we have used that $s_{156}=s_{234}=0$ on the pole, and have multiplied and divided by $\< 4\widetilde{P} \>$ and $[6 \widetilde{P}]$. If we reverse the helicity designation for the internal gluon, we get exactly the same result. Since the result above is already symmetric in $\<34\>$ and $[34]$, we can use it to simultaneously capture both internal gluon helicity assignments. Expanding $F(s,u)$, we consequently have
\be
  \boxed{
  \label{eq:Res34and234}
  \mathbb{Z}_1 \Big|_{s_{34}=s_{234} = 0}
  =
  - g^4 \frac{s_{24}}{s_{34}s_{234}}
  + g^2 \frac{s_{16} s_{24} s_{56}}{s_{34}s_{234}} \,
  f(s_{16}, s_{56})\,.
  }
\ee
As can be seen, this result is totally compatible with~\eqref{eq:Res234} upon taking a further $s_{34}$ pole.

\subsubsection{\texorpdfstring{$s_{34}s_{345}$}{s34s345} Pole Diagram}
By direct computation we similarly find
\be
  \boxed{
  \label{eq:Res34and345}
  \mathbb{Z}_1 \Big|_{s_{34}=s_{345} = 0}
  =
  - g^4 \frac{s_{35}}{s_{34}s_{345}}
  + g^2 \frac{s_{12} s_{16} s_{35}}{s_{34}s_{345}} \,
  f(s_{12}, s_{16})\,.
  }
\ee
Just as before, it captures both helicity options for the internal gluon, and is consistent with the $s_{345}$ pole diagram from~\eqref{eq:Res345}.

\subsubsection{\texorpdfstring{$s_{34}$}{s34} Pole Diagram}\label{sec:s34pole}

Lastly, let us consider the $s_{34}$ pole diagram. Again, we have a choice of whether to set $\<34\>=0$ or $[34]=0$ for the pole, which simply amounts to whether the internal exchanged particle is a positive- or negative-helicity gluon on the 3-particle vertex, respectively. Specifically, introducing $P=P_{34}$, we have
\begin{align}
  \label{s34A}
  \mathbb{Z}_1 \Big|_{s_{34}=0}
  &= \begin{tikzpicture}[baseline={([yshift=-0.1cm]current bounding box.center)}] 
	\node[] (a) at (0,0) {};
    \node[] (a1) at ($(a)+(-0.75,0.75)$) {};
    \node[] (a3) at ($(a)+(-0.75,-0.75)$) {};
	\node[] (b) at ($(a)+(3,0)$) {};
    \node[] (b1) at ($(b)+(0.7,0.85)$) {};
    \node[] (b2) at ($(b)+(0.95,0.3)$) {};
    \node[] (b3) at ($(b)+(0.95,-0.3)$) {};
    \node[] (b4) at ($(b)+(0.7,-0.85)$) {};
	\draw[line width=0.2mm] (a.center) -- (a1.center);
    \node[] at ($(a1)+(-0.2,0.2)$) {$3^{14}$};
    \draw[line width=0.2mm] (a.center) -- (a3.center);
    \node[] at ($(a3)+(-0.3,-0.2)$) {$4^{23}$};
    \draw[line width=0.2mm,decoration={coil, amplitude=1.25mm, segment length=1.75mm, aspect=0.75,pre length=0.75em,post length=0.5em},decorate] ($(a)+(0,-0.05)$) -- ($(b)+(0,-0.05)$);
    \draw[line width=0.2mm] (b.center) -- (b1.center);
    \node[] at ($(b1)+(0.3,0.2)$) {$2^{13}$};
    \draw[line width=0.2mm] (b.center) -- (b2.center);
    \node[] at ($(b2)+(0.35,0.1)$) {$1^{12}$};
    \draw[line width=0.2mm] (b.center) -- (b3.center);
    \node[] at ($(b3)+(0.35,0)$) {$6^{24}$};
    \draw[line width=0.2mm] (b.center) -- (b4.center);
    \node[] at ($(b4)+(0.3,-0.2)$) {$5^{34}$};
    \node[] at ($(a)+(0.95,-0.45)$) {$(-P)^{\pm}$};
    \node[] at ($(b)+(-0.7,-0.425)$) {$P^{\mp}$};
    \filldraw[gray!25] (a.center) circle (0.35cm);
    \draw[line width=0.2mm] (a.center) circle (0.35cm);
    \filldraw[gray!25] (b.center) circle (0.35cm);
    \draw[line width=0.2mm] (b.center) circle (0.35cm);
    \node[] at ($(a)+(0,0)$) {$A_3$};
    \node[] at ($(b)+(0,0)$) {$A_5$};
\end{tikzpicture} \nonumber \\
  &=
  A_3\big[3^{14} \, 4^{23} \, (-P)^\pm\big]  \,
  \frac{1}{s_{34}}  \,
  A_5\big[5^{34} \, 6^{24} \, 1^{12} \, 2^{13} \, P^\mp\big] \,.
\end{align}
Depending on the helicity of the gluon, $A_5\big[5^{34} \, 6^{24} \, 1^{12} \, 2^{13} \, P^\mp\big]$ is either a 5-point NMHV or MHV amplitude, respectively. The MHV amplitude can be directly obtained from the superamplitude in~\reef{eq:superamplitude_A5}--\reef{EFT5ptansatz}, while the NMHV component can be obtained instead via conjugation. In particular, using ${A_5}^*$ to denote conjugation, we have
\begin{align}
  A_5\big[5^{34}\, 6^{24}\, 1^{12}\, 2^{13}\, P^+\big]
  &=
  \frac{\<25\>}
  {\<P2\>\<P5\>} \,
  G_5[5612P]\,, \\
  A_5\big[5^{34}\, 6^{24}\, 1^{12}\, 2^{13}\, P^-\big]
  &= {A_5\big[5^{12}\, 6^{13}\, 1^{34}\, 2^{24}\, P^+ \big]}^*
  =
  \frac{-[25]}
  {[P2][P5]} \,
  {G_5[5612P]}^*\,,
\end{align}
where, importantly, conjugation flips the sign of the Levi–Civita in $G_5$. Taking the case of \reef{s34A} with a negative-helicity gluon on the 3-point vertex, we find 
\begin{align}
\label{s34B}
  \mathbb{Z}_1 \Big|_{[34]=0}
  &=
  g \, \frac{-\<3P\> \<4P\>}{\<34\>} \,
  \frac{1}{s_{34}} \,
  \frac{\<25\>}
  {\<P2\>\<P5\>} \,
  G_5[5612P]
  \nonumber \\
  &=
  -g \, \frac{\<25\> [2|P|3\> [5|P|4\>}
  {s_{34} \<34\> [2|P|2\> [5|P|5\>} \,
  G_5[5612P]
  \nonumber \\
  &=
  -g \, \frac{[24] \<25\> [35]}
  {[34] (s_{23}+s_{24})(s_{35}+s_{45})}  \,
  G_5[5612P] \,,
\end{align}
where we multiply and divide by $[2P][5P]$ to reach the second equality. Now, on the pole $s_{34}=0$, we have
$s_{23} + s_{24} = s_{234}$ and
$s_{35} + s_{45} = s_{345}$. Hence, we can write
\be
  \label{s34C}
  \mathbb{Z}_1 \Big|_{[34]=0}
  = 
  - g \, \frac{[24] \<25\> [35]}
  {[34] s_{234} s_{345}} \,
  G_5[5612P] \,,
\ee
and similarly the $\<34\>=0$ result is obtained by conjugation.

There is a different way of rewriting the $[34]=0$ residue, though, which resonates better with the previous factorization channels. Starting with the first line in  \reef{s34B}, we can use two rounds of the Schouten identity in the numerator. Firstly, we replace
$\<4P\> \<25\>=-\<42\> \<5P\> - \<45\> \<P2\>$ to get
\be
  \label{s34D}
  \mathbb{Z}_1 \Big|_{[34]=0}
  =
  g \, \frac{\<3P\>}{s_{34} \<34\>}
  \bigg(
  \frac{\<24\>}{\<P2\>} 
   +
  \frac{\<45\>}{\<P5\>}
  \bigg) \,
  G_5[5612P]\,.
\ee
Secondly, we use the Schouten identity in both $\<3P\>\<24\>$ and $\<3P\>\<45\>$. This produces two terms with $\<34\>$ that cancel between them, and we are left with
\begin{align}
  \label{s34E}
  \mathbb{Z}_1 \Big|_{[34]=0}
  &=
  -g \, \frac{1}{s_{34} \<34\>}
  \bigg(
  \frac{\<23\>\<4P\>}{\<2P\>} 
   + 
  \frac{\<35\>\<4P\>}{\<5P\>}
  \bigg) \,
  G_5[5612P] \nonumber \\
  &=
  - g \, \frac{1}{s_{34} \<34\>}
  \bigg(
  \frac{\<23\>[2|P|4\>}{[2|P|2\>} 
   + 
  \frac{\<35\>[5|P|4\>}{[5|P|5\>}
  \bigg) \,
  G_5[5612P] \nonumber \\
  &=
  g \, \frac{1}{s_{34}}
  \bigg(
  \frac{s_{23}}{s_{234}} 
   - 
  \frac{s_{35}}{s_{345}}
  \bigg) \,
  G_5[5612P]\,.
\end{align}
Replacing $s_{23} = s_{234} - s_{24}$, this finally simplifies to
\be
  \label{eq:Ressq34}
\boxed{
  \mathbb{Z}_1 \Big|_{[34]=0}
  =
  g \, \bigg(
  \frac{1}{s_{34}} - \frac{s_{24}}{s_{34}s_{234}} 
   - 
  \frac{s_{35}}{s_{34}s_{345}}
  \bigg) \,
  G_5[5612P]\,.
  }
\ee
The $\<34\>=0$ residue is the same, but with the opposite sign for the Levi–Civita term in $G_5[5612P]$. As can be seen, the leading-order term is manifestly the same as in~\eqref{LO}, which was obtained via BCFW recursion.

\section{Comparison to the Open Superstring Amplitude}
\label{app:comparison_string}

In this appendix, we summarize the result for the low-energy expansion of the $n$-point amplitudes in open superstring theory, which can be explicitly compared with our 5- and 6-point generic EFT amplitudes upon choosing the 4-point Veneziano amplitude as the appropriate UV completion. Hence, this comparison serves as a sanity check of our results.\footnote{Note that in the string theory literature, the convention for the Mandelstam variables differs from ours by a minus sign.}

As shown in~\cite{Mafra:2011nv,Mafra:2011nw} (see also~\cite{Stieberger:2006bh,Stieberger:2006te,Broedel:2009nsh,Broedel:2013aza}), the low-energy $\alpha'$-expansion of an $n$-point amplitude in open superstring theory can be written to all orders in terms of Yang–Mills amplitudes as
\begin{equation}
\label{eq: expansion_F_string}
    A_n^{\text{string}}[12\dots n] = \sum_{\sigma \in S_{n-3}}  A_n^{\text{YM}}[1 \, 2_\sigma \, \dots (n-2)_\sigma \, (n-1) \,  n]\, F^\sigma_{n} (\alpha') \,.
\end{equation}
Here, $\sigma$ denotes the $(n-3)!$ permutations of the labels in the canonical order, and the functions $F^\sigma_{n}$ correspond to generalized Euler integrals that encode the $\alpha'$ dependence. In particular, their low-energy expansion takes the form
\begin{equation}
    F^\sigma_{n} (\alpha') = 1 + \alpha'^2 \zeta_2 \, P_{n,2}^\sigma + \alpha'^3 \zeta_3 \, P_{n,3}^\sigma + \alpha'^4 \zeta_2^2 \, P_{n,4}^\sigma + \alpha'^5 \zeta_5 \, P_{n,5}^\sigma + \alpha'^5 \zeta_2 \zeta_3 \, Q_{n,5}^\sigma + \dots \,,
\end{equation}
where $P_{n,i}^\sigma$ and $Q_{n,i}^\sigma$ are polynomials of the Mandelstam variables, which can be found to higher orders in~\cite{Stieberger_website}. Importantly, as already noted in Section~\ref{sec:SYM_nonlinear_constraints}, at order $\alpha'^8$ in the expansion there appears the first multi-zeta value $\zeta_{3,5}$~\cite{Stieberger:2006te,Mafra:2011nw,Schlotterer:2012ny,Broedel:2013aza}, which cannot be algebraically expressed in terms of the single zeta values.

First of all, for the case of $n=4$, there is only one summand in~\eqref{eq: expansion_F_string}, and the function $F_{4}^{(2)}$ is given by
\begin{equation}
    F_{4}^{(2)} = \frac{\Gamma(1+\alpha's) \Gamma(1+\alpha'u)}{\Gamma(1+\alpha's+\alpha'u)} = 1 - \alpha'^2 \zeta_2 \,  s u - \alpha'^3 \zeta_3 \, su(s+u) + \mathcal{O}(\alpha'^4)\,.
\end{equation}
Introducing the Yang–Mills 4-point amplitude into~\eqref{eq: expansion_F_string}, one recovers the Veneziano amplitude.

For $n=5$ we have two permutations, with the corresponding set of functions being
\begin{equation}
\begin{split}
    F_{5}^{(23)} =&\, 1+\alpha'^2 \zeta_2 (s_{12}s_{34}-s_{34}s_{45}-s_{12}s_{15}) -\alpha'^3 \zeta_3 (s_{12}^2 s_{34}+2 s_{12} s_{23} s_{34} \\
&\, +s_{12} s_{34}^2-s_{34}^2 s_{45}-s_{34} s_{45}^2-s_{12}^2 s_{15}-s_{12} s_{15}^2) + \mathcal{O}(\alpha'^4)\,, \\
    F_{5}^{(32)} =&\, \alpha'^2 \zeta_2 \, s_{13} s_{24}-\alpha'^3 \zeta_3 \, s_{13} s_{24} (s_{12}+s_{23}+s_{34}+s_{45}+s_{15}) + \mathcal{O}(\alpha'^4)\,.
\end{split}
\end{equation}
Choosing the basis of Parke–Taylor amplitudes $A_5^{\text{YM}}[--+++]$ and its permutations, we have verified that the corresponding string amplitudes are in complete agreement with our 5-point result from Section~\ref{sec:SYM_5pt_ansatz} when the 4-point Wilson coefficients are fixed to be those from~\eqref{eq:Wilson_coeffs_Veneziano} for the Veneziano amplitude. Concretely, we reproduce the low-energy expansion of the string amplitude up to order $\alpha'^7$ in $F_5^\sigma$, or $p^{13}$ in the amplitude, where our result contains the first free coefficient; recall Table~\ref{tab:solution_ansatz_5pt}.

At 6 point, the six functions are given by
\begin{equation}
\begin{split}
    F_{6}^{(234)} =&\, 1- \alpha'^2 \zeta_2\, (s_{45}s_{56}+s_{12}s_{16}-s_{45}s_{123}-s_{12}s_{345}+s_{123}s_{345}) +\alpha'^3 \zeta_3 \, (2 s_{12} s_{23} s_{45} \\
& +2 s_{12} s_{34} s_{45} +s_{45}^2 s_{56}+s_{45} s_{56}^2+s_{12}^2 s_{16}+s_{12} s_{16}^2-2 s_{34} s_{45} s_{123} -s_{45}^2 s_{123}-s_{45} s_{123}^2\\
& -2 s_{12} s_{45} s_{234}-s_{12}^2 s_{345}-2 s_{12} s_{23} s_{345}+s_{123}^2 s_{345}-s_{12} s_{345}^2+ s_{123} s_{345}^2) + \mathcal{O}(\alpha'^4), \\
F_{6}^{(243)} =& -\alpha'^2 \zeta_2\, s_{35}(s_{34}-s_{56}+s_{123})+\alpha'^3 \zeta_3\, s_{35} (-2 s_{12} s_{23}-2 s_{12} s_{34}+s_{34}^2+s_{34} s_{45}-s_{45} s_{56} \\
& -s_{56}^2 +2 s_{34} s_{123}+s_{45} s_{123}+s_{123}^2+2 s_{12} s_{234}+s_{34} s_{345}-s_{56} s_{345}+s_{123} s_{345}) + \mathcal{O}(\alpha'^4), \\
    F_{6}^{(324)} =& -\alpha'^2 \zeta_2\, s_{13}(s_{23}-s_{16}+s_{345})+\alpha'^3 \zeta_3\, s_{13} (s_{12} s_{23}+s_{23}^2-2 s_{23} s_{45}-2 s_{34} s_{45}-s_{12} s_{16} \\
& -s_{16}^2 +s_{23} s_{123}-s_{16} s_{123}+2 s_{45} s_{234}+s_{12} s_{345}+2 s_{23} s_{345}+s_{123} s_{345}+s_{345}^2) + \mathcal{O}(\alpha'^4), \\
F_{6}^{(342)} =&\, \alpha'^2 \zeta_2\, s_{13}s_{25}+\alpha'^3 \zeta_3\, s_{13}s_{25} (-s_{12}+s_{23}+2 s_{34}-s_{16}-s_{123}-2 s_{234}-s_{345}) + \mathcal{O}(\alpha'^4), \\
F_{6}^{(423)} =&\, \alpha'^2 \zeta_2\,  s_{14}s_{35}+\alpha'^3 \zeta_3\, s_{14}s_{35}(2 s_{23}+s_{34}-s_{45}-s_{56}-s_{123}-2 s_{234}-s_{345}) + \mathcal{O}(\alpha'^4),\\
F_{6}^{(432)} =& -\alpha'^2 \zeta_2\, s_{14}s_{25}+\alpha'^3 \zeta_3\, s_{14}s_{25} (-s_{23}-s_{34}+s_{56}+s_{16}+s_{123}+s_{234}+s_{345}) + \mathcal{O}(\alpha'^4).
\end{split}
\end{equation}
In this case, we can for example choose a basis of Yang–Mills amplitudes made of the split-helicity amplitude $A_6^{\text{YM}}[+---++]$ and its permutations. This choice of basis is particularly convenient, as it preserves the same helicity structure for all amplitudes in the basis, which can thus be simply obtained by relabeling of the external legs. We have verified that the low-energy expansion of the 6-point string amplitude is reproduced by our 6-point EFT amplitude from Sections~\ref{sec:SYM_6pt_ansatz}--\ref{sec:SYM_nonlinear_constraints} (fixing the 4-point Wilson coefficients to be those from the Veneziano amplitude) up to order $\alpha'^7$ in $F_6^\sigma$, or $s^{6}$ in the amplitude, where our result contains the first free parameter; recall Table~\ref{tab:6pt_CTs}.

\section{An Abelian Example: \texorpdfstring{$\mathcal{N}=4$}{N=4} SUSY Dirac-Born-Infeld Theory}
\label{app:DBI}

In this appendix, we consider the case of an $\mathcal{N}=4$ SUSY Abelian theory with higher-derivative EFT corrections, with the aim of investigating if nonlinear constraints can also arise in uncolored theories. In our case, abelian amplitudes can be simply obtained by summing amplitudes from the $\mathcal{N}=4$ SYM EFT over all inequivalent color orderings, a process known as abelianization; see e.g.~\cite{Berman:2024eid}. For instance, we have
\begin{equation}
    A^{\text{Abelian}}[1234] = A_4[1234] + A_4[1342] + A_4[1423] \,.
\end{equation}
Applying it to the 4-point amplitude from~\eqref{eq: scalar_ampls_SYM}, we thus obtain
\begin{align}
\label{eq: 4pt_ampl_DBI}
A^{\text{Abelian}}_4[zz\bar{z}\bar{z}] =&\, s^2 \Big( F(s,u) + F(t,s) + F(u,t) \Big) \nonumber \\
=&\, s^2 \Big( f(s,u) + f(t,s) + f(u,t) \Big)\,,
\end{align}
where the leading-order poles cancel. This can be traced to the fact that all amplitudes with an odd number of external legs vanish in the Abelian theory, including the 3-point vertex, such that the 4-point amplitude becomes completely local. Moreover, as can be seen, the 4-point Abelian amplitude is fully permutation symmetric in the indistinguishable legs, as expected. Therefore, let us define a new $stu$-symmetric function $g(s,t,u)\equiv f(s,u) + f(t,s) + f(u,t)$ that encodes the higher-derivative corrections,
\begin{align}
\label{eq:def_g_DBI}
g(s,t,u)=&\, \pi^2 \sum_{n_1,n_2\geq0 } C_{n_1,n_2}^{(3n_1 + 2 n_2)} {(stu)}^{n_1} {(s^2 + t^2 + u^2)}^{n_2} \nonumber \\
=&\, \pi^2 \Big( C_{0,0}^{(0)} + C_{0,1}^{(2)} \, (s^2 + t^2 + u^2) + C_{1,0}^{(3)} \, stu + C_{0,2}^{(4)} \, (s^2 + t^2 + u^2)^2 + \dots \Big) \,,
\end{align}
with $C_{n_1,n_2}^{(3n_1 + 2 n_2)}$ denoting the new 4-point Wilson coefficients. For convenience, we pull out a factor of $\pi^2$ to render their transcendental weight equal to $3n_1 + 2 n_2$, which also matches the order in $s$ at which they enter in $g(s,t,u)$.

A particular example of a UV completion is obtained by abelianizing the Veneziano amplitude~\eqref{eq:venezianointro},
\be
g(s,t,u)^{\text{super-DBI}} =
-\alpha'^{2} \left[ \frac{
\Gamma(-\alpha' s)
\Gamma(-\alpha'u)}
{\Gamma(1+\alpha't)} + \frac{
\Gamma(-\alpha' t)
\Gamma(-\alpha's)}
{\Gamma(1+\alpha'u)} + \frac{
\Gamma(-\alpha' u)
\Gamma(-\alpha't)}
{\Gamma(1+\alpha's)} \right].
\ee
This corresponds to the $\mathcal{N}=4$ SUSY completion of Dirac-Born-Infeld ($\mathcal{N}=4$ super-DBI) theory, which describes the dynamics of massless fields living on D-branes in open superstring theory. The first Wilson coefficients in this case are given by
\begin{equation}
    C_{0,0}^{(0)} = \frac{\alpha'^2}{2}\,, \quad \enspace C_{0,1}^{(2)} = \frac{\zeta_2}{8} \alpha'^4= \frac{\pi^2}{48} \alpha'^4\,, \quad \enspace C_{1,0}^{(3)} = \frac{\zeta_3}{2} \alpha'^5\,, \quad \enspace C_{0,2}^{(4)} = \frac{3 \zeta_4}{32} \alpha'^6 = \frac{\pi^4}{960} \alpha'^6\,.
\end{equation}
Based on this particular UV completion and a simple power counting, if nonlinear constraints exist among these Wilson coefficients, we would expect them to arise first at order $s^7$ in the 6-point amplitude, where there could be a mixing between $C_{0,2}^{(4)}$ and $(C_{0,1}^{(2)})^2$.

With this goal, we can follow the same procedure as we did for the SYM EFT, and first calculate the 6-point amplitude
\be
   \mathbb{Z}_1 \equiv A_6[z_1 z_2 z_3 
   \bar{z}_3 \bar{z}_1 \bar{z}_2] = A_6[1^{12} \, 2^{13} \, 3^{14} \, 4^{23} \, 5^{34} \, 6^{24}]
\ee
in the EFT expansion of Abelian $\mathcal{N}=4$ SUSY theory, and later subject it to the 6-point NMHV SUSY Ward identity~\eqref{consistency}. To begin with, since 3-point amplitudes vanish in this theory, $\mathbb{Z}_1$ only has factorization channels involving 3-particle poles, which necessarily involve a scalar particle exchange. In total, there are 6 distinct (permutation-inequivalent) factorizations, with poles in $s_{125}$, $s_{134}$, $s_{135}$, $s_{145}$, $s_{234}$ and $s_{345}$, respectively. In this case, one can obtain the entire 6-point amplitude simply by combining the factorization channels without the need for an ansatz,
\begin{align}
    \mathbb{Z}_1 =& - \frac{s_{12} s_{25} s_{36} s_{46}}{s_{125}} 
  g(s_{12},s_{15},s_{25})
  g(s_{34},s_{36},s_{46})  - \frac{s_{13} s_{14} s_{25} s_{56}}{s_{134}} 
  g(s_{13},s_{14},s_{34}) 
  g(s_{25},s_{26},s_{56}) \nonumber \\
  & - \frac{s_{13} s_{24} s_{35} s_{46}}{s_{135}}
  g(s_{13},s_{15},s_{35})
  g(s_{24},s_{26},s_{46}) - \frac{s_{14} s_{23} s_{36} s_{45}}{s_{145}}
  g(s_{14},s_{15},s_{45})
  g(s_{23},s_{26},s_{36}) \nonumber \\
  & - \frac{s_{16} s_{23} s_{24} s_{56}}{s_{234}}
  g(s_{15},s_{16},s_{56})
  g(s_{23},s_{24},s_{34}) - \frac{s_{12} s_{16} s_{35} s_{45}}{s_{345}}
  g(s_{12},s_{16},s_{26}) 
  g(s_{34},s_{35},s_{45}) \,
\end{align}
plus contact terms. 
Nevertheless, solving the 6-point NMHV SUSY Ward identities up to order $\mathcal{O}(s^7)$ does not produce nonlinear constraints on the 4-point Wilson coefficients in this case.

\bibliography{maxSUSYrefs}
\bibliographystyle{JHEP}

\end{document}